\documentclass[twocolumn]{aastex63}
\usepackage{lineno}
\usepackage{amsmath}
\usepackage{amssymb}
\usepackage{epsfig}
\usepackage{natbib}
\usepackage{graphicx}
\usepackage{color}
\usepackage{tabularx}
\usepackage{epsfig}
\usepackage{rotating}
\usepackage{multirow}
\usepackage{subfigure}
\usepackage{ulem}
\usepackage{longtable}
\usepackage{url}
\def\UrlBreaks{\do\A\do\B\do\C\do\D\do\E\do\F\do\G\do\H\do\I\do\J
\do\K\do\L\do\M\do\N\do\O\do\P\do\Q\do\R\do\S\do\T\do\U\do\V
\do\W\do\X\do\Y\do\Z\do\[\do\\\do\]\do\^\do\_\do\`\do\a\do\b
\do\c\do\d\do\e\do\f\do\g\do\h\do\i\do\j\do\k\do\l\do\m\do\n
\do\o\do\p\do\q\do\r\do\s\do\t\do\u\do\v\do\w\do\x\do\y\do\z
\do\.\do\@\do\\\do\/\do\!\do\_\do\|\do\;\do\>\do\]\do\)\do\,
\do\?\do\'\do+\do\=\do\#}

\newcommand       \G            {{\rm G}}
\newcommand       \GBP       {G_{\rm BP}}
\newcommand       \GRP       {G_{\rm RP}}
\newcommand       \mum        {\,{\rm \mu m}}
\newcommand       \Ks           {{ K_{\rm S}}}
\newcommand       \K             {\,{\rm K}}
\newcommand       \Teff           {{ T_{\rm eff}}}

\newcommand{\AV}{A_{V}}
\newcommand{\Av}{A_{V}}

\newcommand{\AKs}{A_{K_{\rm S}}}

\newcommand{\Mr}{M_{r}}
\newcommand{\MJ}{M_{J}}
\newcommand{\MK}{M_{K}}
\newcommand{\MKs}{M_{K_{\rm S}}}
\newcommand{\MG}{M_{G}}
\newcommand{\mKs}{m_{K_{\rm S}}}

\received{2021 May}
\accepted{2021 August}
\submitjournal{ApJ}

\shorttitle{3D parameter Maps of RCs}
\shortauthors{Wang \& Chen}

\begin{document}

\title{3D Parameter Maps of Red Clump Stars in the Milky Way -- Absolute Magnitudes and Intrinsic Colors}

\correspondingauthor{Shu Wang, Xiaodian Chen}
\email{shuwang@nao.cas.cn, chenxiaodian@nao.cas.cn}

\author[0000-0003-4489-9794]{Shu Wang}
\affiliation{CAS Key Laboratory of Optical Astronomy, National Astronomical Observatories, 
Chinese Academy of Sciences, Beijing 100101, China}
\affiliation{Department of Astronomy, China West Normal University, Nanchong, China}

\author[0000-0001-7084-0484]{Xiaodian Chen}
\affiliation{CAS Key Laboratory of Optical Astronomy, National Astronomical Observatories, 
Chinese Academy of Sciences, Beijing 100101, China}
\affiliation{Department of Astronomy, China West Normal University, Nanchong, China} \affiliation{School of Astronomy and Space Science, University of the Chinese Academy of Sciences, Beijing 101408, China}

\begin{abstract}

Red clump stars (RCs) are useful tracers of distances, extinction, chemical abundances, and Galactic structures and kinematics. 
Accurate estimation of the RC parameters---absolute magnitude and intrinsic color---is the basis for obtaining high-precision RC distances. 
By combining astrometric data from {\it Gaia}, spectroscopic data from APOGEE and LAMOST, and multi-band photometric data from {\it Gaia}, APASS, Pan-STARRS1, 2MASS, and {\it WISE} surveys, we use the Gaussian process regression to train machine learners to derive the multi-band absolute magnitudes $M_\lambda$ and intrinsic colors $(\lambda_1-\lambda_2)_0$ for each spectral RC. 
The dependence of $M_\lambda$ on metallicity decreases from optical to infrared bands, while the dependence of $M_\lambda$ on age is relatively similar in each band. 
$(\lambda_1-\lambda_2)_0$ are more affected by metallicity than age. 
The RC parameters are not suitable to be represented by simple constants but are related to the Galactic stellar population structure.
By analyzing the variation of $M_\lambda$ and $(\lambda_1-\lambda_2)_0$ in the spatial distribution, we construct ($R, z$) dependent maps of mean absolute magnitudes and mean intrinsic colors of the Galactic RCs. 
Through external and internal validation, we find that using three-dimensional (3D) parameter maps to determine RC parameters avoids systematic bias and reduces dispersion by about $20\%$ compared to using constant parameters.
Based on {\it Gaia}'s EDR3 parallax, our 3D parameter maps, and extinction--distance profile selection, we obtain a photometric RC sample containing 11 million stars with distance and extinction measurements.

\end{abstract}


\section{Introduction}\label{intro}

Red clump stars (RCs) with almost constant brightness have long been regarded as valuable indicators and are used widely for (1) estimating distances \citep[e.g.,][]{1998ApJ...494L.219P, 1998ApJ...500L.141S, 2000ApJ...539..732A, 2000ApJ...531L..25U, 2002AJ....123..244K, 2007MNRAS.378.1064R, 2007MNRAS.380.1255R, 2012MNRAS.419.1637L, 2020A&A...639A..72W}, 
(2) investigating the reddening and extinction \citep[e.g.,][]{2005ApJ...619..931I, 2009ApJ...707...89G, 2009ApJ...696.1407N, 2009ApJ...707..510Z, 2013ApJ...769...88N, 2016MNRAS.456.2692N, 2017ApJ...848..106W, 2019ApJ...877..116W, 2020NatAs...4..377N}, 
and (3) studying the structures of our Galaxy and nearby dwarf galaxies \citep[e.g.,][]{2002A&A...394..883L, 2007MNRAS.378.1064R, 2010ApJ...724.1491M, 2010ApJ...721L..28N, 2013MNRAS.435.1874W, 2016ApJ...823...30B, 2020MNRAS.491.2104W}.
However, the absolute magnitude of RCs has been reported to depend on the age and metallicity in the literature \citep{1999AJ....118.2321S, 1999AJ....117.1816T, 2001MNRAS.323..109G, 2002MNRAS.337..332S, 2007A&A...463..559V, 2010AJ....140.1038P, 2016ARA&A..54...95G, 2019MNRAS.486.5600O}. 
The use of constant absolute magnitude in studying distances and structures can lead to significantly systematic biases. This problem is mitigated if spectroscopic data is available, since the age of RC can be inferred from elemental abundances, such as [C/N] and [$\alpha$/M] \citep{2016ApJ...823..114N}. Nevertheless, an optimal absolute magnitude is still expected, which can be used for a large number of RCs without spectroscopic information.

\citet{2000ApJ...539..732A} used {\it Hipparcos} RCs to investigate the $K$-band absolute magnitude and found no correlation between $\MK$ and [Fe/H].  
Based on the study of $\sim$ 200 nearby RCs, \citet{2012MNRAS.419.1637L} agreed with the result of \citet{2000ApJ...539..732A}. 
\citet{2001MNRAS.323..109G} and \citet{2016ARA&A..54...95G} summarized that the population effects on RC's absolute magnitude exist in any band. 
\citet{2017ApJ...840...77C} investigated the RC absolute magnitudes from optical to infrared (IR) bands and found a clear trend between absolute magnitudes and ages.
With the wealth of available astrometric data from {\it Gaia} and spectroscopic data from APOGEE \citep[the Apache Point Observatory Galactic Evolution Experiment,][]{2011AJ....142...72E,2017AJ....154...94M}, LAMOST \citep[the Large Sky Area Multi-Object Fiber Spectroscopic Telescope,][]{2012RAA....12.1197C,2012RAA....12..735D,2012RAA....12..723Z}, GALAH \citep[the GALactic Archaeology with HERMES1,][]{2015MNRAS.449.2604D}, there have been many works devoted to studying the RC absolute magnitudes in recent years.
Based on the {\it Gaia} TGAS parallaxes, \citet{2017MNRAS.471..722H} determined the RC absolute magnitudes in 2MASS \citep[The Two Micron All Sky Survey,][]{Cohen2003AJ....126.1090C}, {\it Gaia} $G$, and {\it WISE} \citep[Wide-field Infrared Survey Explorer,][]{2010AJ....140.1868W}
bands. 
\citet{2018A&A...609A.116R} derived RC absolute magnitudes and discussed their dependence on color. 
Later, with {\it Gaia} DR2 and APOGEE data, \citet{2020MNRAS.493.4367C} derived the RC absolute magnitudes in 2MASS and {\it Gaia} $G$ bands and found that the absolute magnitude varies with [$\alpha$/Fe] for the low-$\alpha$ and high-$\alpha$ populations. 
They suggested a more detailed model of the RC absolute magnitudes is necessary for using it precisely.  
\citet{2020ApJ...893..108P} derived the median RC absolute magnitudes for low-$\alpha$ and high-$\alpha$ populations separately. 
\citet{2020ApJS..249...29H} calibrated $\Ks$-band absolute magnitude and its dependency on metallicity and age for LAMOST RCs. 

To better study the absolute magnitude of RC, it is important to have a good understanding of the Galactic structure and the distribution of the stellar population.
RCs' ages are around $1-10$ Gyrs and more concentrated in $1-2$ Gyrs \citep{2016ARA&A..54...95G}. RCs are found mainly in thin and thick disks, and to a lesser extent in the halo. The density of thin and thick disks decreases outwards roughly exponentially with scalelengths around $2.6\pm0.5$ kpc and $2.0\pm0.2$ kpc, and scaleheights around $0.30\pm0.05$ kpc and $0.90\pm0.18$ kpc \citep[and reference therein]{2016ARA&A..54..529B}.
The flare and warp of the stellar disk have been studied by red giants and RCs \citep{2002A&A...394..883L}. Recently, with thousands of classical Cepheids, the flared and warped disk was directly illustrated and well determined \citep{2019NatAs...3..320C, 2019Sci...365..478S}. The amplitudes of flare and warp increase with the Galactocentric distance, and the relationships can be approximated as a power-law equation.

With the availability of large-scale spectroscopic data in recent years, the stellar metallicity and elemental abundance distributions in the Milky Way have been further studied and better understood. 
Based on APOGEE red giants, \citet{2014AJ....147..116H,2015ApJ...808..132H} measured the stellar distribution in the [$\alpha$/Fe]$-$[Fe/H] plane and studied metallicity gradient and metallicity distribution functions in the Milky Way disk.
\citet{2015RAA....15.1209X} derived the radial and vertical metallicity gradients of the Milky Way disk in the anti-center direction by using main sequence turn-off stars in LAMOST. 
Later, \citet{2016ApJ...823...30B} studied the different distributions of RCs with mono-abundance populations and found that low-[$\alpha$/Fe] RCs show flare features in the outer disk.  
\citet{2017MNRAS.471.3057M} further measured the age-metallicity structure and surface-mass density profile of the Milky Way by using red giants. 
\citet{2017A&A...600A..70A} measured the evolution of the radial metallicity gradient in the Milky Way's thin disk. 
Generally, the metallicity decreases with the increase of the Galactocentric distance (only considering $R>4$ kpc), while the gradient $\mathrm{d}{\rm [M/H]}/\mathrm{d}R$ varies with the height.  
The knowledge about the relationship between RC absolute magnitudes and the elemental abundances allow the study of the distribution of RC absolute magnitudes in the Milky Way.

Nowadays, {\it Gaia} early data release 3 (EDR3) is available with about 30\% better distance accuracy than that of DR2 \citep{2018A&A...616A...1G}, and the systematic bias of the parallax has been well constrained by the parallax correction from \citet{2021A&A...649A...4L,2021A&A...649A...2L}. 
Therefore, with a combination of {\it Gaia} EDR3 parallaxes, APOGEE, and LAMOST spectral parameters, it is time to establish a three-dimensional (3D) distribution model of the RC absolute magnitude throughout the Milky Way instead of a constant absolute magnitude. 
To design the model, we consider the main structures of the Milky Way, as well as the distribution of elemental abundances and populations. 
Based on our model, more suitable absolute magnitudes and intrinsic colors can be obtained for RCs at any spatial position and distance without the spectroscopic information. 
In turn, more accurate absolute magnitudes will benefit the study of the Galactic structure and extinction, especially in regions where spectroscopic observations are lacking.

The sketch of this paper is as follows. 
The description of data sets and the spectral RC sample are presented in Section~\ref{data}. 
In Section~\ref{RCabs_int}, we present the method to derive the absolute magnitudes and intrinsic colors of RCs with spectral parameters. 
Section~\ref{3Dmap} presents the methods and results of establishing RCs' 3D absolute magnitude and intrinsic color maps. 
The validation of our 3D parameter maps is discussed in Section~\ref{section4}. 
In Section~\ref{GaiaRC}, we construct a large whole-sky {\it Gaia} RC sample. 
We briefly describe our prospects for this work in Section~\ref{prospect} and summarize our principal conclusions in Section~\ref{Conclusion}.

\section{Data and Sample} \label{data}

\subsection{Data} 

To investigate the multi-band absolute magnitudes of RCs, we used astrometric data from {\it Gaia} EDR3, spectroscopic data from APOGEE and LAMOST, and photometric data from the {\it Gaia}, APASS, Pan-STARRS1 (PS1), 2MASS, and {\it WISE} surveys. 

\subsubsection{Gaia} 

The {\it Gaia} EDR3 provides photometry and astrometry for 1.8 billion sources brighter than 21.0 mag in $G$ band \citep{2021A&A...649A...1G}.
The high-precision photometry contains three broadbands, $G$, $\GBP$, $\GRP$, for more than 1.5 billion stars. 
\citet{2021A&A...649A...4L} investigated the relationship between parallax offset and star's spatial position, magnitude, and color. 
They found an overall zero-point offset of the Gaia EDR3 parallax is about -0.017 mas and provided a formal procedure for offset correction. 
This procedure is very effective in reducing the systematic offset of parallaxes and has been validated by distance tracers such as RCs \citep{2021ApJ...910L...5H} and contact binaries \citep{2021ApJ...911L..20R}. We adopted the correction estimated by this procedure for distance estimation. For our RC sample, the offsets are around -0.03 mas, slightly larger than the mean value.  

\subsubsection{APOGEE} 

APOGEE is a near-IR $H$-band ($15000-17000\rm{\AA}$) spectroscopic survey with high-resolution ($R\sim22,500$). 
The spectra reduction has been described in \citet{2015AJ....150..173N}. 
Stellar atmospheric parameters, such as effective temperature $\Teff$, surface gravity $\log g$, and metallicity [M/H], and chemical abundances are derived by using the APOGEE Stellar Parameters and Chemical Abundance Pipeline \citep[ASPCAP;][]{2016AJ....151..144G}. 
\citet{2018AJ....156..125H} and \citet{2018AJ....156..126J} further analyzed the  precision of these parameters. 
We utilize data from the latest released APOGEE DR16 \citep{2020ApJS..249....3A} with the catalog provided by \citet{2020AJ....160..120J}.  
A detailed description and element-by-element discussion of the reliability of the DR16 results can also be found in \citet{2020AJ....160..120J}.

\subsubsection{LAMOST} 

LAMOST is the Galactic spectroscopic survey that takes 4000 spectra simultaneously with a resolution of $R\sim1,800$ covering the full optical range of $3690-9100\rm{\AA}$.
We adopted $\log g$, $\Teff$, and [Fe/H] from LAMOST DR7 and obtained their chemical abundances [C/Fe], [N/Fe], and [$\alpha$/Fe] from \citet{2019ApJS..245...34X}. 
The typical internal uncertainties are $150$ K in $\Teff$, 0.25 dex in $\log g$, and 0.15 dex in [Fe/H], at a spectral signal-to-noise ratio (S/N) $>10$ \citep{2015RAA....15.1095L}.

\subsubsection{APASS}
The American Association of Variable Star Observers (AAVSO)
Photometric All-Sky Survey (APASS) DR9 provides $B$ and $V$ bands photometric data for stars brighter than $V\sim17$ mag \citep{2016yCat.2336....0H}.

\subsubsection{Pan-STARRS1} 

The PS1 survey images the sky in five optical bands, $g, r, i, z$, and $y$, covering 400 nm to 1$\mum$ \citep{2016arXiv161205560C}. The mean 5\,$\sigma$ limiting magnitudes of point-source in $g, r, i, z$, and $y$ bands are 23.3, 23.2, 23.1, 22.3, and 21.4 mag, respectively.

\subsubsection{2MASS} 

The 2MASS is a whole-sky survey in the near-IR $J H \Ks$ bands \citep{Cohen2003AJ....126.1090C}. The 10\,$\sigma$ limiting magnitudes of the point-source catalog are 15.8, 15.1, and 14.3 mag in the $J$, $H$, and $\Ks$ bands, respectively \citep{2006AJ....131.1163S}.

\subsubsection{{\it WISE}} 
The {\it WISE} survey is a mid-IR whole-sky survey in four bands: $W1$, $W2$, $W3$, and $W4$ bands with centering wavelengths center 3.35, 4.60, 11.56, and 22.09$\mum$, respectively. 
We collected {\it WISE} $W1$, $W2$, and $W3$ bands data from the ALLWISE catalog.

\subsection{Spectral Red Clump Sample}\label{RCsample} 

\begin{figure*}
\centering
\includegraphics[width=\hsize]{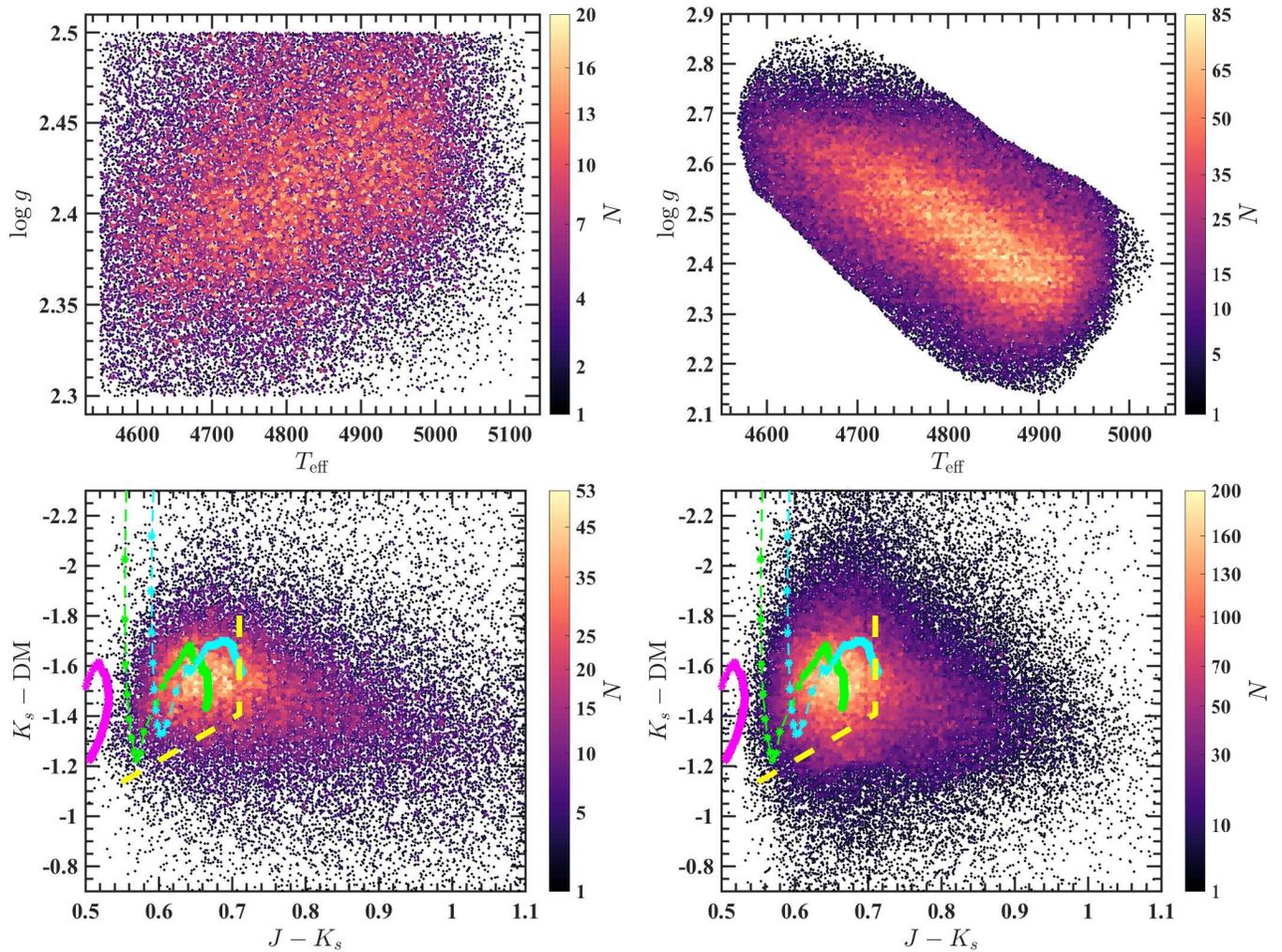}
\caption{
The $\Teff-\log g$ diagrams (top) and $\Ks-{\rm DM}$ vs. ($J-\Ks$) CMDs (bottom) for the selected APOGEE RCs (left) and LAMOST RCs (right). 
The color represents the number density of RCs.
The magenta, green, cyan dotted asterisk lines are the evolutionary tracks from \citet{2001MNRAS.323..109G} with metallicities $Z=0.04, 0.19, 0.30$, respectively. The yellow-dashed lines $(J-\Ks)_0=0.71$ mag and $\MKs+1.7\times(J-\Ks)_0+0.2=0$ mag indicate the boundaries of RCs, whose absolute magnitudes and intrinsic colors are predicted based on spectral parameters (see Section 3). 
}
\label{Fig_RC1}
\end{figure*}

\begin{table*}[ht]
\begin{center}
\caption{\label{tab_sample} Seven Different Red Clump Samples Used In This Work}
\begin{tabular}{p{110pt}p{250pt}p{75pt}p{30pt}}
\hline \hline
Name & Notes &  Sample Size & Section   \\ 
\hline      
APOGEE RC sample  &  spectral RCs, $4550\K \le \Teff \le 5120\K$ and $2.3 \le \log g \le 2.5$, left panels of Figure 1  &  42,947  &  2.2   \\ 
LAMOST RC sample & spectral RCs, in the high-density region of the $\Teff-\log g$ diagram, centered at $\Teff=4800 \K$ and $\log g=2.5$, RCs with $\Ks-{\rm DM}> -1.2$ mag and $\log g>2.6$ are excluded, right panels of Figure 1 &  93,542  &  2.2   \\   
APOGEE RC training sets & subsets of APOGEE RC sample for absolute magnitude training ($\Av<0.1$ mag, $\varpi>1$, and $\sigma_\varpi/\varpi<0.05$), and for intrinsic color training ($\Av<0.1$ mag) &  $\sim 1,300$/7,712 &  3.1,  3.3  \\
LAMOST RC training sets & subsets of LAMOST RC sample for absolute magnitude training ($\Av<0.1$ mag, $\varpi>1$, and $\sigma_\varpi/\varpi<0.05$), for intrinsic color training ($\Av<0.1$ mag) &  $\sim 3,700$/10,797  &  3.1,  3.3  \\
Combined RC sample & a combination of LAMOST, APOGEE, and {\it Gaia} RCs to establish RC 3D parameter maps &  $\sim 156,000$  &  4.1  \\
External test RC sample & photometric RCs from {\it Gaia} with $\sigma_\varpi/\varpi<0.2$, $-2.4<m_G-1.89\times(\GBP-\GRP)-(5\log d+10) <-0.9$ mag, and $M_\Ks+1.7\times(J-\Ks)_0\leq0$ mag  &    &  5.1  \\
{\it Gaia} RC sample & a whole-sky photometric RC sample selected from {\it Gaia} data &  $\sim 11,000,000$  &  6  \\
\hline
\end{tabular}
\end{center}
\end{table*}

The RCs, core-helium-burning evolved stars, cover the range of spectral types G8III--K2III with effective temperatures of 4500--5300K \citep{2016ARA&A..54...95G}. 
These stars stand out in the $\Teff-\log g$ diagram and can be easily identified as a clump. 
Hence, we constructed the spectral RC sample based on the stellar parameters from APOGEE and LAMOST surveys. 

The initial data selection required stars with high-quality data. 
\begin{itemize}
\item[--] For APOGEE data, we required ASPCAPFLAG=0, and S/N$>70$. 
The uncertainties in stellar parameters $\Teff$, $\log g$, and [M/H] are less than 150 K, 0.08 dex, 0.02 dex, respectively.  
We also required that uncertainties in elemental abundances [$\alpha$/M], [C/Fe], and [N/Fe] are less than 0.02 dex, 0.05 dex, 0.05 dex, respectively.   
\item[--] For LAMOST data, we required S/N$\underline{~~}$g$>30$ (Spectral S/N per pixel in $g$ band), and qflag$\underline{~~}$chi2=``good".
The uncertainties of $\Teff$ and $\log g$ are less than $200\K$ and 0.2 dex, respectively.
We also required that uncertainties in elemental abundances [Fe/H], [$\alpha$/Fe], [C/Fe], and [N/Fe] are less than 0.1 dex, 0.04 dex, 0.05 dex, 0.1 dex, respectively.
\end{itemize}
Next, we selected RC candidates from the $\Teff-\log g$ diagram.  
For the APOGEE data, we required that RCs satisfy $4550\K \le \Teff \le 5120\K$ and $2.3 \le \log g \le 2.5$. 
Such a narrow range of $\log g$ effectively removes secondary RC stars (SRCs) and red giants \citep{2019ApJ...877..116W}. 
For the LAMOST data, we first selected stars located in the high-density region of the $\Teff-\log g$ diagram, centered at $\Teff=4800 \K$ and $\log g=2.5$. 
The top panel of Figure~\ref{Fig_RC1} shows the $\Teff-\log g$ plots of APOGEE RCs (left) and LAMOST RCs (right). 
We find that the APOGEE RCs and LAMOST RCs show different correlations between $\log g$ and $\Teff$. This is due to the systematic bias in the $\log g$ calculated by the LAMOST Stellar Parameters Pipeline (LASP) for giant stars. \citet{2015RAA....15.1209X} analyzed the LAMOST spectra based on their pipeline and found that for giants with $2<\log g<3$, their $\log g$ also shows an anti-correlation with LASP \citep[Figure 22 of ][]{2015RAA....15.1209X}. Nevertheless, the systematic deviation of $\log g$ does not affect the selection and internal statistical analysis of the LAMOST RCs. 
Since $\log g$ in LAMOST is not as accurate as in APOGEE, the LAMOST RC candidates selected from the $\Teff-\log g$ diagram contain some contamination, mainly red giants and SRCs. 
Comparing with APOGEE data, we find that 30\% of LAMOST RC candidates do not satisfy the APOGEE RC criterion. If we consider that our APOGEE RC criterion is too strict, the actual contamination rate of LAMOST RC candidates selected based on $\log g$ and $\Teff$ is probably around 20\%. This is consistent with the analysis of \citet{2018ApJ...858L...7T}. 
Then, we tried to remove contamination using the near-IR color-magnitude diagram (CMD). 
Based on the corrected {\it Gaia} parallaxes, we estimated the distance-corrected magnitude in the $\Ks$ band $\Ks-{\rm DM}$ (DM denotes the distance modulus). 
In the plot of $\Ks-{\rm DM}$ versus ($J-\Ks$), some stars are distributed in $\Ks-{\rm DM}>-1.2$ mag, and they deviate from the RCs distribution. 
These stars are mainly distributed in $\log g>2.6$ on the $\Teff-\log g$ diagram, with a high probability of being red giants and SRCs. 
Therefore, we excluded stars with $\Ks-{\rm DM}> -1.2$ mag and $\log g>2.6$ in the selection of LAMOST RCs.

Only APOGEE RCs with $2.3<\log g<2.5, 4550\K<\Teff<5120\K, -0.8<{\rm [M/H]}<0.5, -0.12<{\rm [N/Fe]}<0.5, -0.15<{\rm [C/Fe]}<0.3, -0.05<{\rm [\alpha/M]}<0.3$ and LAMOST RCs with $2.1<\log g<2.9, 4550\K<\Teff<5000\K, -0.9<{\rm [Fe/H]}<0.3, -0.12<{\rm [N/Fe]}<0.5, -0.4<{\rm [C/Fe]}<0.1, -0.1<{\rm [\alpha/Fe]}<0.3$ were selected to ensure that all parameter ranges are covered by the training sets (see Section 3).

The final spectral RC sample contains 42,947 APOGEE RCs and 93,542 LAMOST RCs. 
\citet{2018ApJ...853...20H, 2018ApJ...858L...7T} obtained a sample of RCs based on asteroseismology parameters determined directly from the spectra of APOGEE and LAMOST.  We cross-matched our RC sample with their entire sample and obtained 90,385 common objects. Of these, 83\% were classified as pristine RCs by \citet{2018ApJ...858L...7T}. By examining the period spacing-frequency separation distribution \citep[Figure 1 of][]{2018ApJ...858L...7T}, only ~1\% of our RCs are located in the red giant sequence, while the other 16\% of RCs are located in the transition region. This indicates that our sample is less contaminated. During the comparison, we found that both the ($\log g$, $\Teff$, parallax) cut method and the spectroscopic asteroseismology method miss a certain percentage of RCs to ensure purity.

The bottom panels of Figure~\ref{Fig_RC1} display near-IR bands $\Ks-{\rm DM}$ versus ($J-\Ks$) CMDs. 
The color represents the number density of RCs. 
The distribution of LAMOST RCs in the CMD is similar to that of APOGEE RCs, indicating that the selection criteria we applied to the LAMOST data effectively remove contamination. 
The evolutionary tracks of RCs with different metallicities $Z=0.04, 0.19, 0.30$ \citep{2001MNRAS.323..109G} are plotted as magenta, green, cyan dotted asterisk lines, respectively.  
The yellow-dashed lines $(J-\Ks)_0=0.71$ mag and $\MKs+1.7\times(J-\Ks)_0+0.2=0$ mag indicate the boundaries of RCs, whose absolute magnitude and intrinsic color are predicted based on spectral parameters (see Section 3). These RC boundaries agree with the theoretical evolutionary tracks. Despite the extinction effect, the distribution of most RCs is consistent with the theoretical predictions, which justifies our selection criteria. Note that these criteria were used again in Sections \ref{3Dmapderive} and \ref{result_check} to purify {\it Gaia} RCs.
For clarity, we describe the different RC samples used in this work in Table \ref{tab_sample}.

In the subsequent calculations and predictions of RCs' multi-band absolute magnitudes and intrinsic colors, we also require that the photometric errors $\sigma_m$ and magnitudes of different data satisfy the corresponding criteria:
\begin{enumerate}
\item For {\it Gaia} data, $\sigma_m \le 0.01$ mag and magnitude $\le18.0$ mag in $G$, $\GBP$, $\GRP$ bands.
\item For APASS data, $\sigma_m \le 0.05$ mag in $B$, $V$ bands. 
\item For PS1 data, $\sigma_m \le 0.02$ mag in $g, r, i, z$, $y$ bands. 
\item For 2MASS data, $\sigma_m \le 0.05$ mag and magnitude ranging from 6.0 to 14.0 mag in $J$, $H$, $\Ks$ bands. 
\item For {\it WISE} data, $\sigma_m \le 0.05$ mag in $W1, W2$, $W3$ bands. 
\end{enumerate}

\section{Red Clump Absolute Magnitude and Intrinsic Color} \label{RCabs_int}

This section describes how we derived the absolute magnitude and intrinsic color of RCs based on their spectral parameters. 
A summary of the process is as follows. 
First, we selected RCs from the spectral RC sample (including APOGEE RCs and LAMOST RCs) to construct the training sets. 
Then, we calculated the absolute magnitude and intrinsic color of each training set star. 
After that, we built regression learners based on the training sets to estimate 
$M_\lambda$ (the absolute magnitude in bandpass $\lambda$) and $(\lambda_1-\lambda_2)_0$ (the intrinsic color in bandpass $\lambda_1$ and bandpass $\lambda_2$) from observed parameters, such as $\log {\rm g}$, $T_{\rm eff}$, [M/H] ([Fe/H] for LAMOST data), [C/Fe], [N/Fe], and [$\alpha$/M] ([$\alpha$/Fe] for LAMOST data). 
Finally, we estimated $M_\lambda$ and $(\lambda_1-\lambda_2)_0$ for each spectral RC based on their stellar parameters and spectral parameters. 
Meanwhile, we analyzed the dependence of their $M_\lambda$ and $(\lambda_1-\lambda_2)_0$ on the observed parameters, including position, stellar parameters, and element abundances.

\begin{figure*}
\centering
\includegraphics[width=\hsize]{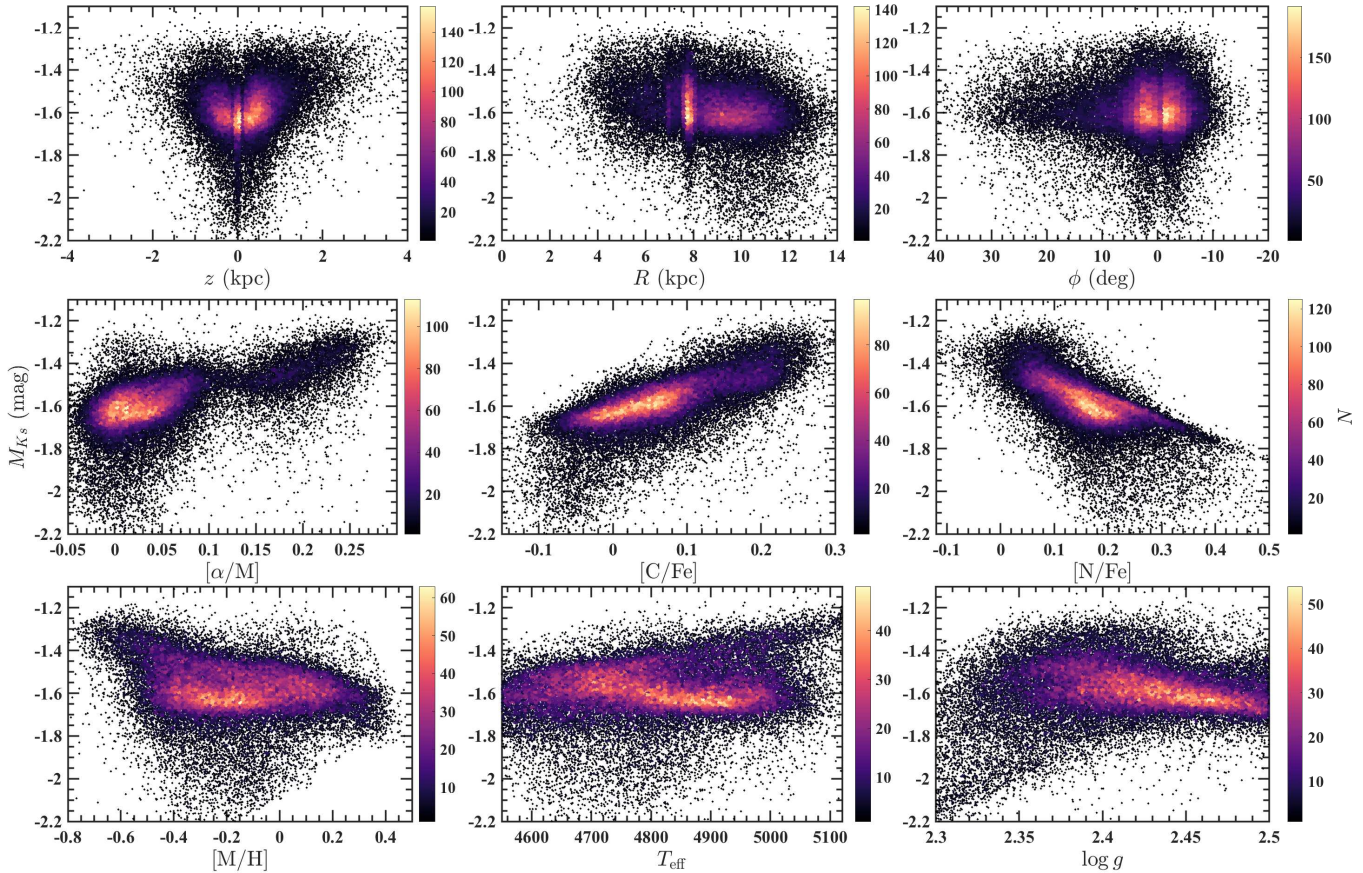}
\caption{The $\MKs$ distributions of APOGEE RCs with respect to the Galactocentric cylindrical coordinates ($z$, $R$, $\phi$), elements abundances ([$\alpha$/M], [C/Fe], [N/Fe]), and stellar parameters ([M/H], $\Teff$, $\log g$). 
The color represents the number density of RCs.
}
\label{Fig_MK}
\end{figure*}

\begin{figure*}
\centering
\includegraphics[width=\hsize]{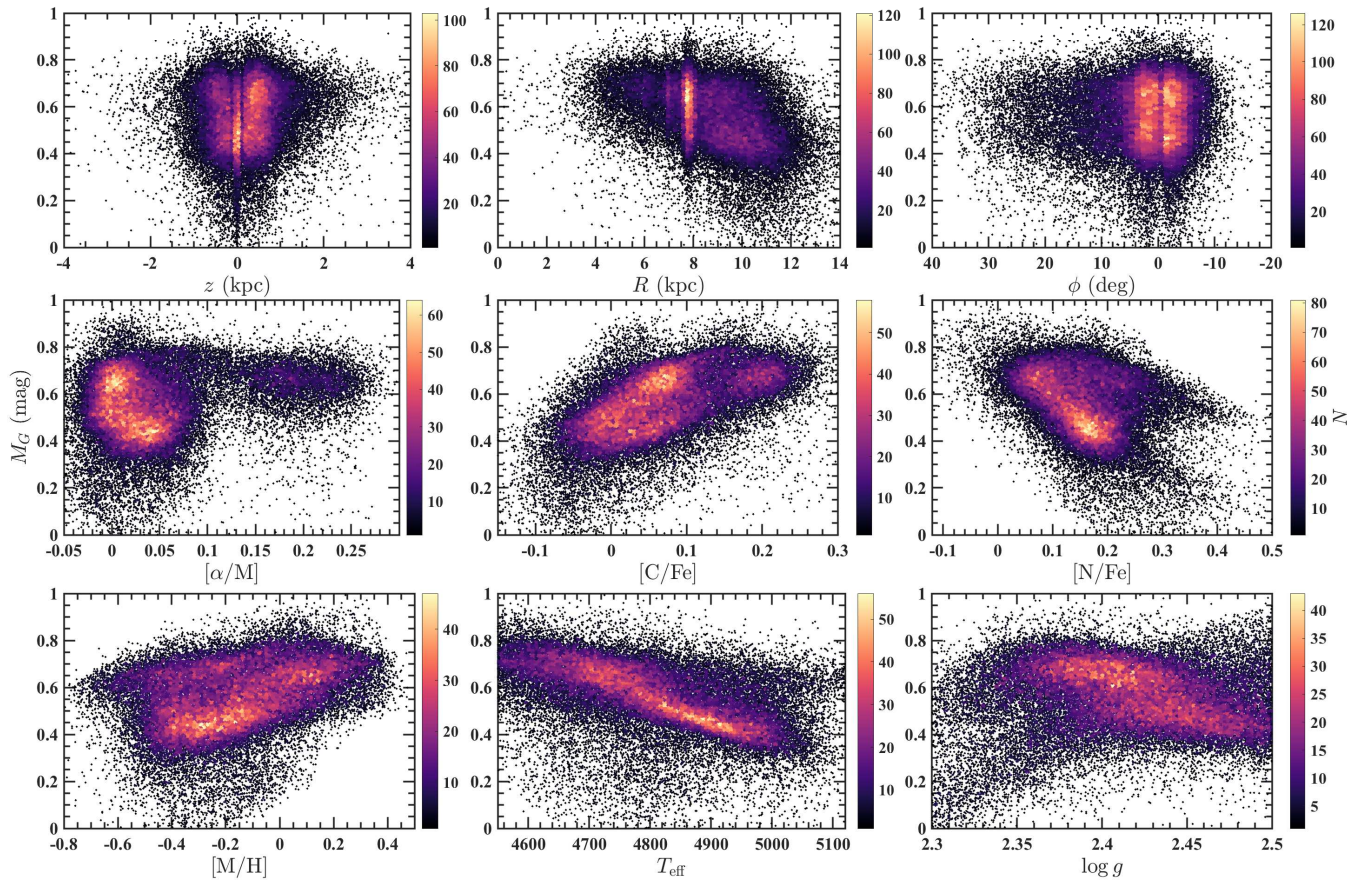}
\caption{The similar distributions of APOGEE RCs as Figure~\ref{Fig_MK} but for the {\it Gaia} $G$-band absolute magnitude $\MG$.}
\label{Fig_MG}
\end{figure*}

\begin{figure*}[ht]
\centering
\includegraphics[width=\hsize]{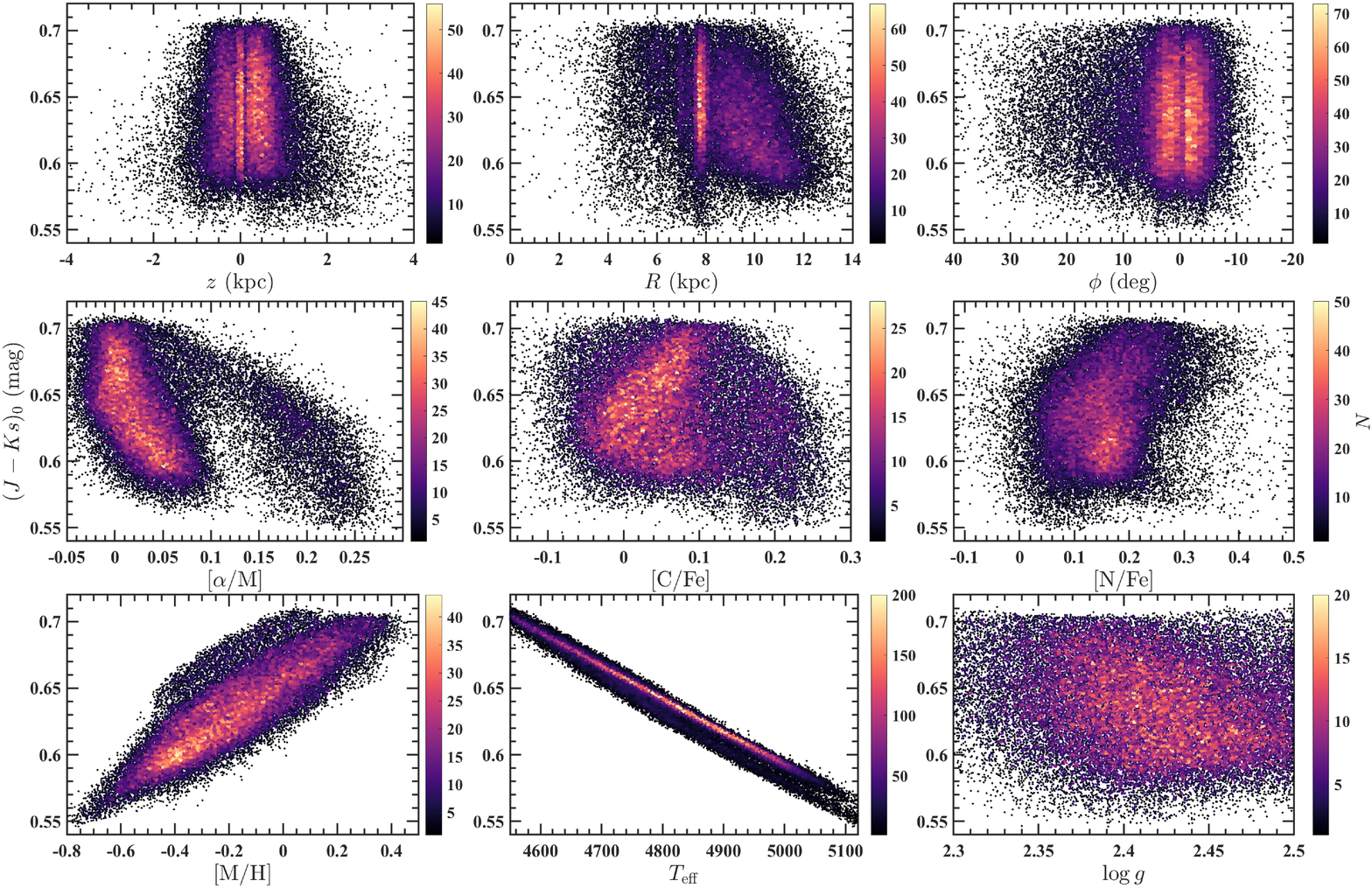}
\caption{The similar distributions of APOGEE RCs as Figure~\ref{Fig_MK} but for the intrinsic color $(J-\Ks)_0$.}
\label{Fig_int}
\end{figure*}

\subsection{Absolute Magnitude Training}\label{RCabs}

Absolute magnitude of an RC star in the $\lambda$ band can be calculated by the equation 
\begin{equation}
M_\lambda=m_\lambda -A_\lambda -5\log d-10 ~.  
\label{equ1}
\end{equation}
The value of apparent magnitude $m_\lambda$ is from the multi-band photometric data. 
The distance information $d$ in kpc can be obtained from the inverse of the corrected {\it Gaia} EDR3 parallax $\varpi-\varpi_{\rm offset}$. 
The extinction $A_\lambda$ is estimated with the combination of color excess and the extinction conversion factor. 
The accuracy of $M_\lambda$ measurements worsens as the increasing distance and extinction uncertainties. 
Therefore, we use RCs with reliable absolute magnitudes to build regression learners to learn the relationship between absolute magnitudes and spectral parameters. The regression learners are then applied to estimate the absolute magnitudes of all RCs with spectral parameters. 
This avoids, when the parallax and extinction uncertainties are large, the large errors caused by using equation ~ref{equ1} to calculate absolute magnitudes.

To avoid loss of parameter features, a machine learning algorithm, Gaussian Process regression (GPR) learner, is used to train and predict $M_\lambda$ of RCs. 
The GPR learner is a kernel-based nonparametric probabilistic learner. 
It uses a Gaussian Process prior for regression analysis and is widely used in the machine learning community \citep{2006gpml.book.....R}. 
In our tests, considering a small training set ($10^3-10^4$) available for training learners, the GPR learner can predict the RC absolute magnitudes more accurately than learners such as regression ensembles and support vector products. 
Here, we use the Matlab Regression Learner application to train regression learners and estimate RC's $M_\lambda$.
For the choice of the correlation function (called the kernel), we chose the Mat\'ern correlation with $\nu=5/2$.

For different databases (APOGEE and LAMOST) or different bands, we trained $M_\lambda$ separately. 
As a result, there are 34 trainers in total for 17 bands and two databases. 
It includes our constructed Weisenheit magnitude \citep{1982ApJ...253..575M} $W_{G,\GBP,\GRP}=m_G-1.89\times(\GBP-\GRP)$ for {\it Gaia} bands, which is less affected by extinction and metallicity. The coefficient 1.89 is from \citet{2019ApJ...877..116W} and suitable for RCs. 
A brief description of the process is given below. We first built a training set for each band. 
For each RC in the training set, we calculated $M_\lambda$ by equation~\ref{equ1}. 
Then based on the training set, we trained the GPR learners to establish the relationship between $M_\lambda$ and the spectral parameters. 
We performed 10-fold cross-validation to generate regression learners and evaluate them simultaneously. 
Finally, these generated GPR learners were used to estimate $M_\lambda$ of APOGEE RCs and LAMOST RCs.

To build the training sets, we set the following criteria: low extinction ($\Av<0.1$ mag) and accurate {\it Gaia} parallax ($\varpi>1$, $\sigma_\varpi/\varpi<0.05$).
This is to ensure the reliability of $M_\lambda$ for the training RCs since the accuracy of $M_\lambda$ estimated by equation~\ref{equ1} decreases with increasing distance and extinction uncertainties. 
The extinction $\Av$ is converted from color excess $E(\GBP-\GRP)$ by $\Av=2.394\times E(\GBP-\GRP)$, where the extinction coefficient 2.394 is from \citet{2019ApJ...877..116W}. 
For APOGEE RCs, $E(\GBP-\GRP)$ can be roughly calculated by using empirical formulae between intrinsic colors and stellar parameters \citep[Table 1 of ][]{2019ApJ...877..116W}.   
For LAMOST RCs, we established the linear relations between the stellar parameters of LAMOST and APOGEE based on more than 6,000 RCs with both APOGEE and LAMOST data. We then used the empirical intrinsic color formulae to determine $E(\GBP-\GRP)$. 
Since only low-extinction objects are used, the error of  $M_\lambda$ propagated from the extinction error is negligible.
We also required that RCs have good astrometric solutions by ${\rm RUWE}<1.4$ \citep{2021ApJ...907L..33S}, where RUWE is the renormalized unit weight error.
After selection, the training sets for absolute magnitudes contain around 1,300 APOGEE RCs and 3,700 LAMOST RCs (see Table~\ref{tab_sample}). These numbers vary slightly in different bands.

With the training sets, we trained GPR learners to establish the relationships between $M_\lambda$ and parameters $\log g$, $\Teff$, [M/H], [C/Fe], [N/Fe], and [$\alpha$/M]. 
Simultaneously, we performed 10-fold cross-validation to test the resulting GPR learners. 
This was done by dividing the training set into 10 equal-sized random folds: 9 folds were trained to derive the GPR learners, and the remaining one was used to validate it. 
This process was repeated 10 times. The root mean square errors (RMSEs) in validation are $\sim0.10$ mag in {\it Gaia} bands and 2MASS bands, $\sim0.12$ mag in {\it WISE} bands and APASS bands, and $0.12-0.18$ mag in PS1 bands. The best model appears in 2MASS bands for the APOGEE sample, where we think that its $0.09$ mag RMSE is close to the intrinsic scatter of RC absolute magnitude. 
Learners perform slightly worse in PS1 bands, mainly due to the effect of saturation, and we need to compromise between maintaining the number of the training set and the photometric quality. 
To avoid overfitting at the parameter edges, we selected samples whose difference between the predicted and training values is less than one RMSE for re-training to obtain the final learners.
Finally, we built 34 $M_\lambda$ learners $M_{\lambda,\rm{pred}}$ for 17 bands and two databases.

\subsection{Absolute Magnitude Estimation}\label{RCabs2}

We applied the 34 learners to estimate $M_\lambda$ for all RCs in APOGEE (42,947 RCs) and LAMOST (93,542 RCs). To check the possible offsets between the two databases, we estimated the absolute magnitude differences based on 6,252 RCs with both APOGEE and LAMOST observations. The differences in $M_\lambda$ are around or less than $0.005\pm0.090$ mag. These small systematic offsets imply that the combination of the two databases for subsequent analysis is reliable. 

We take $\Ks$ and $G$ bands as examples to understand the importance of each parameter in estimating $M_\lambda$.
Figures \ref{Fig_MK} and \ref{Fig_MG} show the distributions of $\MKs$ and $\MG$ of APOGEE RCs with respect to different observed parameters, including the position in Galactic cylindrical coordinates ($z$, $R$, $\phi$), elements abundances ([$\alpha$/M], [C/Fe], [N/Fe]), and stellar parameters ([M/H], $\Teff$, $\log g$). 
The color represents the number density of RCs. 
According to the high-density regions in the figures, we find that $\MKs$ and $\MG$ are both strongly correlated to [$\alpha$/M], [C/Fe], [N/Fe] (middle panels of Figures~\ref{Fig_MK} and ~\ref{Fig_MG}) that characterize the RC age. As [C/Fe] increases and [N/Fe] decreases, the RC gradually becomes fainter in both $\Ks$ and $G$ bands. As [M/H] increases, RC gradually becomes fainter in the $G$ band. Compared to $\MG$, $\MKs$ shows a very weak correlation with [M/H]. Overall, $M_\lambda$ becomes fainter with increasing age and metallicity. The relationship between $M_\lambda$ and metallicity weakened with increasing wavelengths from ultraviolet to IR.

For parameters characterizing the Galactic structure, $M_\lambda$ almost does not vary with the Galactocentric angle $\phi$, mainly because APOGEE does not cover much of RCs in the warped disk. 
In the IR band, $\MKs$ becomes fainter with increasing height from the Galactic plane $\,| z \,|$ and varies little with Galactocentric radius $R$. Due to the metallicity effect, the absolute magnitude of optical bands, e.g., $MG$, becomes significantly brighter with the increase of $R$.  $\MG$ does not vary with increasing $\,| z \,|$ because the fainting of $\MG$ with age counteracts the brightening of $\MG$ with the decrease of metallicity. Combining the relationship between $M_\lambda$ and ($R, z$), and the relationship between $M_\lambda$ and metallicity and elemental abundances, we verified that [M/H] decreases with increasing $R$ and $ \,| z \,|$, and that the stellar age becomes older with increasing $ \,| z \,|$ in regions $ \,| z \,|<4$ kpc, $4<R<14$ kpc.

Both $\MKs$ and $\MG$ are insensitive to $\log g$ because the accuracy of $\log g$ prevents seeing more details. At longer wavelengths, the dependence of $M_\lambda$ on $\Teff$ is weaker. To establish RCs' $M_\lambda$ maps, a detailed two-dimensional analysis of the variation of $M_\lambda$ with $(R, z)$ is described in Section~\ref{3Dmap}.

\subsection{Intrinsic Colors}\label{RCint}

Although intrinsic colors $(\lambda_1-\lambda_2)_0$ can be obtained from the difference between two predicted absolute magnitudes $(M_{\lambda_1}-M_{\lambda_2})$, we still trained intrinsic color learners independently for $(J-\Ks)_0$ and $(\GBP-\GRP)_0$. They can be used to testify the absolute magnitude learners and for subsequent extinction calculations. 
We built the RC training sets for intrinsic colors by selecting RCs with low extinction $\Av<0.1$ mag. 
The training sets contain 7,712 APOGEE RCs and 10,797 LAMOST RCs (see Table~\ref{tab_sample}). 
The intrinsic color of each RC in the training set was estimated by the difference between the observed color and the extinction mentioned in Section~\ref{RCabs}.
We then trained the GPR learners to establish the relationships between intrinsic colors and observed spectral parameters. 
RMSEs of the cross-validation are $\sim0.012$ mag and $\sim0.024$ mag for $(J-\Ks)_0$ and $(\GBP-\GRP)_0$, which are much smaller than that of absolute magnitude learners. Therefore, the spectral parameters are better to calculate the intrinsic color compared to the absolute magnitude. The difference between intrinsic colors estimated by the intrinsic color learner, i.e., $(J-\Ks)_0$, and the two absolute magnitude learners, i.e., $\MJ-\MKs$, is less than 0.001 mag, which demonstrates the intrinsic consistency between the learners.

Figure~\ref{Fig_int} shows the distributions of the estimated $(J-\Ks)_0$ of APOGEE RCs with different observed parameters. The variations of $(J-\Ks)_0$ on stellar parameters $\Teff$ and [M/H] are significant. The weak relationships between $(J-\Ks)_0$ and elemental abundances [$\alpha$/M], [C/Fe], and [N/Fe] suggest that age has little effect on the intrinsic color. This also indicates that the effect of age on the absolute magnitude is relatively consistent in different bands. $(J-\Ks)_0$ is also not sensitive to $\phi$ and $\log g$. In $(R, z)$ space, $(J-\Ks)_0$ becomes bluer as the increase of $\,| z \,|$ and $R$.

\subsection{Usage of our GPR learners}\label{UGPR}

To facilitate users to use our learners to estimate the absolute magnitudes and intrinsic colors for RCs from the APOGEE or LAMOST parameters, we assembled the learners as an application `RC2021' and stored it in an external repository\footnote{doi:\url{https://doi.org/10.5281/zenodo.5140055}}. To avoid external interpolation, the best applicable ranges for our application are $2.3<\log g<2.5, 4550\K<\Teff<5120\K, -0.8<{\rm [M/H]}<0.5, -0.12<{\rm [N/Fe]}<0.5, -0.15<{\rm [C/Fe]}<0.3, -0.05<{\rm [\alpha/M]}<0.3$ for APOGEE parameters and $2.1<\log g<2.9, 4550\K<\Teff<5000\K, -0.9<{\rm [Fe/H]}<0.3, -0.12<{\rm [N/Fe]}<0.5, -0.4<{\rm [C/Fe]}<0.1, -0.1<{\rm [\alpha/Fe]}<0.3$ for LAMOST parameters. The supplementary figures showing the parameter distributions of the training sets were attached with the application.

\section{Red Clump 3D Parameter Maps}\label{3Dmap}

For RCs with spectral parameters, their absolute magnitude and intrinsic color can be better derived from the spectral parameters without simply adopting constant mean values. However, to apply to a larger sample of RCs without spectroscopic observations, we would like to build three-dimensional maps of the absolute magnitude and intrinsic color associated with the location of RCs in the Milky Way.
This section describes how we built models to derive 3D absolute magnitude and intrinsic color maps based on our knowledge of the RCs distribution and the Galactic structure. 
With these maps, we can obtain the multi-band mean absolute magnitudes and intrinsic colors of RCs at any location ($l, b, d$) when the spectral data are not available.

\subsection{Building 3D Map Model}\label{3Dmapderive}

\begin{figure*}[ht]
\centering
\includegraphics[width=\hsize]{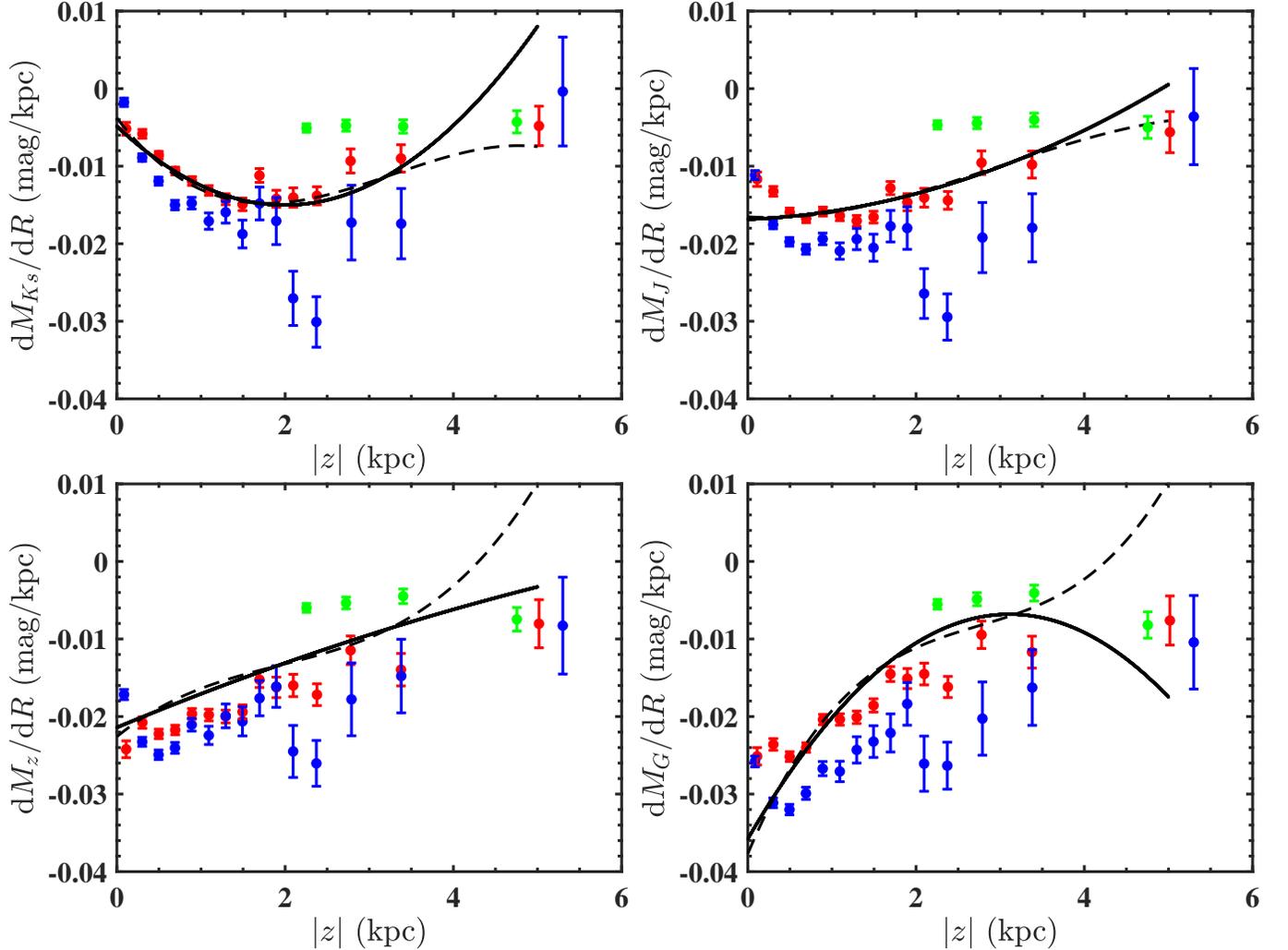}
\caption{The distributions of $\mathrm{d}{M_\lambda}/\mathrm{d}R$ to $\,| z \,|$, where $\lambda$ are $\Ks, J, z, G$ bands, respectively, from the top left to the bottom right.
Red, blue, and green dots are gradients in each $\,| z \,|$ bin for LAMOST RCs, APOGEE RCs, and {\it Gaia} RCs, respectively. 
The black-solid lines and black-dashed lines are from our 3D maps (Section \ref{results}) by assuming the functional form of quadratic and cubic polynomials, respectively.} 
\label{Fig_zr}
\end{figure*}

To establish 3D parameter maps, we further analyzed the variation of $M_{\lambda,\rm{pred}}$ (from absolute magnitude learners) with the Galactocentric radius $R$, the height $z$, and the Galactocentric angle $\phi$, following the results discussed in Sections~\ref{RCabs2} and \ref{RCint}. First, we find that in the nonparametric fit, $M_{\lambda,\rm{pred}}$ varies very little with different $\phi$. This is reasonable because the non-axisymmetric warp and overdensity are not significant in the spectral RC sample. Therefore, we omitted the parameter $\phi$ in building the map model. As $\,| z\,|$ increases, the fainting of $M_{\lambda,\rm{pred}}$ can be approximated by a first- or second-order polynomial.
The variation of $M_{\lambda,\rm{pred}}$ with $R$ is more complicated. At different $\,| z \,|$, RCs contain different proportions of thin-disk, thick-disk, and halo components, so the gradients $\mathrm{d}{M_\lambda}/\mathrm{d}R$ are variable. This is consistent with the current understanding of the Galactic disk structure \citep{2016ApJ...823...30B}.
At $\,| z \,| <$ 0.2 kpc, RCs are mostly thin-disk components and $M_{\lambda,\rm{pred}}$ is only related to the metallicity. In IR bands, such as the $\Ks$ band, the gradient $\mathrm{d}{\MKs}/\mathrm{d}R$ is close to zero.
At 0.2 kpc $< \,| z \,| <$ 2 kpc, RCs consist of thick-disk stars, thin-disk stars, and a small number of halo components. When $R$ is small, the thick-disk RCs dominate. 
When $R$ becomes larger, the proportion of thin-disk RCs increases due to the effect of the flared disk. So the gradients of $\mathrm{d}{\MKs}/\mathrm{d}R$ are noticeable in these $\,| z \,|$ ranges. At $\,| z \,| >$2 kpc, RCs are gradually dominated by halo components and $\mathrm{d}{M_\lambda}/\mathrm{d}R$ is close to zero again.
Therefore, we expect that the variation of $\mathrm{d}{\MKs}/\mathrm{d}R$ with $\,| z \,|$ can be simplified and described by a quadratic polynomial.

Based on the above analysis, we set up a function $M_{\lambda,{\rm map}} = f_\lambda(R, z)$ dependent on $R$ and $z$ to predict the absolute magnitude distribution of the Galactic RCs. The form of the function $M_{\lambda,{\rm map}} = f_\lambda(R, z)$ is
\begin{eqnarray}
M_{\lambda,{\rm map}} & = & f_\lambda(R,z) 
       \nonumber\\
      & = &  a_1\,(R-8.0)\,z^2+ a_2\,(R-8.0)\,|z| 
       \nonumber\\
      & \; &   +a_3\,(R-8.0) +a_4\,|z| +a_5 +a_6\,R^2\,|z| ~.
\label{equ2}
\end{eqnarray}
The first three terms denote the quadratic polynomials of $\mathrm{d}{M_\lambda}/\mathrm{d}R$ with $\,| z \,|$. 
$a_4\,|z|$ shows the linear relationship between $M_\lambda$ and ${\,| z \,|}$. We also tried an individual quadratic term $z^2$ and found that it could be ignored. $a_5$ is the zero point of $M_\lambda$.
When $R=8.0$ and $z=0$, i.e., $M_\lambda=a_5$, it is the RC absolute magnitude in the solar neighborhood. 
We also considered the flared disk denoted by the last term $a_6\,R^2\,|z|$. The functional forms for intrinsic colors $(J-\Ks)_0$ and $(\GBP-\GRP)_0$ are the same as equation~\ref{equ2}.

\begin{table*}[ht]
\begin{center}
\caption{\label{tab2} Coefficients of Absolute Magnitude and Intrinsic Color Maps $f_\lambda(R,z)$}
\hspace{1in}
\begin{tabular}{lcccccccc}
\hline \hline
\multicolumn{9}{c}{$f_\lambda(R,z)= a_1\,(R-8.0)\,z^2+ a_2\,(R-8.0)\,|z|+a_3\,(R-8.0) +a_4\,|z| +a_5 +a_6\,R^2\,|z|$} \\
\hline
Survey & $M_\lambda$ & $a_1$ & $a_2$ & $a_3$&  $a_4$ & $a_5$ &  $a_6$  & $\sigma$   \\ 
\hline   
{\it Gaia} & $G$                         &-0.0030 & 0.0300  &-0.0358 &0.0582  & 0.570  & -0.00071 &0.159    \\  
           & $\GBP$                      &-0.0051 & 0.0458  &-0.0476 &0.0549  & 1.103  & -0.00101 &0.180    \\  
           & $\GRP$                      &-0.0016 & 0.0187  &-0.0283 &0.0590  & -0.114 & -0.00045 &0.146    \\  
           & $W_{G,\GBP,\GRP}$$^*$     &0.0027  & -0.0166 &-0.0042 &0.0654  & -1.727 & 0.00033  &0.134    \\  
APASS      & $B$                         &-0.0107 & 0.0886  &-0.0722 &0.0705  & 1.928  & -0.00203 &0.232    \\  
           & $V$                         &-0.0067 & 0.0497  &-0.0472 &0.0610  & 0.857  & -0.00101 &0.175    \\  
Pan-STARRS & $g$                         &-0.0080 & 0.0701  &-0.0591 &0.0563  & 1.332  & -0.00166 &0.216    \\  
           & $r$                         &-0.0032 & 0.0298  &-0.0369 &0.0353  & 0.593  & -0.00064 &0.185    \\  
           & $i$                         &-0.0017 & 0.0146  &-0.0290 &0.0247  & 0.302  & -0.00021 &0.178    \\  
           & $z$                         &-0.0002 & 0.0015  &-0.0214 &0.0198  & 0.140  & 0.00019  &0.168    \\  
           & $y$                         &-0.0004 & 0.0102  &-0.0224 &0.0573  & 0.057  & -0.00030 &0.160    \\  
2MASS      & $J$                         &0.0006  & 0.0013  &-0.0169 &0.0610  & -0.984 & -0.00005 &0.134    \\  
           & $H$                         &0.0013  & -0.0049 &-0.0110 &0.0609  & -1.504 & 0.00007  &0.136    \\  
           & $\Ks$                       &0.0026  & -0.0147 &-0.0049 &0.0583  & -1.628 & 0.00028  &0.133    \\  
{\it WISE} & $W1$                        &0.0028  & -0.0165 &-0.0043 &0.0648  & -1.724 & 0.00034  &0.141    \\  
           & $W2$                        &0.0012  & -0.0022 &-0.0126 &0.0736  & -1.596 & -0.00004 &0.141    \\  
           & $W3$                        &0.0028  & -0.0165 &-0.0103 &0.0648  & -1.705 & 0.00048  &0.105    \\  
\hline                                                                                                          
{\it Gaia} & $(\GBP-\GRP)_0$             &-0.0039 & 0.0289  &-0.0184 &-0.0090 & 1.244  & -0.00061 &0.061    \\  
2MASS      & $(J-\Ks)_0$                 &-0.0020 & 0.0159  &-0.0104 &0.0003  & 0.656  & -0.00036 &0.040    \\  
   
\hline
\end{tabular}
\tablenotetext{}{Note $*$: Weisenheit magnitude $W_{G,\GBP,\GRP}=m_G-1.89\times(\GBP-\GRP)$.}
\end{center}
\end{table*}
%

\begin{figure*}[ht]
\centering
\subfigure{\includegraphics[width=3.5in]{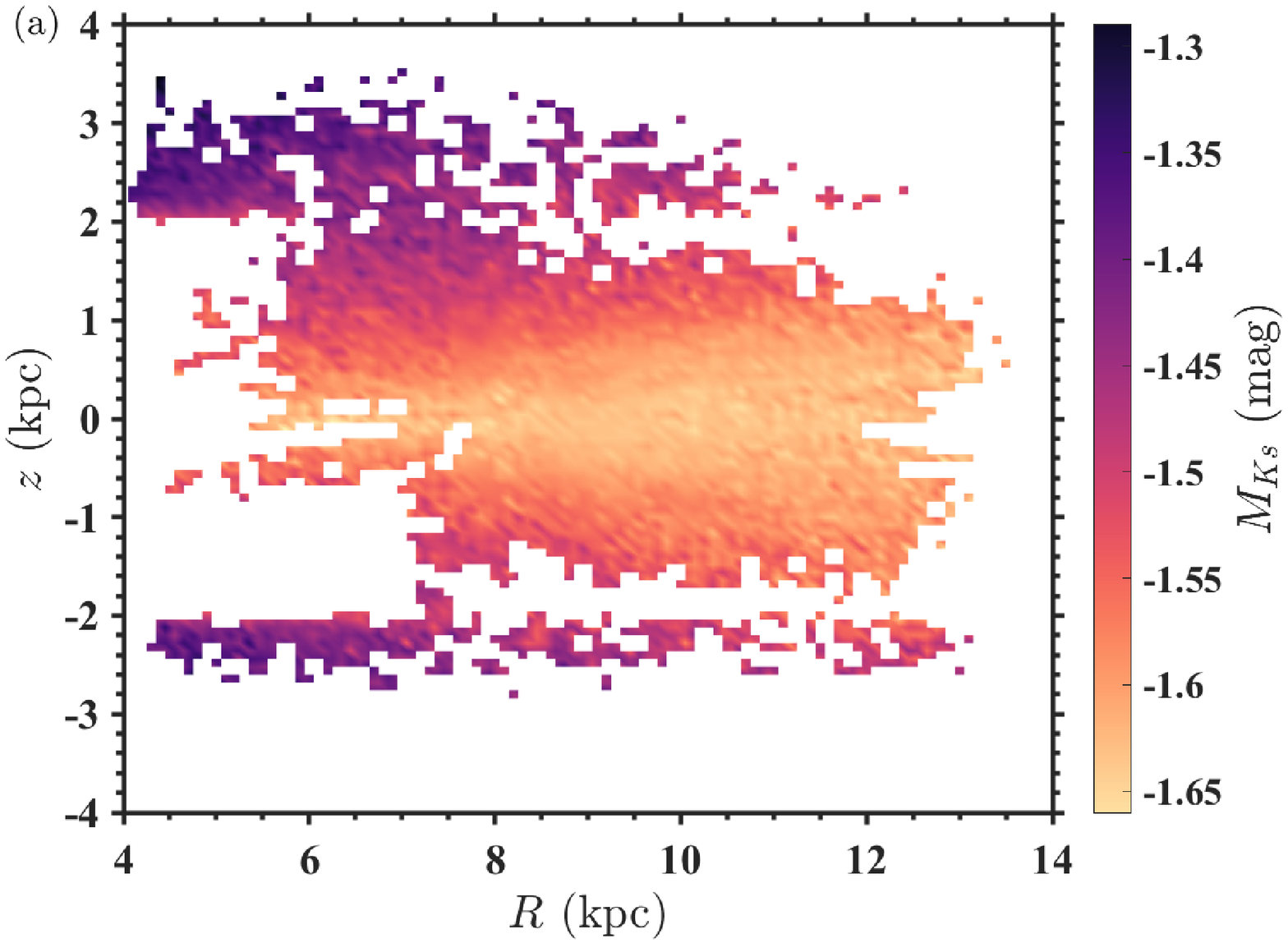}}
\subfigure{\includegraphics[width=3.5in]{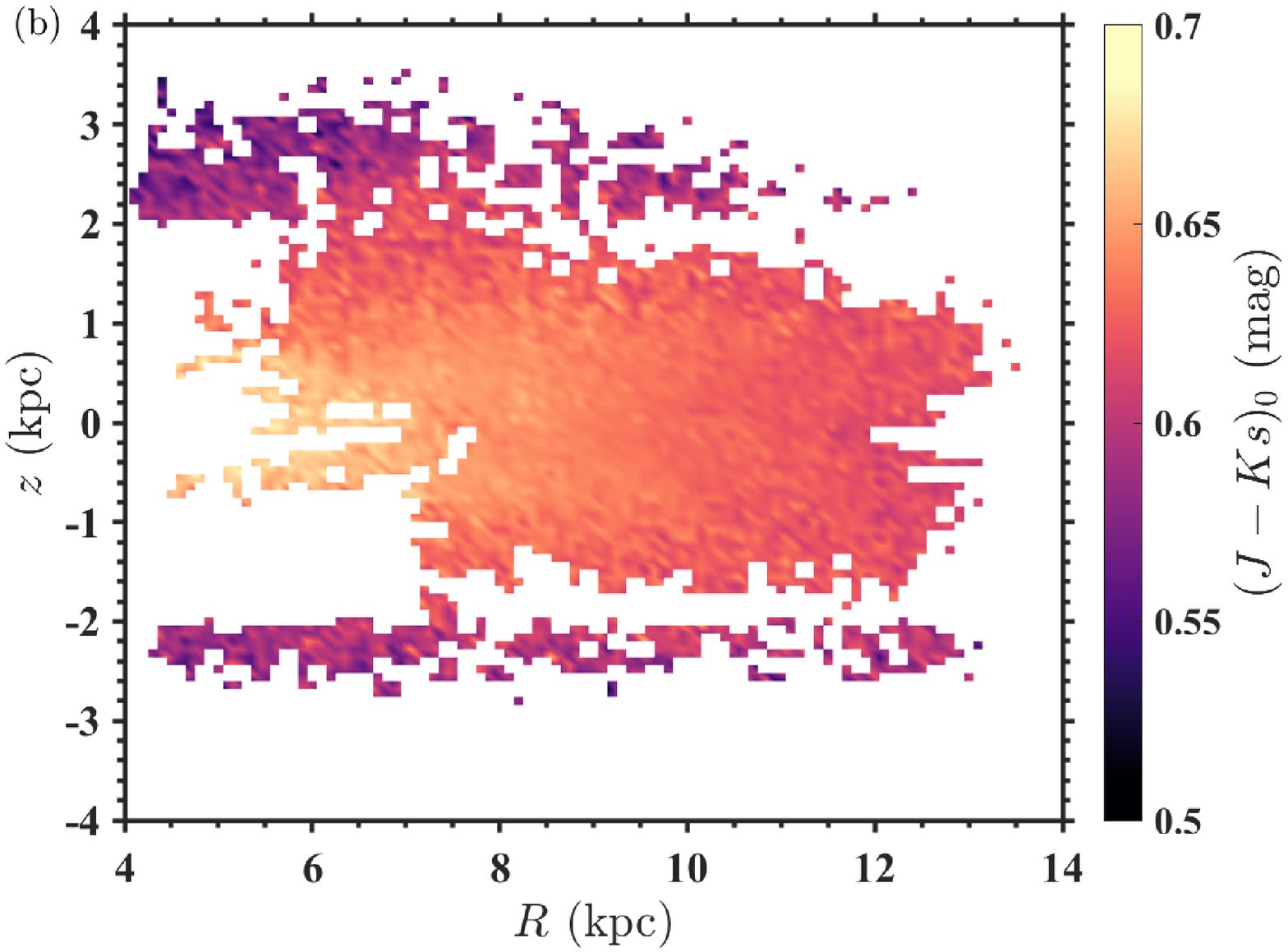}}
\quad
\subfigure{\includegraphics[width=3.5in]{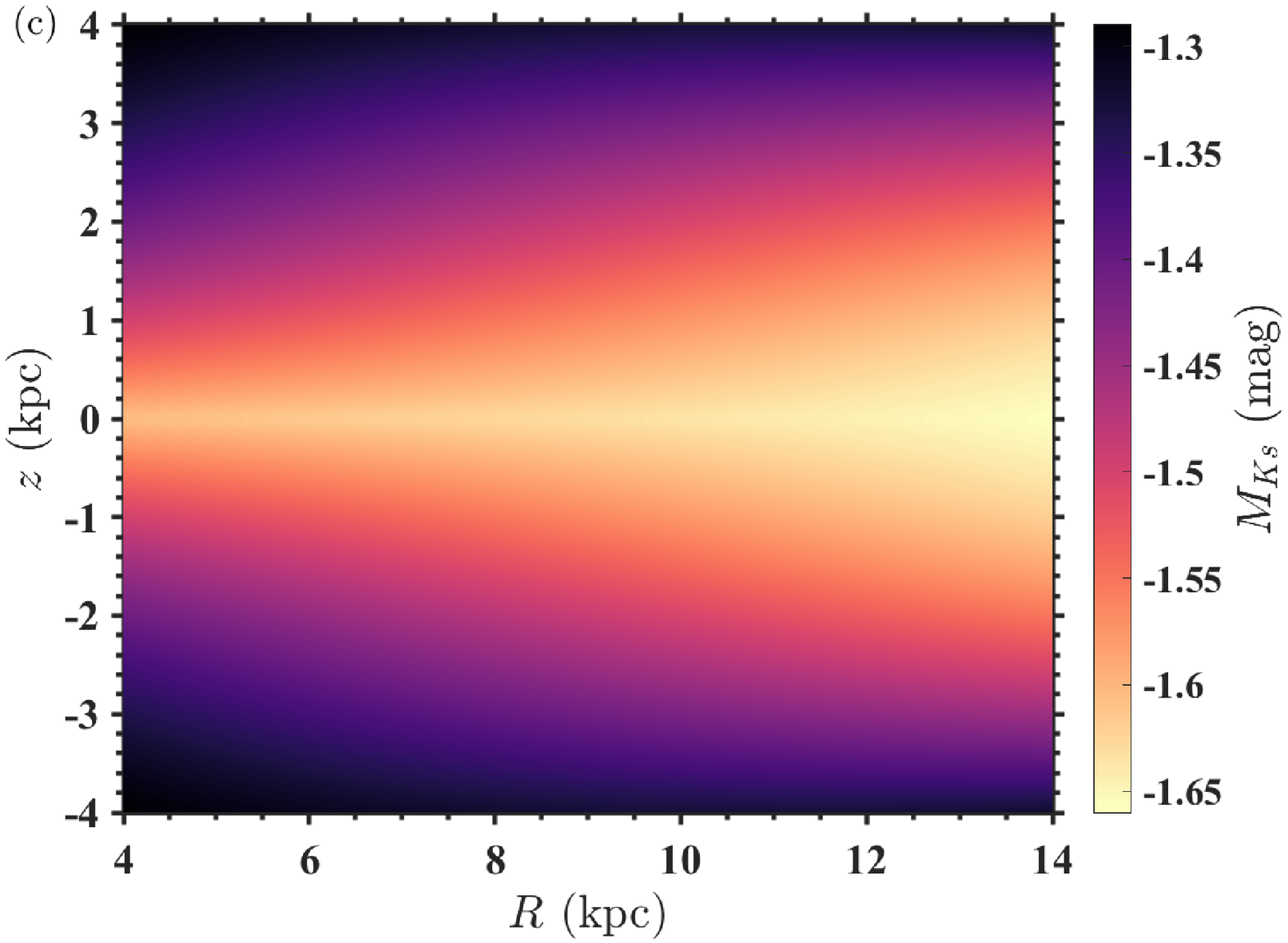}}
\subfigure{\includegraphics[width=3.5in]{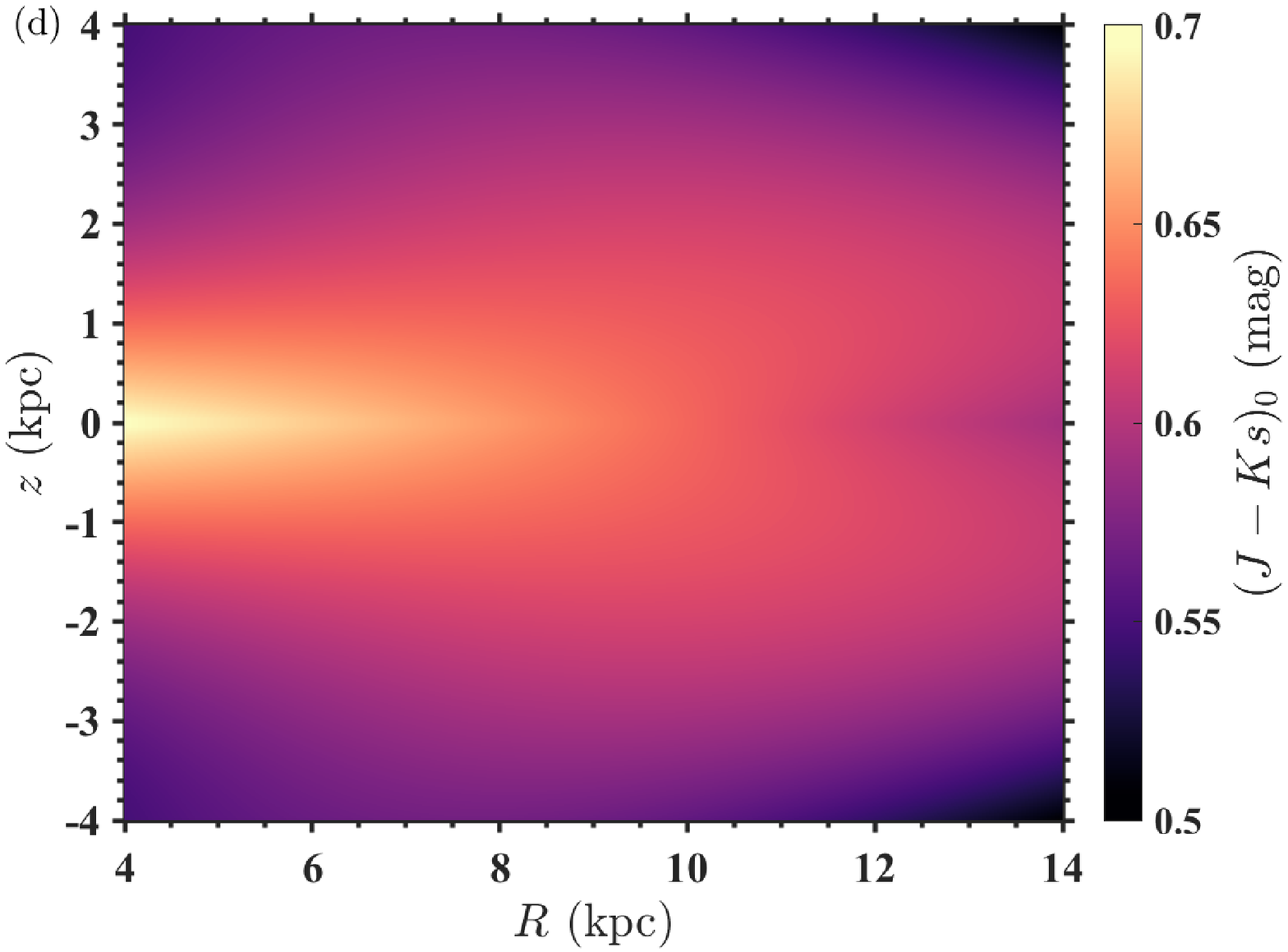}}
\caption{The absolute magnitude $\MKs$ and intrinsic color $(J-\Ks)_0$ distributions of RCs in the $R-z$ plane. The colorbars denote the values of $\MKs$ and $(J-\Ks)_0$. $\MKs$ and $(J-\Ks)_0$ derived from observations are plotted in panels (a) and (b), while $\MKs$ and $(J-\Ks)_0$ from our parameter maps are plotted in panels (c) and (d).}\label{Fig_jk_kmag}
\end{figure*}

To verify the functional form of the quadratic polynomial in observations, we investigated how $\mathrm{d}{M_\lambda}/\mathrm{d}R$ varies with $\,| z \,|$ in different bands. 
Figure~\ref{Fig_zr} shows the distributions of gradients $\mathrm{d}{M_\lambda}/\mathrm{d}R$ to $\,| z \,|$ in the$\Ks, J, z, G$ bands, respectively. 
The red, blue, and greed dots are gradients of LAMOST RCs, APOGEE RCs, and {\it Gaia} RCs, respectively. The black solid and dashed lines are not fitting lines but are taken from our final maps $M_{\lambda,\rm{map}}$ (see Section \ref{results}) by assuming quadratic and cubic polynomials in equation \ref{equ2}, respectively. From this figure, we find the quadratic and cubic polynomials agree well in $|z|<4$ kpc. The number of RCs with $|z|>4$ kpc is too small to constrain the maps. We therefore adopted a quadratic polynomial between $\mathrm{d}{M_\lambda}/\mathrm{d}R$ and $\,| z \,|$ in equation \ref{equ2} to establish our maps $M_{\lambda,\rm{map}}$. We also tried a two-exponent model and found that the quadratic polynomial model is more appropriate and more convenient to use.

In Figure~\ref{Fig_zr}, the trend of the gradients varies from one band to another. We can see that it changes from a concave function in IR bands to a convex function in optical bands. 
In the $\Ks$ band, the gradients show a decreasing trend followed by an increasing trend. 
When $\,| z \,|$ is small (dominated by thin-disk RCs) or large (dominated by halo RCs), $ \mathrm{d}{\MKs}/\mathrm{d}R$ are close to zero. 
At $\,| z \,| \sim$ 2 kpc, $\,| \mathrm{d}{\MKs}/\mathrm{d}R \,|$ is larger.  
The variation of $\mathrm{d}{\MKs}/\mathrm{d}R$ with $\,| z \,|$ satisfies an approximate quadratic polynomial.

The gradients of the LAMOST and APOGEE RCs show similar trends, which coincide with each other considering the uncertainty. Considering the similar gradients and small absolute magnitude bias, we combined the spectral RC samples of APOGEE and LAMOST to determine the maps. 
The combined sample effectively reduces the selection effect caused by the different coverage of LAMOST (northern sky, mainly low extinction regions) and APOGEE (disk dominated). 
At $|z|>2$ kpc, the gradient uncertainties of LAMOST and APOGEE RCs become larger due to the sharp decrease in the number of RCs. 
In these regions, the gradient uncertainties of {\it Gaia} RCs are small. The inclusion of {\it Gaia} RCs in the combined sample can avoid the external interpolation at $|z|>2$ kpc. Moreover, {\it Gaia} RCs show a small gradient at $|z|>3$ kpc, consistent with our understanding of the Galactic population. The following three paragraphs explain the details of how we constructed the combined RC sample.

To obtain the parameters of $M_{\lambda,{\rm map}} = f_\lambda(R, z)$, $a_n (n=1, 2, ..., 6)$, we performed Markov chain Monte Carlo (MCMC) simulations to fit equation~\ref{equ2} to $M_{\lambda,\rm{pred}}$ of RCs. To reduce the selection effect and completeness problem, we assembled an RC sample with RCs from APOGEE, LAMOST, and {\it Gaia}. 
First, we selected all APOGEE RCs and supplemented them with LAMOST RCs for boxes in $R-z$ space not covered by APOGEE. The space of $4<R<14$ kpc and $-4<z<4$ kpc was divided into $100\times100$ boxes. 
For boxes containing both APOGEE and LAMOST RCs, we also supplemented the LAMOST RCs with an average absolute magnitude similar to that of the APOGEE RCs.
Since the parameters of LAMOST are less accurate than those of APOGEE, the absolute magnitude distribution of LAMOST RCs is wider than that of APOGEE RCs in the same box. 
We selected LAMOST RCs according to the APOGEE RCs' absolute magnitude distribution. 
In each box, we required $ \,| M_{\lambda,\rm LA}-\langle M_{\lambda,\rm AP}\rangle \,| < 1 \sigma_{\lambda, \rm AP}$.  $M_{\lambda, \rm LA}$ denotes the absolute magnitudes of LAMOST RCs, $\langle M_{\lambda,\rm AP}\rangle$ and $\sigma_{\lambda, \rm AP}$ are the average absolute magnitude and the standard deviation of APOGEE RCs in each box, respectively. 
We also tested and found that adjusting the threshold from $1 \sigma_{\lambda, \rm AP}$ to 0 or $3 \sigma_{\lambda, \rm AP}$ hardly affect our final maps.

For the boxes not covered by APOGEE or LAMOST, we selected RCs from {\it Gaia} as a supplement.
The selected {\it Gaia} RCs satisfy the following criteria:  $\,| z \,|>2$ kpc and $\sigma_\varpi/\varpi<0.2$, and the extinction-corrected $G$-band absolute magnitude $-2.4<m_G-1.89\times(\GBP-\GRP)-(5\log d+10) <-0.9$ mag. These RCs are far from the disk, so there is almost no contamination from giants or dwarfs. We also excluded SRCs and red giants by the cuts of $0.4<(J-\Ks)_0<0.71$ mag and $M_{\Ks}+1.7\times(J-\Ks)_0<0$ mag. The second cut is 0.2 mag ($10\%$ in parallax) looser than the yellow-dashed line used in Figure \ref{Fig_RC1} to avoid introducing bias in absolute magnitude determination. A tighter selection cut, while improving the purity of the sample, would lose true RCs whose parallaxes are underestimated. 
We also only adopted RCs with declination in the range of $-36^\circ$ to $90^\circ$, and the extinction of these RCs can be read from the extinction map of \citet{2019ApJ...887...93G}. 
The multi-band absolute magnitudes, intrinsic colors $(J-\Ks)_0$ and $(\GBP-\GRP)_0$ were estimated.

The combined RC sample used to establish 3D maps of absolute magnitudes and intrinsic colors contains $\sim156,000$ RCs including $\sim42,000$ APOGEE RCs, $\sim67,000$ LAMOST RCs, and $\sim47,000$ {\it Gaia} RCs. 
MCMC simulations were first run in the $\Ks$ band, and we excluded $\sim13,000$ spectral RCs and $\sim29,000$ {\it Gaia} RCs with $|M_{\Ks,{\rm map}}-M_{\Ks,{\rm pred}}|>0.15$ mag to avoid overfitting at the edges of the model. Based on the remaining RCs, the polynomial coefficients $a_n$ of both absolute magnitude and intrinsic color functions were determined by MCMC simulations.

The multi-band coefficients are listed in Table~\ref{tab2}. The last column of Table~\ref{tab2} shows the RMSEs of absolute magnitudes and intrinsic colors between our 3D parameter maps and observed values for $\sim156,000$ RCs. The RMSEs are close to uncertainties of absolute magnitudes and intrinsic colors estimated by spectral parameters mentioned in Sections~\ref{RCabs} and ~\ref{RCint}. Based on the RMSEs, the 3D parameter maps can predict multi-band absolute magnitudes and intrinsic colors with an accuracy of $20\%$ better than the constant values. The overall offsets between our 3D parameter maps and observed values are less than 0.005 mag. The error analysis shows that our 3D parameter maps are reliable.

\subsection{Results of 3D Maps}\label{results}

From Table~\ref{tab2}, the average absolute magnitudes and intrinsic colors of RCs at any spatial location can be estimated from the 3D parameter maps with $M_\lambda = f_\lambda(l, b, {\rm distance}) = f_\lambda(R, z)$. The maps are suitable for RCs located at $4 <R<14$ kpc, $-4< z<4$ kpc. To better understand this map, we discuss some specific cases. 
\begin{enumerate}
\item We set $z=0$, equation~\ref{equ2} simplifies to $M_\lambda=a_3(R-8.0)+a_5$, which represents the variation of $M_\lambda$ in the middle plane. The slope $a_3$ becomes flat from optical bands to IR bands. 
Thus, in the $\Ks$ band or {\it WISE} bands, a constant absolute magnitude can be used to approximate calculate the distances of thin-disk RCs. 
\item We set $R=8$ kpc, equation~\ref{equ2} simplifies to $M_\lambda=a_4\,| z\,|+a_5+64a_6\,| z\,|$, which represents the variation of absolute magnitudes with $\,| z\,|$ at $R=8$ kpc. We find that the gradients increase toward long wavelength bands, and in the $V, r$, and $G$ bands, the gradients are close to zero. Thus, the constant absolute magnitudes in these bands can be used to estimate the distances of RCs in the long cylinder along $z$ with $R\sim8.0$ kpc. 
\item We set both $R=8$ kpc and $z=0$ kpc, equation~\ref{equ2} simplifies to $M_\lambda=a_5$, representing the multi-band absolute magnitudes of RCs in the solar neighborhood. 
\end{enumerate}

Figure~\ref{Fig_jk_kmag} shows $\MKs$ and $(J-\Ks)_0$ distributions of RCs in the $R-z$ plane from observations (panels a, b) and our maps (panels c, d). The observed average $\MKs$ and $(J-\Ks)_0$ are estimated from at least 4 RCs in each $0.1\times0.08$ kpc$^2$ bin. The corresponding $M_G$ and $(\GBP-\GRP)_0$ distributions are shown in Appendix Figure~\ref{Fig_bprp_Gmag}.  We find that our parameter maps reproduce the observed features well. In the $\MKs$ distribution, the flared disk is evident. Young RCs with brighter $\Ks$-band absolute magnitude can exist at higher heights $\,| z\,|$ when $R>8$ kpc. In the thin disk, the variation of $\MKs$ is small, varying by about 0.05 mag from $R=4$ kpc to $R=14$ kpc. At $\,| z\,| \sim 4$ kpc, the gradient of $M_\Ks$ with $R$ is small, and RCs are old halo stars. The gradients of $\mathrm{d}M_\lambda/\mathrm{d}\,| z\,|$ are different at different $R$. 

For {\it WISE} bands, the distributions of absolute magnitudes in the $R-z$ plane are similar to that of the $\Ks$ band. At shorter wavelengths, the slopes $\mathrm{d}M_\lambda/\mathrm{d}\rm{[M/H]}$ become steeper, making the distribution of $M_\lambda$ more distorted relative to $\MKs$. At a given age, metal-poor RCs are brighter. 
The one dex of metallicity change leads to an absolute magnitude change of 0.2 mag in the $J$ band and 0.5 mag in the $V$ band. 
Thus, we see a significant gradient between $M_G$ and $R$ in the thin disk (Appendix Figure~\ref{Fig_bprp_Gmag}). In the $M_G$ (also $M_V$ and $M_r$) distribution, the variation of $M_G$ with $\,| z\,|$ is not significant. Consider that the metallicity decreases and the age increases as $\,| z\,|$ increases. The former makes $M_\lambda$ brighter, while the latter makes $M_\lambda$ fainter, and these two effects counteract each other in the $V, r, G$ bands.

\section{Discussion}\label{section4}
\subsection{Validation of 3D Parameter Maps}\label{result_check}

\begin{figure*}[ht]
\centering
\includegraphics[width=\hsize]{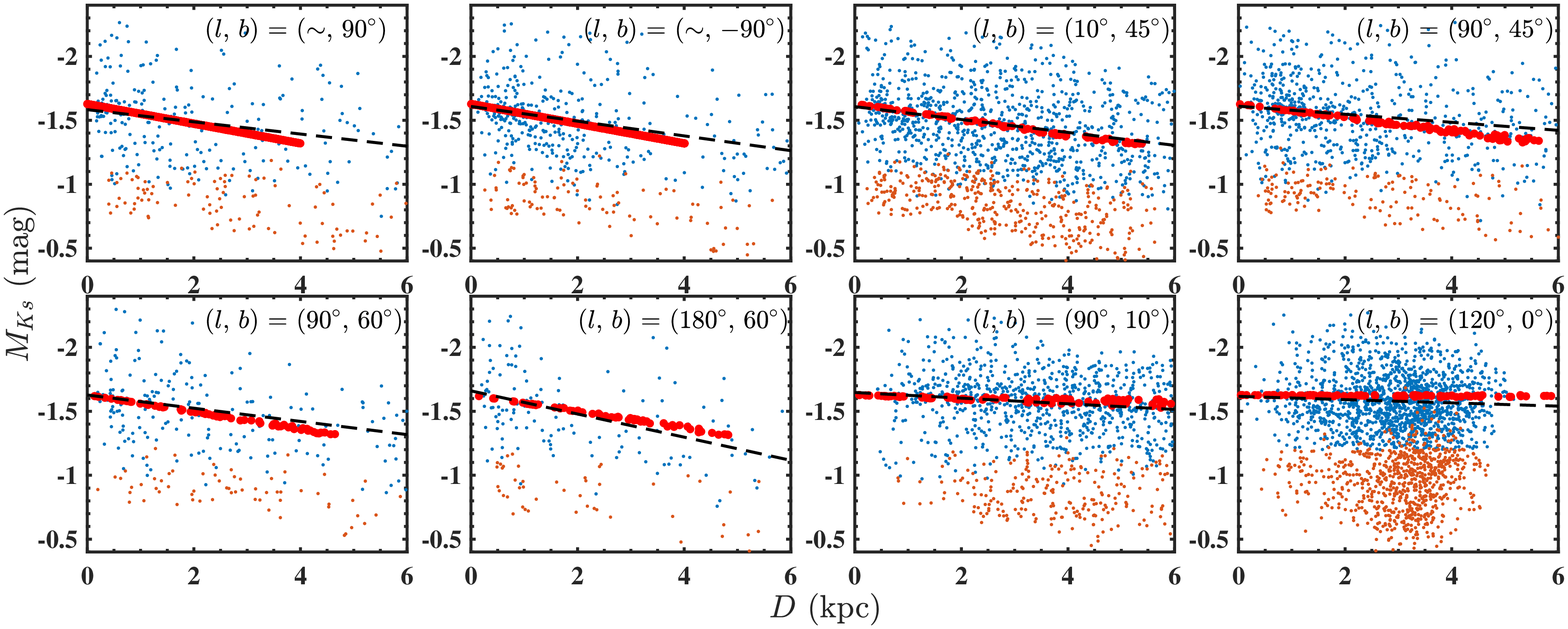}
\caption{The absolute magnitude $\MKs$ distributions of RCs along distance $D$ at different lines of sight.
Blue dots are distribution of the external test RC sample (spectral RCs have been excluded), and orange dots are the contamination. 
Large red dots indicate the predicted mean absolute magnitudes from our 3D maps. The black dashed lines are linear fits of the blue dots, indicating the variation of the mean absolute magnitude with distance.
The main contribution of the scatter of observed absolute magnitude is the uncertainties of the {\it Gaia} parallax. }
\label{Fig_pred_k}
\end{figure*}

\begin{figure*}[ht]
\centering
\includegraphics[width=\hsize]{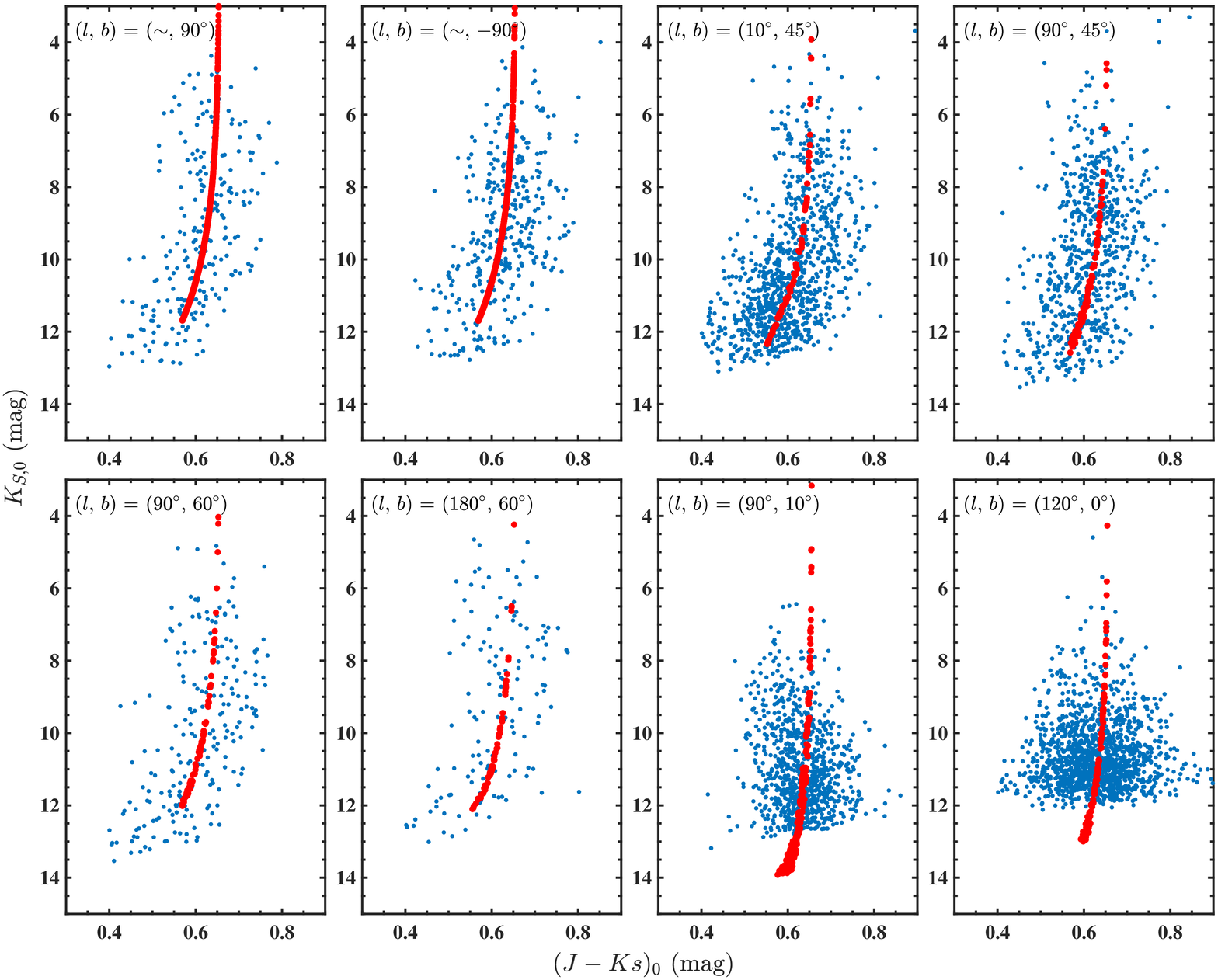}
\caption{The extinction-corrected CMDs of RCs in 2MASS bands $\Ks_{,0}$ vs. $(J-\Ks)_0$ at different lines of sight. Blue dots are distributions of the external test RC sample, while red dots show the distributions of predicted mean magnitudes and mean colors from our 3D maps.}
\label{Fig_pred_jk}
\end{figure*}

We constructed an external test RC sample (Table~\ref{tab_sample}) using {\it Gaia} data to assess the validity of the 3D parameter maps by comparing the observed results with the model predictions. 
The selection criteria for the test RC sample are $\sigma_\varpi/\varpi<0.2$ and  $-2.4 <m_G-1.89\times(\GBP-\GRP)-(5\log d+10) <-0.9$ mag. 
The contamination was excluded by $M_\Ks+1.7\times(J-\Ks)_0>0$ (see Sections \ref{RCsample} and \ref{3Dmapderive}). 
The spectral RCs from APOGEE and LAMOST, which are the main contributors to our 3D maps, were excluded to make the test RC sample more independent.

Eight typical regions with a $10^\circ$ box size ($2^\circ$ for $|b|\leq10^\circ$) at different Galactic longitudes and latitudes were selected for examination. Figure~\ref{Fig_pred_k} is an example showing a comparison of predicted and observed values of absolute magnitudes in the $\Ks$ band at different distances. The $r$-band and $W_{G,\GBP,\GRP}$-band figures are shown in Appendix Figures~\ref{Fig_pred_r} and \ref{Fig_pred_wg}.
The dots are the observed values of RCs (blue) and contamination (orange) with distances converted from the {\it Gaia} parallaxes and absolute magnitudes from $\MKs=m_\Ks -A_\Ks -5\log d - 10$. 
The extinction $A_\Ks$ was estimated based on Green's 3D extinction map \citep{2019ApJ...887...93G} and Wang's extinction coefficients \citep{2019ApJ...877..116W}.  
The black dashed lines are linear fits of the observed RC absolute magnitudes. 
The red dots are the mean absolute magnitudes predicted by our 3D absolute magnitude maps, which are consistent with the observed absolute magnitudes. 
At high latitudes, both predictions and observations show a slight increase of $\MKs$ with distance.
In contrast, at low latitudes, $\MKs$ hardly vary with distance.
 
The $M_\lambda$ distributions (Figures~\ref{Fig_pred_k}, \ref{Fig_pred_r}, and \ref{Fig_pred_wg}) show that the mean RC absolute magnitudes are not uniform across the different spatial locations of the Milky Way as a result of the effects of age and metallicity. 
The predictions of our 3D parameter maps (red dots) are in good agreement with the observed mean absolute magnitude distributions of RCs (black-dashed lines) at different spatial locations.
The scatter of the observed absolute magnitudes (blue dots) is mainly due to the $20\%$ uncertainty of the {\it Gaia} parallax.

We also examined our intrinsic color maps.  
Figure~\ref{Fig_pred_jk} is the 2MASS CMDs used to compare the observed results (blue dots) to our predictions (red dots). Similar figures for PS1 bands (Figure~\ref{Fig_pred_gr}) and {\it Gaia} bands (Figure~\ref{Fig_pred_bprp}) are shown in the Appendix.
For the observed results, the intrinsic color and extinction-corrected magnitude are determined by $(J-\Ks)_0=(J-\Ks)-E(J-\Ks)$ and $\Ks_{,0}=m_\Ks-A_\Ks$, respectively. 
Based on the 3D intrinsic color map, we derived the predicted intrinsic color. 
The predicted extinction-corrected magnitude is equal to the absolute magnitude from the 3D absolute magnitude map plus the distance modulus. 
We find that the observations are distributed around the predicted values.
The significant intrinsic color dispersion at the Galactic plane, such as the region of ($l, b$)=(120$^\circ$, 0$^\circ$), is due to the unidentified contamination or large extinction uncertainties. 
To summarize, the consistency of predictions and observations validates that our 3D parameter maps are reliable.

\subsection{Comparison with Previous Results}

The RC absolute magnitude, as the main parameter of RCs, has been investigated in many works.
In this section, we briefly summarize the measurements of the RC absolute magnitude in the literature and compare them with our measurements. 

In the last century, the $I$ band was often used to measure the distance of RCs, considering the constant value of $M_I$ \citep{1998ApJ...494L.219P}. 
However, the $I$-band RC absolute magnitude is still somewhat affected by the metallicity. The band used to measure RC distances has gradually shifted from $I$ band to near-IR $\Ks$ band because the measurement of distances in the $\Ks$ band is less affected by both extinction and metallicity. \citet{2000ApJ...539..732A} obtained the $K$-band absolute magnitude $-1.61\pm0.03$ mag, which was supported by the following measurements: $-1.61\pm0.04$ mag \citep{2002AJ....123.1603G}, $-1.613\pm0.015$ mag \citep{2012MNRAS.419.1637L}, $-1.626\pm0.057$ mag \citep{2017ApJ...840...77C}, $-1.61\pm0.01$ mag \citep{2017MNRAS.471..722H}, $-1.606\pm0.009$ mag \citep{2018A&A...609A.116R}, $-1.622\pm0.004$ mag \citep{2020MNRAS.493.4367C}. These values are in agreement with each other.
A slightly fainter $\MKs$ has also been reported, such as $-1.57\pm0.05$ mag \citep{2007A&A...463..559V}, $-1.54\pm0.04$ mag  \citep{2008A&A...488..935G}, $-1.53\pm0.01$ mag  \citep{2014MNRAS.441.1105F}, $-1.55\pm0.08$ mag  \citep{2015RAA....15.1166W}. Compared to the $\Ks$ band, the reported absolute magnitudes have greater variability in optical bands.
Recently, \citet{2020ApJ...893..108P} defined high-$\alpha$ and low-$\alpha$ populations in the [$\alpha$/Fe]$-$[Fe/H] plane and provided their absolute magnitudes separately. 
\citet{2020ApJS..249...29H} considered the effects of metallicity and age on $\MKs$. They used a third-order polynomial to describe the relationship between $MKs$ and age for RCs with four different metallicities. \citet{2021MNRAS.500.2590G} studied the variation of RC's mode absolute magnitude with $\,| z\,|$ in the solar neighborhood.

According to our maps, RC parameters (absolute magnitude and intrinsic color) are not represented by a constant or simple multiple constants in the vast majority of cases. When comparing absolute magnitudes obtained from different works, it is only meaningful to select RCs from similar regions or similar samples for comparison. The reason why the $\Ks$-band absolute magnitudes happen to be consistent in the past works is that the selected samples are nearby RCs or younger RCs. At the solar position, our 3D map gives an absolute magnitude of $1.628$ mag in the $Ks$ band, which is consistent with previously reported values. However, if an RC sample contains a higher proportion of thick-disk or halo RCs, the determined $\Ks$-band mean absolute magnitude will be fainter.
$M_\lambda$ is affected by both age and metallicity. The dependence of $M_\lambda$ on metallicity decreases from optical to IR bands, while the dependence of $M_\lambda$ on age is relatively similar in each band. Intrinsic colors are more affected by metallicity than age.  Considering that the distributions of metallicity and age can be inferred from the Galactic structure, we recommend using 3D parameter maps rather than a constant value to represent RC's absolute magnitude and intrinsic color. Position-dependent parameters would be more appropriate for RC distance measurements and comparisons between different works.

Our 3D absolute magnitude and intrinsic color maps, which take into account the population effects of the main structures of the Milky Way and the distribution of elemental abundances, can better represent the properties of RCs in the Milky Way.

\section{{\it Gaia} Red Clump sample}\label{GaiaRC}

In this section, we introduce how to use the 3D parameter maps to determine distances of RCs, and we also provide a whole-sky photometric RC sample based on the {\it Gaia} parallaxes. 

To obtain RC parameters based on 3D maps, we discuss the following three cases.
\begin{enumerate}
\item With known distances from external measurements, the absolute magnitude and the intrinsic color of an RC can be obtained directly from the 3D parameter map by ${\rm Value}_\lambda = f_\lambda(l, b, {\rm distance}) = f_\lambda(R, z)$. 
\item In the lack of distance information, iterations are needed to obtain the best distance and absolute amplitude from the apparent magnitude. 
\item If extinction is also considered, multi-band iterations are required to obtain the best distance, extinction, and absolute magnitude simultaneously. Alternatively, the extinction can be obtained from an external 3D extinction map.
\end{enumerate}

\begin{figure*}[ht]
\centering
\includegraphics[width=\hsize]{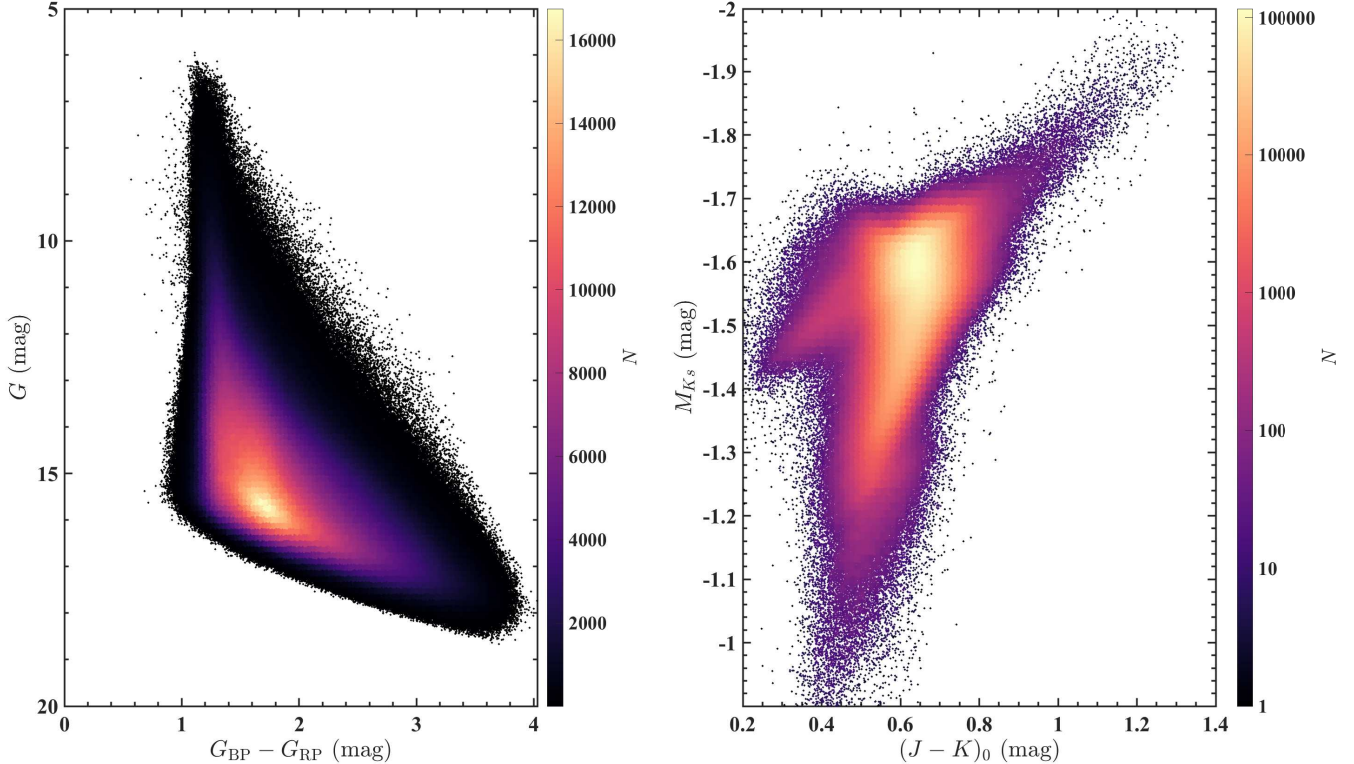}
\caption{The CMDs of 11 million RCs in optical bands $G$ vs. $(\GBP-\GRP)$ (left) and near-IR bands $\MKs$ vs. $(J-\Ks)_0$ (right), respectively. Most of RCs are concentrated around [$\MKs$, $(J-\Ks)_0$]=[-1.580, 0.645]. The color represents the number density of RCs.}
\label{Fig_RCsample}
\end{figure*}

In Section~\ref{result_check}, we have selected an external test RC sample from the {\it Gaia} catalog based on the roughly estimated Weisenheit absolute magnitude $M_{W_{G,\GBP,\GRP}}$. 
Here, we only adopted the criterion $-2.4<m_G-1.89\times(\GBP-\GRP)-(5\log d+10) <-0.9$ mag and obtained an initial sample of 15 million RC candidates. This sample contains contamination of dwarfs, SRCs, red giants, and blue giants. We calculated the distance of each candidate under the assumption of being an RC. 
The specific procedure is as follows.
First, We considered an initial distance modulus $({\rm DM})_0$ and an initial $V$-band extinction $(\AV)_0$: 
$({\rm DM})_0=\mKs-(-1.56)$ and $(\AV)_0=[(\GBP-\GRP)-1.18]\times2.394$, where the value of $-1.56$ is the mean value of $\MKs$, 1.18 is the mean value of intrinsic color $(\GBP-\GRP)_0$, and $2.394$ is the extinction coefficient. 
Then we created a set of [${\rm DM}$, $\Av$] with 100$\times$100 meshes to include all possible distance modulus and extinction. 
${\rm DM}$ is in the range of [$({\rm DM})_0-5\sigma_{{\rm DM}}-0.2 (\AV)_0$, $({\rm DM})_0+5\sigma_{{\rm DM}}$], where $\sigma_{{\rm DM}}$, the scatter of ${\rm DM}$, is assumed to be 0.14 mag. 
Since $\AKs/\AV$ was found ranging from 0.08 to 0.12, we took a larger factor of 0.2 to avoid a possible underestimation of extinction. 
As $(\GBP-\GRP)$ can deviate from the mean value by 0.5 mag, we set $\AV$ in the range of [$(\AV)_0-0.5\times 2.394$, $(\AV)_0+0.5\times 2.394$].  
After that, we calculated the distance modulus at each given band by the equation ${\rm DM}_\lambda=m_\lambda-M_\lambda-A_\lambda$. 
For each [${\rm DM}$, $\Av$] value in the meshes, the $M_\lambda$ is inferred from our 3D absolute magnitude map with the distance converted from ${\rm DM}$, and $A_\lambda$ is $\AV$ multiplied by the extinction coefficient $C_{\rm ext, \lambda}$ taken from \citep{2019ApJ...877..116W}. The dimension of ${\rm DM}_\lambda$ is $15\ {\rm million}\times6\times100\times100$. We used six bands from {\it Gaia} and 2MASS to determine the multi-band mean distance moduli $\langle{\rm DM}_\lambda \rangle$ and its standard deviation $\sigma_{{\rm DM}_\lambda}$. For bands with no photometry or poor-quality photometry, we set their weights to zero. At last, we obtained the optimal distance modulus and extinction for 15 million RC candidates by searching the lowest $\sigma_{{\rm DM}_\lambda}$ in $100\times100$ meshes.

As for contamination, blue giants can be eliminated by the color information. SRCs and dwarfs are fainter than RCs, as well as some red giants are brighter than RCs, both of which can be removed by better distance information. Here we combine extinction and distance to improve the purity of the sample.
If a dwarf or SRC is mistaken for an RC, the distance is overestimated. If a giant is mistaken for an RC, the distance is underestimated. Assuming that the RC sample is pure, the distribution of extinction with distance should display a concentrated monotonic non-decreasing trend in each line of sight. 
Contamination well below the extinction--distance profile are dwarfs and a fraction of SRCs, and those well above are giants. 
To remove contamination, we determined the extinction--distance profile for RC candidates in each line of sight. These candidates were divided into solid angle bins, where each bin contained 1,000 stars. For each bin, we obtained the distribution of $\langle{\rm DM}\rangle$ vs. $\langle \AV \rangle$ and fitted this distribution by the GPR. 
RC candidates that lie outside the fitting line by $2\sigma$ (95\%) are considered as contamination and excluded.
After completing the sample purification, we obtained the final {\it Gaia} photometric RC sample containing 11 million RCs listed in Table~\ref{tab_sample}.

Figure \ref{Fig_RCsample} shows the density distributions of 11 million RCs in the ($\GBP-\GRP$, $\G$) CMD and the extinction-corrected ($(J-\Ks)_0$, $\MKs$) CMD. The RC $\MKs$ is distributed between -2.0 mag and -0.9 mag, and most of the RCs concentrated around [$\MKs$, $(J-\Ks)_0$]=[-1.600, 0.645]. There are also about 50,000 RCs with abnormal intrinsic colors, i.e., $(J-\Ks)_0>0.8$ mag or $(J-\Ks)_0<0.4$ mag, which may be the result of inaccurate extinction estimation or are not genuine RCs. The catalog for 11 million RCs is available in the repository \footnote{doi:\url{https://doi.org/10.5281/zenodo.5140055}}, which includes information of position, distance modulus, extinction, corrected distance modulus, corrected extinction, and uncertainty of distance modulus. The corrected extinction and distance modulus are determined by using a uniform mean extinction at each distance and line of sight based on the extinction--distance profile. The corrected extinction avoids negative extinction and is more suitable for RCs in the low extinction regions ($\AV<0.5$ mag). However, it loses the differential extinction information and is less suitable for RCs in high extinction regions ($\AV>1.0$ mag). The distance modulus uncertainty is the root of the sum of squares of the internal standard deviation and the map RMSE. We adopted an RMSE of 0.14 mag based on Table \ref{tab2}.
The catalog also contains a `flag' column to distinguish between high and low probability RCs. The high probability RCs satisfying $\MKs+1.7\times(J-\Ks)_0<0$ are marked with flag$=1$. Low probability RCs are more inclined to be contaminated by faint red giants and SRCs.
For high probability RCs, the distances converted from high-quality {\it Gaia} parallaxes ($\sigma_\varpi/\varpi<0.1$) are in good agreement with our distances. For $\sim188,000$ RCs with $\sigma_\varpi/\varpi<0.05, \varpi>1$, the distance modulus difference is ${\rm DM_{map}}-{\rm DM}_{\it Gaia}=0.022\pm0.226$ mag, while for $\sim892,000$ RCs with $\sigma_\varpi/\varpi<0.1, \varpi>0.5$, the distance modulus difference is ${\rm DM_{map}}-{\rm DM}_{\it Gaia}=0.033\pm0.229$ mag. The distance modulus difference reflects the purity of the sample, and an RC sample heavily contaminated by red giants and SRCs will have a larger distance modulus difference. Therefore, the criterion of $\MKs+1.7\times(J-\Ks)_0<0$ is very effective for selecting high-probability RC samples.

Comparing to the photometric RC sample from \citet{2020MNRAS.495.3087L}, we find that 61\% of their best sample ($\sim405,000$, $20\%$ contamination rate) and 46\% of their main sample (2.6 million, $33\%$ contamination rate) are included in our sample. The vast majority of RCs missed in our sample are due to our absolute magnitude selection, which is limited by the accuracy of the {\it Gaia} parallax. 
If our selection criteria were applied to their RC sample, 83\% of their RCs were included in our sample. The other 15\% of their RCs were excluded by our extinction-distance selection. The inclusion rate of our sample decreases as the purity of their sample decreases, which also hints that our sample successfully avoids some contamination.

\section{Prospect}\label{prospect}

In this section, we discuss future work that can be done with RC's 3D parameter maps.
The first thing we want to know is how much the warped disk affects our 3D parameter maps. 
In the outer disk, a warped disk has been found through OB-type stars, Cepheids, RCs, and red giants \citep{2018MNRAS.481L..21P, 2019NatAs...3..320C, 2019A&A...627A.150R}. Unlike Cepheids, the warp model traced by other tracers is not well established by the current surveys. The warp was found in precession \citep{2019NatAs...3..320C} and evolved with time \citep{2020NatAs...4..590P,2020ApJ...897..119W}. The density of different stars in the warp also varies. Considering these reasons, it is not appropriate to use Cepheid's warp model for RC. We expect to obtain the number density of warp RCs in the line of sight where warp's $z$-height reaches its maximum or minimum value through the future database. By comparing them with the number density of thick-disk RCs in the same region, we can evaluate the effect of warp on our 3D parameter maps.

The bulge region has not been included in our 3D parameter maps due to the small number of RCs with spectra and poorly understood structure. The bulge traced by RCs shows an X-shaped structure \citep{2010ApJ...721L..28N,2010ApJ...724.1491M}, which is not supported by other tracers such as RR Lyrae \citep{2015ApJ...811..113P}, Mira \citep{2017ApJ...836..218L}. The kinematic information and element abundances reveal that the bulge possesses two components with different age distributions \citep{2019MNRAS.490.4740B,2020ApJ...900....4S}, and the younger component having a more pronounced X-shaped structure \citep{2020MNRAS.492.3128G}. With $6000-7000$ bulge stars in APOGEE DR16, the metallicity and age distributions in the bulge were studied in more detail \citep{2020ApJ...901..109H,2020arXiv200712915Q}. However, most of these stars are red giants with luminosity brighter than RCs. A recent work studying of RCs in the bulge with spectral information was based on hundreds of RCs \citep{2021ApJ...907...47L}. Therefore, to build 3D absolute magnitude maps of RCs in the bulge, we need to wait for a more in-depth IR spectroscopic survey. Alternatively, we can use red giants to build the maps after establishing the correlation between red giants and RCs in terms of elemental abundance distributions and density distribution.

The 3D parameter maps of RCs in the Milky Way are helpful to future work on the 3D structure of the Magellanic Clouds (MCs). Based on RCs in the Survey of the MAgellanic Stellar History \citep[SMASH]{2018ApJ...866...90C} and the Optical Gravitational Lensing Experiment OGLE \citep{2021ApJS..252...23S}, the 3D structure of the MCs was seen in more detail. 
The [$\alpha$/H] and [Fe/H] distributions in the MCs were better studied by $\sim3600$ red giants with APOGEE spectra \citep{2020ApJ...895...88N}. 
\citet{2021ApJ...910..121N} investigated the color--metallicity relation of RCs in the MCs, which is an important parameter to anchor the distance of RCs. 
From our 3D parameter maps, age is the most important parameter for deriving the RC distance. 
With more spectral data in the future, the age distribution of RCs in the MCs and its association with the 3D structure of MCs will be investigated.

In contrast to RCs, the absolute magnitudes of red giants are related not only to their age and metallicity but also to their intrinsic color. We expect to build 3D absolute magnitude maps of red giants in the future. Compared to RCs, many red giants are brighter and can measure farther distances. Besides, red giants can track older and more metal-poor environments than RCs. A better understanding of the dependence of absolute magnitudes on age and metallicity for red giants, especially for the tip of the red giant branch, will help the optimization of the Hubble constant \citep{2019ApJ...882...34F} and the discovery of new structures in the Galactic halo \citep{2021arXiv210409515C}. With millions of spectral red giants, the Milky Way is one of the best places to study the absolute magnitude of red giants.

\section{Summary}\label{Conclusion}

To make RCs a more accurate and convenient distance indicator, we investigated the effects of age and metallicity on their parameters, including multi-band absolute magnitudes and intrinsic colors. We established position-dependent 3D maps for RCs' mean absolute magnitudes and intrinsic colors based on a combination of spectroscopic, astrometric, and photometric data from {\it Gaia}, APOGEE, LAMOST, APASS, PS1, 2MASS, and {\it WISE} surveys. 
The main results of this work are as follows.

1. For 42,947 APOGEE RCs and 93,542 LAMOST RCs, we selected nearby RCs with low extinction to build machine learners between the absolute magnitude $M_\lambda$ and the intrinsic color $(\lambda_1-\lambda_2)_0$ with the spectral parameters by Gaussian process regression. Then $M_\lambda$ and $(\lambda_1-\lambda_2)_0$ were calculated for all spectral RCs. 
We find that $M_\lambda$ becomes fainter with increasing age and metallicity, and the metallicity dependence decreases with increasing wavelength. 
The intrinsic colors are strongly related to metallicity and effective temperature. 
In contrast, the age effect on intrinsic colors is small, implying that the age effect on $M_\lambda$ is almost consistent in different bands. 

2. We further analyzed the variation of $M_\lambda$ and $(\lambda_1-\lambda_2)_0$ in spatial distribution and constructed an ($R, z$) dependence function to predict the $M_\lambda$ and $(\lambda_1-\lambda_2)_0$ distribution of the Galactic RCs. 
In establishing the function, we considered the main Galactic structure, elemental abundance distribution, and population distribution of the Milky Way.
Finally, we presented for the first time the 3D maps of RC's multi-band absolute magnitudes and intrinsic colors in a volume of $4 <R<14$ kpc, $0<\,| z \,|<4$ kpc. 
For RCs in the thin disk, the variation of $M_\lambda$ with the Galactocentric radius $R$ is not as apparent in IR bands as in optical bands.  In the solar neighborhood, the absolute magnitude of RCs in the $z$-direction is more consistent in optical bands.
The $\Ks$-band absolute magnitude of our 3D maps for RCs in the solar neighborhood is consistent with previously reported values.

3. Based on {\it Gaia}'s EDR3 parallax, our 3D parameter maps, and the extinction--distance profile selection, we obtained the largest all-sky photometric RC sample to date, containing 11 million stars.

4. With our 3D parameter maps, more appropriate mean absolute magnitudes and mean intrinsic colors for RCs without spectral information can be obtained based on spatial positions ($l,b,d$). 
In studies that take RCs as indicators, such as the 3D structure of the Milky Way, the 3D extinction maps, and the extinction law, we recommend using position-dependent absolute magnitudes (intrinsic colors) rather than a constant value to reduce the systematic uncertainty.

\acknowledgments{We thank the referee for very insightful and helpful suggestions/comments. 
This work is supported by the National Key Research and Development Program of China, grant 2019YFA0405504. 
It is also supported by the National Natural Science Foundation of China (NSFC) through the projects 12003046, 11903045, 12173047, 12133002, and 11973001. 
We acknowledge the science research grants from the China Manned Space Project with No. CMS-CSST-220221-A09.
This work has made use of data from the surveys by LAMOST, {\it Gaia}, APOGEE, APASS, Pan-STARRS1, 2MASS, and {\it WISE}.
Guoshoujing Telescope (the Large Sky Area Multi-Object Fiber Spectroscopic Telescope, LAMOST) is a National Major Scientific Project built by the Chinese Academy of Sciences. Funding for the project has been provided by the National Development and Reform Commission. LAMOST is operated and managed by the National Astronomical Observatories, Chinese Academy of Sciences. 
APOGEE survey is part of Sloan Digital Sky Survey (SDSS) IV. 
SDSS-IV acknowledges support and resources from the Center for High Performance Computing  at the University of Utah. SDSS-IV is managed by the Astrophysical Research Consortium for the Participating Institutions of the SDSS Collaboration (\url{https://www.sdss.org}). 
This work has made use of data from the European Space Agency (ESA) mission
{\it Gaia} (\url{https://www.cosmos.esa.int/gaia}), processed by the {\it Gaia}
Data Processing and Analysis Consortium (DPAC,
\url{https://www.cosmos.esa.int/web/gaia/dpac/consortium}). Funding for the DPAC
has been provided by national institutions, in particular the institutions
participating in the {\it Gaia} Multilateral Agreement.
This research has made use of the APASS database, located at the AAVSO web site. Funding for APASS has been provided by the Robert Martin Ayers Sciences Fund.
This work has made use of Pan-STARRS1 data (https://outerspace.stsci. edu/display/PANSTARRS). Fourteen organizations in six nations (plus two funding organizations) supported the Pan-STARRS1 survey. 
The Two Micron All Sky Survey is a joint project of the University of Massachusetts and the Infrared Processing and Analysis Center/California Institute of Technology, funded by the NASA and the NSF. 
The Wide-field Infrared Survey Explorer is a joint project of the University of California, Los Angeles, and the Jet Propulsion Laboratory/California Institute of Technology, funded by the NASA.
}

\bibliography{reference}{}

\begin{thebibliography}{}
\expandafter\ifx\csname natexlab\endcsname\relax\def\natexlab#1{#1}\fi
\providecommand{\url}[1]{\href{#1}{#1}}
\providecommand{\dodoi}[1]{doi:~\href{http://doi.org/#1}{\nolinkurl{#1}}}
\providecommand{\doeprint}[1]{\href{http://ascl.net/#1}{\nolinkurl{http://ascl.net/#1}}}
\providecommand{\doarXiv}[1]{\href{https://arxiv.org/abs/#1}{\nolinkurl{https://arxiv.org/abs/#1}}}

\bibitem[{{Ahumada} {et~al.}(2020){Ahumada}, {Prieto}, {Almeida}, {Anders},
  {Anderson}, {Andrews}, {Anguiano}, {Arcodia}, {Armengaud}, {Aubert}, {Avila},
  {Avila-Reese}, {Badenes}, {Balland}, {Barger}, {Barrera-Ballesteros}, {Basu},
  {Bautista}, {Beaton}, {Beers}, {Benavides}, {Bender}, {Bernardi}, {Bershady},
  {Beutler}, {Bidin}, {Bird}, {Bizyaev}, {Blanc}, {Blanton}, {Boquien},
  {Borissova}, {Bovy}, {Brandt}, {Brinkmann}, {Brownstein}, {Bundy}, {Bureau},
  {Burgasser}, {Burtin}, {Cano-D{\'\i}az}, {Capasso}, {Cappellari}, {Carrera},
  {Chabanier}, {Chaplin}, {Chapman}, {Cherinka}, {Chiappini}, {Doohyun Choi},
  {Chojnowski}, {Chung}, {Clerc}, {Coffey}, {Comerford}, {Comparat}, {da
  Costa}, {Cousinou}, {Covey}, {Crane}, {Cunha}, {Ilha}, {Dai}, {Damsted},
  {Darling}, {Davidson}, {Davies}, {Dawson}, {De}, {de la Macorra}, {De Lee},
  {Queiroz}, {Deconto Machado}, {de la Torre}, {Dell'Agli}, {du Mas des
  Bourboux}, {Diamond-Stanic}, {Dillon}, {Donor}, {Drory}, {Duckworth},
  {Dwelly}, {Ebelke}, {Eftekharzadeh}, {Davis Eigenbrot}, {Elsworth},
  {Eracleous}, {Erfanianfar}, {Escoffier}, {Fan}, {Farr},
  {Fern{\'a}ndez-Trincado}, {Feuillet}, {Finoguenov}, {Fofie},
  {Fraser-McKelvie}, {Frinchaboy}, {Fromenteau}, {Fu}, {Galbany}, {Garcia},
  {Garc{\'\i}a-Hern{\'a}ndez}, {Oehmichen}, {Ge}, {Maia}, {Geisler}, {Gelfand},
  {Goddy}, {Gonzalez-Perez}, {Grabowski}, {Green}, {Grier}, {Guo}, {Guy},
  {Harding}, {Hasselquist}, {Hawken}, {Hayes}, {Hearty}, {Hekker}, {Hogg},
  {Holtzman}, {Horta}, {Hou}, {Hsieh}, {Huber}, {Hunt}, {Chitham}, {Imig},
  {Jaber}, {Angel}, {Johnson}, {Jones}, {J{\"o}nsson}, {Jullo}, {Kim},
  {Kinemuchi}, {Kirkpatrick}, {Kite}, {Klaene}, {Kneib}, {Kollmeier}, {Kong},
  {Kounkel}, {Krishnarao}, {Lacerna}, {Lan}, {Lane}, {Law}, {Le Goff}, {Leung},
  {Lewis}, {Li}, {Lian}, {Lin}, {Long}, {Longa-Pe{\~n}a}, {Lundgren}, {Lyke},
  {Ted Mackereth}, {MacLeod}, {Majewski}, {Manchado}, {Maraston}, {Martini},
  {Masseron}, {Masters}, {Mathur}, {McDermid}, {Merloni}, {Merrifield},
  {M{\'e}sz{\'a}ros}, {Miglio}, {Minniti}, {Minsley}, {Miyaji}, {Mohammad},
  {Mosser}, {Mueller}, {Muna}, {Mu{\~n}oz-Guti{\'e}rrez}, {Myers}, {Nadathur},
  {Nair}, {Nandra}, {do Nascimento}, {Nevin}, {Newman}, {Nidever}, {Nitschelm},
  {Noterdaeme}, {O'Connell}, {Olmstead}, {Oravetz}, {Oravetz}, {Osorio},
  {Pace}, {Padilla}, {Palanque-Delabrouille}, {Palicio}, {Pan}, {Pan},
  {Parker}, {Paviot}, {Peirani}, {Ram{\'r}ez}, {Penny}, {Percival},
  {Perez-Fournon}, {P{\'e}rez-R{\`a}fols}, {Petitjean}, {Pieri},
  {Pinsonneault}, {Poovelil}, {Povick}, {Prakash}, {Price-Whelan}, {Raddick},
  {Raichoor}, {Ray}, {Rembold}, {Rezaie}, {Riffel}, {Riffel}, {Rix}, {Robin},
  {Roman-Lopes}, {Rom{\'a}n-Z{\'u}{\~n}iga}, {Rose}, {Ross}, {Rossi},
  {Rowlands}, {Rubin}, {Salvato}, {S{\'a}nchez}, {S{\'a}nchez-Menguiano},
  {S{\'a}nchez-Gallego}, {Sayres}, {Schaefer}, {Schiavon}, {Schimoia},
  {Schlafly}, {Schlegel}, {Schneider}, {Schultheis}, {Schwope}, {Seo},
  {Serenelli}, {Shafieloo}, {Shamsi}, {Shao}, {Shen}, {Shetrone}, {Shirley},
  {Aguirre}, {Simon}, {Skrutskie}, {Slosar}, {Smethurst}, {Sobeck}, {Sodi},
  {Souto}, {Stark}, {Stassun}, {Steinmetz}, {Stello}, {Stermer},
  {Storchi-Bergmann}, {Streblyanska}, {Stringfellow}, {Stutz}, {Su{\'a}rez},
  {Sun}, {Taghizadeh-Popp}, {Talbot}, {Tayar}, {Thakar}, {Theriault}, {Thomas},
  {Thomas}, {Tinker}, {Tojeiro}, {Toledo}, {Tremonti}, {Troup}, {Tuttle},
  {Unda-Sanzana}, {Valentini}, {Vargas-Gonz{\'a}lez}, {Vargas-Maga{\~n}a},
  {V{\'a}zquez-Mata}, {Vivek}, {Wake}, {Wang}, {Weaver}, {Weijmans}, {Wild},
  {Wilson}, {Wilson}, {Wolthuis}, {Wood-Vasey}, {Yan}, {Yang}, {Y{\`e}che},
  {Zamora}, {Zarrouk}, {Zasowski}, {Zhang}, {Zhao}, {Zhao}, {Zheng}, {Zheng},
  {Zhu}, \& {Zou}}]{2020ApJS..249....3A}
{Ahumada}, R., {Prieto}, C.~A., {Almeida}, A., {et~al.} 2020, \apjs, 249, 3,
  \dodoi{10.3847/1538-4365/ab929e}

\bibitem[{{Alves}(2000)}]{2000ApJ...539..732A}
{Alves}, D.~R. 2000, \apj, 539, 732, \dodoi{10.1086/309278}

\bibitem[{{Anders} {et~al.}(2017){Anders}, {Chiappini}, {Minchev}, {Miglio},
  {Montalb{\'a}n}, {Mosser}, {Rodrigues}, {Santiago}, {Baudin}, {Beers}, {da
  Costa}, {Garc{\'\i}a}, {Garc{\'\i}a-Hern{\'a}ndez}, {Holtzman}, {Maia},
  {Majewski}, {Mathur}, {Noels-Grotsch}, {Pan}, {Schneider}, {Schultheis},
  {Steinmetz}, {Valentini}, \& {Zamora}}]{2017A&A...600A..70A}
{Anders}, F., {Chiappini}, C., {Minchev}, I., {et~al.} 2017, \aap, 600, A70,
  \dodoi{10.1051/0004-6361/201629363}

\bibitem[{{Bland-Hawthorn} \& {Gerhard}(2016)}]{2016ARA&A..54..529B}
{Bland-Hawthorn}, J., \& {Gerhard}, O. 2016, \araa, 54, 529,
  \dodoi{10.1146/annurev-astro-081915-023441}

\bibitem[{{Bovy} {et~al.}(2019){Bovy}, {Leung}, {Hunt}, {Mackereth},
  {Garc{\'\i}a-Hern{\'a}ndez}, \& {Roman-Lopes}}]{2019MNRAS.490.4740B}
{Bovy}, J., {Leung}, H.~W., {Hunt}, J. A.~S., {et~al.} 2019, \mnras, 490, 4740,
  \dodoi{10.1093/mnras/stz2891}

\bibitem[{{Bovy} {et~al.}(2016){Bovy}, {Rix}, {Schlafly}, {Nidever},
  {Holtzman}, {Shetrone}, \& {Beers}}]{2016ApJ...823...30B}
{Bovy}, J., {Rix}, H.-W., {Schlafly}, E.~F., {et~al.} 2016, \apj, 823, 30,
  \dodoi{10.3847/0004-637X/823/1/30}

\bibitem[{{Chambers} {et~al.}(2016){Chambers}, {Magnier}, {Metcalfe},
  {Flewelling}, {Huber}, {Waters}, {Denneau}, {Draper}, {Farrow}, {Finkbeiner},
  {Holmberg}, {Koppenhoefer}, {Price}, {Rest}, {Saglia}, {Schlafly}, {Smartt},
  {Sweeney}, {Wainscoat}, {Burgett}, {Chastel}, {Grav}, {Heasley}, {Hodapp},
  {Jedicke}, {Kaiser}, {Kudritzki}, {Luppino}, {Lupton}, {Monet}, {Morgan},
  {Onaka}, {Shiao}, {Stubbs}, {Tonry}, {White}, {Ba{\~n}ados}, {Bell},
  {Bender}, {Bernard}, {Boegner}, {Boffi}, {Botticella}, {Calamida},
  {Casertano}, {Chen}, {Chen}, {Cole}, {Deacon}, {Frenk}, {Fitzsimmons},
  {Gezari}, {Gibbs}, {Goessl}, {Goggia}, {Gourgue}, {Goldman}, {Grant},
  {Grebel}, {Hambly}, {Hasinger}, {Heavens}, {Heckman}, {Henderson}, {Henning},
  {Holman}, {Hopp}, {Ip}, {Isani}, {Jackson}, {Keyes}, {Koekemoer}, {Kotak},
  {Le}, {Liska}, {Long}, {Lucey}, {Liu}, {Martin}, {Masci}, {McLean}, {Mindel},
  {Misra}, {Morganson}, {Murphy}, {Obaika}, {Narayan}, {Nieto-Santisteban},
  {Norberg}, {Peacock}, {Pier}, {Postman}, {Primak}, {Rae}, {Rai}, {Riess},
  {Riffeser}, {Rix}, {R{\"o}ser}, {Russel}, {Rutz}, {Schilbach}, {Schultz},
  {Scolnic}, {Strolger}, {Szalay}, {Seitz}, {Small}, {Smith}, {Soderblom},
  {Taylor}, {Thomson}, {Taylor}, {Thakar}, {Thiel}, {Thilker}, {Unger},
  {Urata}, {Valenti}, {Wagner}, {Walder}, {Walter}, {Watters}, {Werner},
  {Wood-Vasey}, \& {Wyse}}]{2016arXiv161205560C}
{Chambers}, K.~C., {Magnier}, E.~A., {Metcalfe}, N., {et~al.} 2016, arXiv
  e-prints, arXiv:1612.05560.
\newblock \doarXiv{1612.05560}

\bibitem[{{Chan} \& {Bovy}(2020)}]{2020MNRAS.493.4367C}
{Chan}, V.~C., \& {Bovy}, J. 2020, \mnras, 493, 4367,
  \dodoi{10.1093/mnras/staa571}

\bibitem[{{Chen} {et~al.}(2019){Chen}, {Wang}, {Deng}, {de Grijs}, {Liu}, \&
  {Tian}}]{2019NatAs...3..320C}
{Chen}, X., {Wang}, S., {Deng}, L., {et~al.} 2019, Nature Astronomy, 3, 320,
  \dodoi{10.1038/s41550-018-0686-7}

\bibitem[{{Chen} {et~al.}(2017){Chen}, {Casagrande}, {Zhao}, {Bovy}, {Silva
  Aguirre}, {Zhao}, \& {Jia}}]{2017ApJ...840...77C}
{Chen}, Y.~Q., {Casagrande}, L., {Zhao}, G., {et~al.} 2017, \apj, 840, 77,
  \dodoi{10.3847/1538-4357/aa6d0f}

\bibitem[{{Choi} {et~al.}(2018){Choi}, {Nidever}, {Olsen}, {Blum}, {Besla},
  {Zaritsky}, {van der Marel}, {Bell}, {Gallart}, {Cioni}, {Johnson}, {Vivas},
  {Saha}, {de Boer}, {No{\"e}l}, {Monachesi}, {Massana}, {Conn},
  {Martinez-Delgado}, {Mu{\~n}oz}, \& {Stringfellow}}]{2018ApJ...866...90C}
{Choi}, Y., {Nidever}, D.~L., {Olsen}, K., {et~al.} 2018, \apj, 866, 90,
  \dodoi{10.3847/1538-4357/aae083}

\bibitem[{{Cohen} {et~al.}(2003){Cohen}, {Wheaton}, \&
  {Megeath}}]{Cohen2003AJ....126.1090C}
{Cohen}, M., {Wheaton}, W.~A., \& {Megeath}, S.~T. 2003, \aj, 126, 1090,
  \dodoi{10.1086/376474}

\bibitem[{{Conroy} {et~al.}(2021){Conroy}, {Naidu}, {Garavito-Camargo},
  {Besla}, {Zaritsky}, {Bonaca}, \& {Johnson}}]{2021arXiv210409515C}
{Conroy}, C., {Naidu}, R.~P., {Garavito-Camargo}, N., {et~al.} 2021, Nature,
  592, 534.
\newblock \doarXiv{2104.09515}

\bibitem[{{Cui} {et~al.}(2012){Cui}, {Zhao}, {Chu}, {Li}, {Li}, {Zhang}, {Su},
  {Yao}, {Wang}, {Xing}, {Li}, {Zhu}, {Wang}, {Gu}, {Luo}, {Xu}, {Zhang},
  {Liu}, {Zhang}, {Yang}, {Cao}, {Chen}, {Chen}, {Chen}, {Chen}, {Chu}, {Feng},
  {Gong}, {Hou}, {Hu}, {Hu}, {Hu}, {Jia}, {Jiang}, {Jiang}, {Jiang}, {Jin},
  {Li}, {Li}, {Li}, {Liu}, {Liu}, {Lu}, {Mao}, {Men}, {Qi}, {Qi}, {Shi},
  {Tang}, {Tao}, {Wang}, {Wang}, {Wang}, {Wang}, {Wang}, {Wang}, {Wang},
  {Wang}, {Wang}, {Wang}, {Wang}, {Wang}, {Xu}, {Xu}, {Yang}, {Yu}, {Yuan},
  {Yuan}, {Zhai}, {Zhang}, {Zhang}, {Zhang}, {Zhao}, {Zhou}, {Zhou}, {Zhu}, \&
  {Zou}}]{2012RAA....12.1197C}
{Cui}, X.-Q., {Zhao}, Y.-H., {Chu}, Y.-Q., {et~al.} 2012, Research in Astronomy
  and Astrophysics, 12, 1197, \dodoi{10.1088/1674-4527/12/9/003}

\bibitem[{{De Silva} {et~al.}(2015){De Silva}, {Freeman}, {Bland-Hawthorn},
  {Martell}, {de Boer}, {Asplund}, {Keller}, {Sharma}, {Zucker}, {Zwitter},
  {Anguiano}, {Bacigalupo}, {Bayliss}, {Beavis}, {Bergemann}, {Campbell},
  {Cannon}, {Carollo}, {Casagrande}, {Casey}, {Da Costa}, {D'Orazi}, {Dotter},
  {Duong}, {Heger}, {Ireland}, {Kafle}, {Kos}, {Lattanzio}, {Lewis}, {Lin},
  {Lind}, {Munari}, {Nataf}, {O'Toole}, {Parker}, {Reid}, {Schlesinger},
  {Sheinis}, {Simpson}, {Stello}, {Ting}, {Traven}, {Watson}, {Wittenmyer},
  {Yong}, \& {{\v{Z}}erjal}}]{2015MNRAS.449.2604D}
{De Silva}, G.~M., {Freeman}, K.~C., {Bland-Hawthorn}, J., {et~al.} 2015,
  \mnras, 449, 2604, \dodoi{10.1093/mnras/stv327}

\bibitem[{{Deng} {et~al.}(2012){Deng}, {Newberg}, {Liu}, {Carlin}, {Beers},
  {Chen}, {Chen}, {Christlieb}, {Grillmair}, {Guhathakurta}, {Han}, {Hou},
  {Lee}, {L{\'e}pine}, {Li}, {Liu}, {Pan}, {Sellwood}, {Wang}, {Wang}, {Yang},
  {Yanny}, {Zhang}, {Zhang}, {Zheng}, \& {Zhu}}]{2012RAA....12..735D}
{Deng}, L.-C., {Newberg}, H.~J., {Liu}, C., {et~al.} 2012, Research in
  Astronomy and Astrophysics, 12, 735, \dodoi{10.1088/1674-4527/12/7/003}

\bibitem[{{Eisenstein} {et~al.}(2011){Eisenstein}, {Weinberg}, {Agol},
  {Aihara}, {Allende Prieto}, {Anderson}, {Arns}, {Aubourg}, {Bailey},
  {Balbinot}, {Barkhouser}, {Beers}, {Berlind}, {Bickerton}, {Bizyaev},
  {Blanton}, {Bochanski}, {Bolton}, {Bosman}, {Bovy}, {Brandt}, {Breslauer},
  {Brewington}, {Brinkmann}, {Brown}, {Brownstein}, {Burger}, {Busca},
  {Campbell}, {Cargile}, {Carithers}, {Carlberg}, {Carr}, {Chang}, {Chen},
  {Chiappini}, {Comparat}, {Connolly}, {Cortes}, {Croft}, {Cunha}, {da Costa},
  {Davenport}, {Dawson}, {De Lee}, {Porto de Mello}, {de Simoni}, {Dean},
  {Dhital}, {Ealet}, {Ebelke}, {Edmondson}, {Eiting}, {Escoffier}, {Esposito},
  {Evans}, {Fan}, {Femen{\'\i}a Castell{\'a}}, {Dutra Ferreira}, {Fitzgerald},
  {Fleming}, {Font-Ribera}, {Ford}, {Frinchaboy}, {Garc{\'\i}a P{\'e}rez},
  {Gaudi}, {Ge}, {Ghezzi}, {Gillespie}, {Gilmore}, {Girardi}, {Gott}, {Gould},
  {Grebel}, {Gunn}, {Hamilton}, {Harding}, {Harris}, {Hawley}, {Hearty},
  {Hennawi}, {Gonz{\'a}lez Hern{\'a}ndez}, {Ho}, {Hogg}, {Holtzman},
  {Honscheid}, {Inada}, {Ivans}, {Jiang}, {Jiang}, {Johnson}, {Jordan},
  {Jordan}, {Kauffmann}, {Kazin}, {Kirkby}, {Klaene}, {Knapp}, {Kneib},
  {Kochanek}, {Koesterke}, {Kollmeier}, {Kron}, {Lampeitl}, {Lang}, {Lawler},
  {Le Goff}, {Lee}, {Lee}, {Leisenring}, {Lin}, {Liu}, {Long}, {Loomis},
  {Lucatello}, {Lundgren}, {Lupton}, {Ma}, {Ma}, {MacDonald}, {Mack},
  {Mahadevan}, {Maia}, {Majewski}, {Makler}, {Malanushenko}, {Malanushenko},
  {Mandelbaum}, {Maraston}, {Margala}, {Maseman}, {Masters}, {McBride},
  {McDonald}, {McGreer}, {McMahon}, {Mena Requejo}, {M{\'e}nard},
  {Miralda-Escud{\'e}}, {Morrison}, {Mullally}, {Muna}, {Murayama}, {Myers},
  {Naugle}, {Neto}, {Nguyen}, {Nichol}, {Nidever}, {O'Connell}, {Ogando},
  {Olmstead}, {Oravetz}, {Padmanabhan}, {Paegert}, {Palanque-Delabrouille},
  {Pan}, {Pandey}, {Parejko}, {P{\^a}ris}, {Pellegrini}, {Pepper}, {Percival},
  {Petitjean}, {Pfaffenberger}, {Pforr}, {Phleps}, {Pichon}, {Pieri}, {Prada},
  {Price-Whelan}, {Raddick}, {Ramos}, {Reid}, {Reyle}, {Rich}, {Richards},
  {Rieke}, {Rieke}, {Rix}, {Robin}, {Rocha-Pinto}, {Rockosi}, {Roe},
  {Rollinde}, {Ross}, {Ross}, {Rossetto}, {S{\'a}nchez}, {Santiago}, {Sayres},
  {Schiavon}, {Schlegel}, {Schlesinger}, {Schmidt}, {Schneider}, {Sellgren},
  {Shelden}, {Sheldon}, {Shetrone}, {Shu}, {Silverman}, {Simmerer}, {Simmons},
  {Sivarani}, {Skrutskie}, {Slosar}, {Smee}, {Smith}, {Snedden}, {Stassun},
  {Steele}, {Steinmetz}, {Stockett}, {Stollberg}, {Strauss}, {Szalay},
  {Tanaka}, {Thakar}, {Thomas}, {Tinker}, {Tofflemire}, {Tojeiro}, {Tremonti},
  {Vargas Maga{\~n}a}, {Verde}, {Vogt}, {Wake}, {Wan}, {Wang}, {Weaver},
  {White}, {White}, {Wilson}, {Wisniewski}, {Wood-Vasey}, {Yanny}, {Yasuda},
  {Y{\`e}che}, {York}, {Young}, {Zasowski}, {Zehavi}, \&
  {Zhao}}]{2011AJ....142...72E}
{Eisenstein}, D.~J., {Weinberg}, D.~H., {Agol}, E., {et~al.} 2011, \aj, 142,
  72, \dodoi{10.1088/0004-6256/142/3/72}

\bibitem[{{Francis} \& {Anderson}(2014)}]{2014MNRAS.441.1105F}
{Francis}, C., \& {Anderson}, E. 2014, \mnras, 441, 1105,
  \dodoi{10.1093/mnras/stu631}

\bibitem[{{Freedman} {et~al.}(2019){Freedman}, {Madore}, {Hatt}, {Hoyt},
  {Jang}, {Beaton}, {Burns}, {Lee}, {Monson}, {Neeley}, {Phillips}, {Rich}, \&
  {Seibert}}]{2019ApJ...882...34F}
{Freedman}, W.~L., {Madore}, B.~F., {Hatt}, D., {et~al.} 2019, \apj, 882, 34,
  \dodoi{10.3847/1538-4357/ab2f73}

\bibitem[{{Gaia Collaboration} {et~al.}(2018){Gaia Collaboration}, {Brown},
  {Vallenari}, {Prusti}, {de Bruijne}, {Babusiaux}, {Bailer-Jones}, {Biermann},
  {Evans}, {Eyer}, {Jansen}, {Jordi}, {Klioner}, {Lammers}, {Lindegren},
  {Luri}, {Mignard}, {Panem}, {Pourbaix}, {Randich}, {Sartoretti}, {Siddiqui},
  {Soubiran}, {van Leeuwen}, {Walton}, {Arenou}, {Bastian}, {Cropper},
  {Drimmel}, {Katz}, {Lattanzi}, {Bakker}, {Cacciari}, {Casta{\~n}eda},
  {Chaoul}, {Cheek}, {De Angeli}, {Fabricius}, {Guerra}, {Holl}, {Masana},
  {Messineo}, {Mowlavi}, {Nienartowicz}, {Panuzzo}, {Portell}, {Riello},
  {Seabroke}, {Tanga}, {Th{\'e}venin}, {Gracia-Abril}, {Comoretto},
  {Garcia-Reinaldos}, {Teyssier}, {Altmann}, {Andrae}, {Audard},
  {Bellas-Velidis}, {Benson}, {Berthier}, {Blomme}, {Burgess}, {Busso},
  {Carry}, {Cellino}, {Clementini}, {Clotet}, {Creevey}, {Davidson}, {De
  Ridder}, {Delchambre}, {Dell'Oro}, {Ducourant},
  {Fern{\'a}ndez-Hern{\'a}ndez}, {Fouesneau}, {Fr{\'e}mat}, {Galluccio},
  {Garc{\'\i}a-Torres}, {Gonz{\'a}lez-N{\'u}{\~n}ez}, {Gonz{\'a}lez-Vidal},
  {Gosset}, {Guy}, {Halbwachs}, {Hambly}, {Harrison}, {Hern{\'a}ndez},
  {Hestroffer}, {Hodgkin}, {Hutton}, {Jasniewicz}, {Jean-Antoine-Piccolo},
  {Jordan}, {Korn}, {Krone-Martins}, {Lanzafame}, {Lebzelter}, {L{\"o}ffler},
  {Manteiga}, {Marrese}, {Mart{\'\i}n-Fleitas}, {Moitinho}, {Mora}, {Muinonen},
  {Osinde}, {Pancino}, {Pauwels}, {Petit}, {Recio-Blanco}, {Richards},
  {Rimoldini}, {Robin}, {Sarro}, {Siopis}, {Smith}, {Sozzetti}, {S{\"u}veges},
  {Torra}, {van Reeven}, {Abbas}, {Abreu Aramburu}, {Accart}, {Aerts},
  {Altavilla}, {{\'A}lvarez}, {Alvarez}, {Alves}, {Anderson}, {Andrei},
  {Anglada Varela}, {Antiche}, {Antoja}, {Arcay}, {Astraatmadja}, {Bach},
  {Baker}, {Balaguer-N{\'u}{\~n}ez}, {Balm}, {Barache}, {Barata}, {Barbato},
  {Barblan}, {Barklem}, {Barrado}, {Barros}, {Barstow}, {Bartholom{\'e}
  Mu{\~n}oz}, {Bassilana}, {Becciani}, {Bellazzini}, {Berihuete}, {Bertone},
  {Bianchi}, {Bienaym{\'e}}, {Blanco-Cuaresma}, {Boch}, {Boeche}, {Bombrun},
  {Borrachero}, {Bossini}, {Bouquillon}, {Bourda}, {Bragaglia}, {Bramante},
  {Breddels}, {Bressan}, {Brouillet}, {Br{\"u}semeister}, {Brugaletta},
  {Bucciarelli}, {Burlacu}, {Busonero}, {Butkevich}, {Buzzi}, {Caffau},
  {Cancelliere}, {Cannizzaro}, {Cantat-Gaudin}, {Carballo}, {Carlucci},
  {Carrasco}, {Casamiquela}, {Castellani}, {Castro-Ginard}, {Charlot},
  {Chemin}, {Chiavassa}, {Cocozza}, {Costigan}, {Cowell}, {Crifo}, {Crosta},
  {Crowley}, {Cuypers}, {Dafonte}, {Damerdji}, {Dapergolas}, {David}, {David},
  {de Laverny}, {De Luise}, {De March}, {de Martino}, {de Souza}, {de Torres},
  {Debosscher}, {del Pozo}, {Delbo}, {Delgado}, {Delgado}, {Di Matteo},
  {Diakite}, {Diener}, {Distefano}, {Dolding}, {Drazinos}, {Dur{\'a}n},
  {Edvardsson}, {Enke}, {Eriksson}, {Esquej}, {Eynard Bontemps}, {Fabre},
  {Fabrizio}, {Faigler}, {Falc{\~a}o}, {Farr{\`a}s Casas}, {Federici},
  {Fedorets}, {Fernique}, {Figueras}, {Filippi}, {Findeisen}, {Fonti},
  {Fraile}, {Fraser}, {Fr{\'e}zouls}, {Gai}, {Galleti}, {Garabato},
  {Garc{\'\i}a-Sedano}, {Garofalo}, {Garralda}, {Gavel}, {Gavras}, {Gerssen},
  {Geyer}, {Giacobbe}, {Gilmore}, {Girona}, {Giuffrida}, {Glass}, {Gomes},
  {Granvik}, {Gueguen}, {Guerrier}, {Guiraud}, {Guti{\'e}rrez-S{\'a}nchez},
  {Haigron}, {Hatzidimitriou}, {Hauser}, {Haywood}, {Heiter}, {Helmi}, {Heu},
  {Hilger}, {Hobbs}, {Hofmann}, {Holland}, {Huckle}, {Hypki}, {Icardi},
  {Jan{\ss}en}, {Jevardat de Fombelle}, {Jonker}, {Juh{\'a}sz}, {Julbe},
  {Karampelas}, {Kewley}, {Klar}, {Kochoska}, {Kohley}, {Kolenberg},
  {Kontizas}, {Kontizas}, {Koposov}, {Kordopatis}, {Kostrzewa-Rutkowska},
  {Koubsky}, {Lambert}, {Lanza}, {Lasne}, {Lavigne}, {Le Fustec}, {Le
  Poncin-Lafitte}, {Lebreton}, {Leccia}, {Leclerc}, {Lecoeur-Taibi},
  {Lenhardt}, {Leroux}, {Liao}, {Licata}, {Lindstr{\o}m}, {Lister}, {Livanou},
  {Lobel}, {L{\'o}pez}, {Managau}, {Mann}, {Mantelet}, {Marchal}, {Marchant},
  {Marconi}, {Marinoni}, {Marschalk{\'o}}, {Marshall}, {Martino}, {Marton},
  {Mary}, {Massari}, {Matijevi{\v{c}}}, {Mazeh}, {McMillan}, {Messina},
  {Michalik}, {Millar}, {Molina}, {Molinaro}, {Moln{\'a}r}, {Montegriffo},
  {Mor}, {Morbidelli}, {Morel}, {Morris}, {Mulone}, {Muraveva}, {Musella},
  {Nelemans}, {Nicastro}, {Noval}, {O'Mullane}, {Ord{\'e}novic},
  {Ord{\'o}{\~n}ez-Blanco}, {Osborne}, {Pagani}, {Pagano}, {Pailler},
  {Palacin}, {Palaversa}, {Panahi}, {Pawlak}, {Piersimoni}, {Pineau}, {Plachy},
  {Plum}, {Poggio}, {Poujoulet}, {Pr{\v{s}}a}, {Pulone}, {Racero}, {Ragaini},
  {Rambaux}, {Ramos-Lerate}, {Regibo}, {Reyl{\'e}}, {Riclet}, {Ripepi}, {Riva},
  {Rivard}, {Rixon}, {Roegiers}, {Roelens}, {Romero-G{\'o}mez}, {Rowell},
  {Royer}, {Ruiz-Dern}, {Sadowski}, {Sagrist{\`a} Sell{\'e}s}, {Sahlmann},
  {Salgado}, {Salguero}, {Sanna}, {Santana-Ros}, {Sarasso}, {Savietto},
  {Schultheis}, {Sciacca}, {Segol}, {Segovia}, {S{\'e}gransan}, {Shih},
  {Siltala}, {Silva}, {Smart}, {Smith}, {Solano}, {Solitro}, {Sordo}, {Soria
  Nieto}, {Souchay}, {Spagna}, {Spoto}, {Stampa}, {Steele},
  {Steidelm{\"u}ller}, {Stephenson}, {Stoev}, {Suess}, {Surdej}, {Szabados},
  {Szegedi-Elek}, {Tapiador}, {Taris}, {Tauran}, {Taylor}, {Teixeira},
  {Terrett}, {Teyssandier}, {Thuillot}, {Titarenko}, {Torra Clotet}, {Turon},
  {Ulla}, {Utrilla}, {Uzzi}, {Vaillant}, {Valentini}, {Valette}, {van Elteren},
  {Van Hemelryck}, {van Leeuwen}, {Vaschetto}, {Vecchiato}, {Veljanoski},
  {Viala}, {Vicente}, {Vogt}, {von Essen}, {Voss}, {Votruba}, {Voutsinas},
  {Walmsley}, {Weiler}, {Wertz}, {Wevers}, {Wyrzykowski}, {Yoldas},
  {{\v{Z}}erjal}, {Ziaeepour}, {Zorec}, {Zschocke}, {Zucker}, {Zurbach}, \&
  {Zwitter}}]{2018A&A...616A...1G}
{Gaia Collaboration}, {Brown}, A.~G.~A., {Vallenari}, A., {et~al.} 2018, \aap,
  616, A1, \dodoi{10.1051/0004-6361/201833051}

\bibitem[{{Gaia Collaboration} {et~al.}(2021){Gaia Collaboration}, {Brown},
  {Vallenari}, {Prusti}, {de Bruijne}, {Babusiaux}, {Biermann}, {Creevey},
  {Evans}, {Eyer}, {Hutton}, {Jansen}, {Jordi}, {Klioner}, {Lammers},
  {Lindegren}, {Luri}, {Mignard}, {Panem}, {Pourbaix}, {Randich}, {Sartoretti},
  {Soubiran}, {Walton}, {Arenou}, {Bailer-Jones}, {Bastian}, {Cropper},
  {Drimmel}, {Katz}, {Lattanzi}, {van Leeuwen}, {Bakker}, {Cacciari},
  {Casta{\~n}eda}, {De Angeli}, {Ducourant}, {Fabricius}, {Fouesneau},
  {Fr{\'e}mat}, {Guerra}, {Guerrier}, {Guiraud}, {Jean-Antoine Piccolo},
  {Masana}, {Messineo}, {Mowlavi}, {Nicolas}, {Nienartowicz}, {Pailler},
  {Panuzzo}, {Riclet}, {Roux}, {Seabroke}, {Sordo}, {Tanga}, {Th{\'e}venin},
  {Gracia-Abril}, {Portell}, {Teyssier}, {Altmann}, {Andrae}, {Bellas-Velidis},
  {Benson}, {Berthier}, {Blomme}, {Brugaletta}, {Burgess}, {Busso}, {Carry},
  {Cellino}, {Cheek}, {Clementini}, {Damerdji}, {Davidson}, {Delchambre},
  {Dell'Oro}, {Fern{\'a}ndez-Hern{\'a}ndez}, {Galluccio}, {Garc{\'\i}a-Lario},
  {Garcia-Reinaldos}, {Gonz{\'a}lez-N{\'u}{\~n}ez}, {Gosset}, {Haigron},
  {Halbwachs}, {Hambly}, {Harrison}, {Hatzidimitriou}, {Heiter},
  {Hern{\'a}ndez}, {Hestroffer}, {Hodgkin}, {Holl}, {Jan{\ss}en}, {Jevardat de
  Fombelle}, {Jordan}, {Krone-Martins}, {Lanzafame}, {L{\"o}ffler}, {Lorca},
  {Manteiga}, {Marchal}, {Marrese}, {Moitinho}, {Mora}, {Muinonen}, {Osborne},
  {Pancino}, {Pauwels}, {Petit}, {Recio-Blanco}, {Richards}, {Riello},
  {Rimoldini}, {Robin}, {Roegiers}, {Rybizki}, {Sarro}, {Siopis}, {Smith},
  {Sozzetti}, {Ulla}, {Utrilla}, {van Leeuwen}, {van Reeven}, {Abbas}, {Abreu
  Aramburu}, {Accart}, {Aerts}, {Aguado}, {Ajaj}, {Altavilla}, {{\'A}lvarez},
  {{\'A}lvarez Cid-Fuentes}, {Alves}, {Anderson}, {Anglada Varela}, {Antoja},
  {Audard}, {Baines}, {Baker}, {Balaguer-N{\'u}{\~n}ez}, {Balbinot}, {Balog},
  {Barache}, {Barbato}, {Barros}, {Barstow}, {Bartolom{\'e}}, {Bassilana},
  {Bauchet}, {Baudesson-Stella}, {Becciani}, {Bellazzini}, {Bernet}, {Bertone},
  {Bianchi}, {Blanco-Cuaresma}, {Boch}, {Bombrun}, {Bossini}, {Bouquillon},
  {Bragaglia}, {Bramante}, {Breedt}, {Bressan}, {Brouillet}, {Bucciarelli},
  {Burlacu}, {Busonero}, {Butkevich}, {Buzzi}, {Caffau}, {Cancelliere},
  {C{\'a}novas}, {Cantat-Gaudin}, {Carballo}, {Carlucci}, {Carnerero},
  {Carrasco}, {Casamiquela}, {Castellani}, {Castro-Ginard}, {Castro Sampol},
  {Chaoul}, {Charlot}, {Chemin}, {Chiavassa}, {Cioni}, {Comoretto}, {Cooper},
  {Cornez}, {Cowell}, {Crifo}, {Crosta}, {Crowley}, {Dafonte}, {Dapergolas},
  {David}, {David}, {de Laverny}, {De Luise}, {De March}, {De Ridder}, {de
  Souza}, {de Teodoro}, {de Torres}, {del Peloso}, {del Pozo}, {Delbo},
  {Delgado}, {Delgado}, {Delisle}, {Di Matteo}, {Diakite}, {Diener},
  {Distefano}, {Dolding}, {Eappachen}, {Edvardsson}, {Enke}, {Esquej}, {Fabre},
  {Fabrizio}, {Faigler}, {Fedorets}, {Fernique}, {Fienga}, {Figueras},
  {Fouron}, {Fragkoudi}, {Fraile}, {Franke}, {Gai}, {Garabato},
  {Garcia-Gutierrez}, {Garc{\'\i}a-Torres}, {Garofalo}, {Gavras}, {Gerlach},
  {Geyer}, {Giacobbe}, {Gilmore}, {Girona}, {Giuffrida}, {Gomel}, {Gomez},
  {Gonzalez-Santamaria}, {Gonz{\'a}lez-Vidal}, {Granvik},
  {Guti{\'e}rrez-S{\'a}nchez}, {Guy}, {Hauser}, {Haywood}, {Helmi}, {Hidalgo},
  {Hilger}, {H{\l}adczuk}, {Hobbs}, {Holland}, {Huckle}, {Jasniewicz},
  {Jonker}, {Juaristi Campillo}, {Julbe}, {Karbevska}, {Kervella}, {Khanna},
  {Kochoska}, {Kontizas}, {Kordopatis}, {Korn}, {Kostrzewa-Rutkowska},
  {Kruszy{\'n}ska}, {Lambert}, {Lanza}, {Lasne}, {Le Campion}, {Le Fustec},
  {Lebreton}, {Lebzelter}, {Leccia}, {Leclerc}, {Lecoeur-Taibi}, {Liao},
  {Licata}, {Lindstr{\o}m}, {Lister}, {Livanou}, {Lobel}, {Madrero Pardo},
  {Managau}, {Mann}, {Marchant}, {Marconi}, {Marcos Santos}, {Marinoni},
  {Marocco}, {Marshall}, {Martin Polo}, {Mart{\'\i}n-Fleitas}, {Masip},
  {Massari}, {Mastrobuono-Battisti}, {Mazeh}, {McMillan}, {Messina},
  {Michalik}, {Millar}, {Mints}, {Molina}, {Molinaro}, {Moln{\'a}r},
  {Montegriffo}, {Mor}, {Morbidelli}, {Morel}, {Morris}, {Mulone}, {Munoz},
  {Muraveva}, {Murphy}, {Musella}, {Noval}, {Ord{\'e}novic}, {Orr{\`u}},
  {Osinde}, {Pagani}, {Pagano}, {Palaversa}, {Palicio}, {Panahi}, {Pawlak},
  {Pe{\~n}alosa Esteller}, {Penttil{\"a}}, {Piersimoni}, {Pineau}, {Plachy},
  {Plum}, {Poggio}, {Poretti}, {Poujoulet}, {Pr{\v{s}}a}, {Pulone}, {Racero},
  {Ragaini}, {Rainer}, {Raiteri}, {Rambaux}, {Ramos}, {Ramos-Lerate}, {Re
  Fiorentin}, {Regibo}, {Reyl{\'e}}, {Ripepi}, {Riva}, {Rixon}, {Robichon},
  {Robin}, {Roelens}, {Rohrbasser}, {Romero-G{\'o}mez}, {Rowell}, {Royer},
  {Rybicki}, {Sadowski}, {Sagrist{\`a} Sell{\'e}s}, {Sahlmann}, {Salgado},
  {Salguero}, {Samaras}, {Sanchez Gimenez}, {Sanna}, {Santove{\~n}a},
  {Sarasso}, {Schultheis}, {Sciacca}, {Segol}, {Segovia}, {S{\'e}gransan},
  {Semeux}, {Shahaf}, {Siddiqui}, {Siebert}, {Siltala}, {Slezak}, {Smart},
  {Solano}, {Solitro}, {Souami}, {Souchay}, {Spagna}, {Spoto}, {Steele},
  {Steidelm{\"u}ller}, {Stephenson}, {S{\"u}veges}, {Szabados}, {Szegedi-Elek},
  {Taris}, {Tauran}, {Taylor}, {Teixeira}, {Thuillot}, {Tonello}, {Torra},
  {Torra}, {Turon}, {Unger}, {Vaillant}, {van Dillen}, {Vanel}, {Vecchiato},
  {Viala}, {Vicente}, {Voutsinas}, {Weiler}, {Wevers}, {Wyrzykowski}, {Yoldas},
  {Yvard}, {Zhao}, {Zorec}, {Zucker}, {Zurbach}, \&
  {Zwitter}}]{2021A&A...649A...1G}
---. 2021, \aap, 649, A1, \dodoi{10.1051/0004-6361/202039657}

\bibitem[{{Gao} {et~al.}(2009){Gao}, {Jiang}, \& {Li}}]{2009ApJ...707...89G}
{Gao}, J., {Jiang}, B.~W., \& {Li}, A. 2009, \apj, 707, 89,
  \dodoi{10.1088/0004-637X/707/1/89}

\bibitem[{{Garc{\'\i}a P{\'e}rez} {et~al.}(2016){Garc{\'\i}a P{\'e}rez},
  {Allende Prieto}, {Holtzman}, {Shetrone}, {M{\'e}sz{\'a}ros}, {Bizyaev},
  {Carrera}, {Cunha}, {Garc{\'\i}a-Hern{\'a}ndez}, {Johnson}, {Majewski},
  {Nidever}, {Schiavon}, {Shane}, {Smith}, {Sobeck}, {Troup}, {Zamora},
  {Weinberg}, {Bovy}, {Eisenstein}, {Feuillet}, {Frinchaboy}, {Hayden},
  {Hearty}, {Nguyen}, {O'Connell}, {Pinsonneault}, {Wilson}, \&
  {Zasowski}}]{2016AJ....151..144G}
{Garc{\'\i}a P{\'e}rez}, A.~E., {Allende Prieto}, C., {Holtzman}, J.~A.,
  {et~al.} 2016, \aj, 151, 144, \dodoi{10.3847/0004-6256/151/6/144}

\bibitem[{{Girardi}(2016)}]{2016ARA&A..54...95G}
{Girardi}, L. 2016, \araa, 54, 95, \dodoi{10.1146/annurev-astro-081915-023354}

\bibitem[{{Girardi} \& {Salaris}(2001)}]{2001MNRAS.323..109G}
{Girardi}, L., \& {Salaris}, M. 2001, \mnras, 323, 109,
  \dodoi{10.1046/j.1365-8711.2001.04084.x}

\bibitem[{{Gontcharov} \& {Mosenkov}(2021)}]{2021MNRAS.500.2590G}
{Gontcharov}, G.~A., \& {Mosenkov}, A.~V. 2021, \mnras, 500, 2590,
  \dodoi{10.1093/mnras/staa2761}

\bibitem[{{Grady} {et~al.}(2020){Grady}, {Belokurov}, \&
  {Evans}}]{2020MNRAS.492.3128G}
{Grady}, J., {Belokurov}, V., \& {Evans}, N.~W. 2020, \mnras, 492, 3128,
  \dodoi{10.1093/mnras/stz3617}

\bibitem[{{Green} {et~al.}(2019){Green}, {Schlafly}, {Zucker}, {Speagle}, \&
  {Finkbeiner}}]{2019ApJ...887...93G}
{Green}, G.~M., {Schlafly}, E., {Zucker}, C., {Speagle}, J.~S., \&
  {Finkbeiner}, D. 2019, \apj, 887, 93, \dodoi{10.3847/1538-4357/ab5362}

\bibitem[{{Grocholski} \& {Sarajedini}(2002)}]{2002AJ....123.1603G}
{Grocholski}, A.~J., \& {Sarajedini}, A. 2002, \aj, 123, 1603,
  \dodoi{10.1086/339027}

\bibitem[{{Groenewegen}(2008)}]{2008A&A...488..935G}
{Groenewegen}, M.~A.~T. 2008, \aap, 488, 935,
  \dodoi{10.1051/0004-6361:200810201}

\bibitem[{{Hasselquist} {et~al.}(2020){Hasselquist}, {Zasowski}, {Feuillet},
  {Schultheis}, {Nataf}, {Anguiano}, {Beaton}, {Beers}, {Cohen}, {Cunha},
  {Fern{\'a}ndez-Trincado}, {Garc{\'\i}a-Hern{\'a}ndez}, {Geisler}, {Holtzman},
  {Johnson}, {Lane}, {Majewski}, {Moni Bidin}, {Nitschelm}, {Roman-Lopes},
  {Schiavon}, {Smith}, \& {Sobeck}}]{2020ApJ...901..109H}
{Hasselquist}, S., {Zasowski}, G., {Feuillet}, D.~K., {et~al.} 2020, \apj, 901,
  109, \dodoi{10.3847/1538-4357/abaeee}

\bibitem[{{Hawkins} {et~al.}(2017){Hawkins}, {Leistedt}, {Bovy}, \&
  {Hogg}}]{2017MNRAS.471..722H}
{Hawkins}, K., {Leistedt}, B., {Bovy}, J., \& {Hogg}, D.~W. 2017, \mnras, 471,
  722, \dodoi{10.1093/mnras/stx1655}

\bibitem[{{Hawkins} {et~al.}(2018){Hawkins}, {Ting}, \&
  {Walter-Rix}}]{2018ApJ...853...20H}
{Hawkins}, K., {Ting}, Y.-S., \& {Walter-Rix}, H. 2018, \apj, 853, 20,
  \dodoi{10.3847/1538-4357/aaa08a}

\bibitem[{{Hayden} {et~al.}(2014){Hayden}, {Holtzman}, {Bovy}, {Majewski},
  {Johnson}, {Allende Prieto}, {Beers}, {Cunha}, {Frinchaboy}, {Garc{\'\i}a
  P{\'e}rez}, {Girardi}, {Hearty}, {Lee}, {Nidever}, {Schiavon}, {Schlesinger},
  {Schneider}, {Schultheis}, {Shetrone}, {Smith}, {Zasowski}, {Bizyaev},
  {Feuillet}, {Hasselquist}, {Kinemuchi}, {Malanushenko}, {Malanushenko},
  {O'Connell}, {Pan}, \& {Stassun}}]{2014AJ....147..116H}
{Hayden}, M.~R., {Holtzman}, J.~A., {Bovy}, J., {et~al.} 2014, \aj, 147, 116,
  \dodoi{10.1088/0004-6256/147/5/116}

\bibitem[{{Hayden} {et~al.}(2015){Hayden}, {Bovy}, {Holtzman}, {Nidever},
  {Bird}, {Weinberg}, {Andrews}, {Majewski}, {Allende Prieto}, {Anders},
  {Beers}, {Bizyaev}, {Chiappini}, {Cunha}, {Frinchaboy},
  {Garc{\'\i}a-Her{\'n}andez}, {Garc{\'\i}a P{\'e}rez}, {Girardi}, {Harding},
  {Hearty}, {Johnson}, {M{\'e}sz{\'a}ros}, {Minchev}, {O'Connell}, {Pan},
  {Robin}, {Schiavon}, {Schneider}, {Schultheis}, {Shetrone}, {Skrutskie},
  {Steinmetz}, {Smith}, {Wilson}, {Zamora}, \&
  {Zasowski}}]{2015ApJ...808..132H}
{Hayden}, M.~R., {Bovy}, J., {Holtzman}, J.~A., {et~al.} 2015, \apj, 808, 132,
  \dodoi{10.1088/0004-637X/808/2/132}

\bibitem[{{Henden} {et~al.}(2016){Henden}, {Templeton}, {Terrell}, {Smith},
  {Levine}, \& {Welch}}]{2016yCat.2336....0H}
{Henden}, A.~A., {Templeton}, M., {Terrell}, D., {et~al.} 2016, VizieR Online
  Data Catalog, II/336

\bibitem[{{Holtzman} {et~al.}(2018){Holtzman}, {Hasselquist}, {Shetrone},
  {Cunha}, {Allende Prieto}, {Anguiano}, {Bizyaev}, {Bovy}, {Casey},
  {Edvardsson}, {Johnson}, {J{\"o}nsson}, {Meszaros}, {Smith}, {Sobeck},
  {Zamora}, {Chojnowski}, {Fernandez-Trincado}, {Garcia-Hernandez}, {Majewski},
  {Pinsonneault}, {Souto}, {Stringfellow}, {Tayar}, {Troup}, \&
  {Zasowski}}]{2018AJ....156..125H}
{Holtzman}, J.~A., {Hasselquist}, S., {Shetrone}, M., {et~al.} 2018, \aj, 156,
  125, \dodoi{10.3847/1538-3881/aad4f9}

\bibitem[{{Huang} {et~al.}(2021){Huang}, {Yuan}, {Beers}, \&
  {Zhang}}]{2021ApJ...910L...5H}
{Huang}, Y., {Yuan}, H., {Beers}, T.~C., \& {Zhang}, H. 2021, \apjl, 910, L5,
  \dodoi{10.3847/2041-8213/abe69a}

\bibitem[{{Huang} {et~al.}(2020){Huang}, {Sch{\"o}nrich}, {Zhang}, {Wu},
  {Chen}, {Wang}, {Xiang}, {Wang}, {Yuan}, {Li}, {Sun}, {Li}, \&
  {Liu}}]{2020ApJS..249...29H}
{Huang}, Y., {Sch{\"o}nrich}, R., {Zhang}, H., {et~al.} 2020, \apjs, 249, 29,
  \dodoi{10.3847/1538-4365/ab994f}

\bibitem[{{Indebetouw} {et~al.}(2005){Indebetouw}, {Mathis}, {Babler}, {Meade},
  {Watson}, {Whitney}, {Wolff}, {Wolfire}, {Cohen}, {Bania}, {Benjamin},
  {Clemens}, {Dickey}, {Jackson}, {Kobulnicky}, {Marston}, {Mercer},
  {Stauffer}, {Stolovy}, \& {Churchwell}}]{2005ApJ...619..931I}
{Indebetouw}, R., {Mathis}, J.~S., {Babler}, B.~L., {et~al.} 2005, \apj, 619,
  931, \dodoi{10.1086/426679}

\bibitem[{{J{\"o}nsson} {et~al.}(2018){J{\"o}nsson}, {Allende Prieto},
  {Holtzman}, {Feuillet}, {Hawkins}, {Cunha}, {M{\'e}sz{\'a}ros},
  {Hasselquist}, {Fern{\'a}ndez-Trincado}, {Garc{\'\i}a-Hern{\'a}ndez},
  {Bizyaev}, {Carrera}, {Majewski}, {Pinsonneault}, {Shetrone}, {Smith},
  {Sobeck}, {Souto}, {Stringfellow}, {Teske}, \&
  {Zamora}}]{2018AJ....156..126J}
{J{\"o}nsson}, H., {Allende Prieto}, C., {Holtzman}, J.~A., {et~al.} 2018, \aj,
  156, 126, \dodoi{10.3847/1538-3881/aad4f5}

\bibitem[{{J{\"o}nsson} {et~al.}(2020){J{\"o}nsson}, {Holtzman}, {Allende
  Prieto}, {Cunha}, {Garc{\'\i}a-Hern{\'a}ndez}, {Hasselquist}, {Masseron},
  {Osorio}, {Shetrone}, {Smith}, {Stringfellow}, {Bizyaev}, {Edvardsson},
  {Majewski}, {M{\'e}sz{\'a}ros}, {Souto}, {Zamora}, {Beaton}, {Bovy}, {Donor},
  {Pinsonneault}, {Poovelil}, \& {Sobeck}}]{2020AJ....160..120J}
{J{\"o}nsson}, H., {Holtzman}, J.~A., {Allende Prieto}, C., {et~al.} 2020, \aj,
  160, 120, \dodoi{10.3847/1538-3881/aba592}

\bibitem[{{Kim} {et~al.}(2002){Kim}, {Kim}, {Lee}, {Sarajedini}, \&
  {Geisler}}]{2002AJ....123..244K}
{Kim}, M., {Kim}, E., {Lee}, M.~G., {Sarajedini}, A., \& {Geisler}, D. 2002,
  \aj, 123, 244, \dodoi{10.1086/324639}

\bibitem[{{Laney} {et~al.}(2012){Laney}, {Joner}, \&
  {Pietrzy{\'n}ski}}]{2012MNRAS.419.1637L}
{Laney}, C.~D., {Joner}, M.~D., \& {Pietrzy{\'n}ski}, G. 2012, \mnras, 419,
  1637, \dodoi{10.1111/j.1365-2966.2011.19826.x}

\bibitem[{{Lim} {et~al.}(2021){Lim}, {Lee}, {Koch}, {Hong}, {Johnson}, {Kim},
  {Chung}, {Mateo}, \& {Bailey}}]{2021ApJ...907...47L}
{Lim}, D., {Lee}, Y.-W., {Koch}, A., {et~al.} 2021, \apj, 907, 47,
  \dodoi{10.3847/1538-4357/abd08d}

\bibitem[{{Lindegren} {et~al.}(2021{\natexlab{a}}){Lindegren}, {Bastian},
  {Biermann}, {Bombrun}, {de Torres}, {Gerlach}, {Geyer}, {Hern{\'a}ndez},
  {Hilger}, {Hobbs}, {Klioner}, {Lammers}, {McMillan}, {Ramos-Lerate},
  {Steidelm{\"u}ller}, {Stephenson}, \& {van Leeuwen}}]{2021A&A...649A...4L}
{Lindegren}, L., {Bastian}, U., {Biermann}, M., {et~al.} 2021{\natexlab{a}},
  \aap, 649, A4, \dodoi{10.1051/0004-6361/202039653}

\bibitem[{{Lindegren} {et~al.}(2021{\natexlab{b}}){Lindegren}, {Klioner},
  {Hern{\'a}ndez}, {Bombrun}, {Ramos-Lerate}, {Steidelm{\"u}ller}, {Bastian},
  {Biermann}, {de Torres}, {Gerlach}, {Geyer}, {Hilger}, {Hobbs}, {Lammers},
  {McMillan}, {Stephenson}, {Casta{\~n}eda}, {Davidson}, {Fabricius},
  {Gracia-Abril}, {Portell}, {Rowell}, {Teyssier}, {Torra}, {Bartolom{\'e}},
  {Clotet}, {Garralda}, {Gonz{\'a}lez-Vidal}, {Torra}, {Abbas}, {Altmann},
  {Anglada Varela}, {Balaguer-N{\'u}{\~n}ez}, {Balog}, {Barache}, {Becciani},
  {Bernet}, {Bertone}, {Bianchi}, {Bouquillon}, {Brown}, {Bucciarelli},
  {Busonero}, {Butkevich}, {Buzzi}, {Cancelliere}, {Carlucci}, {Charlot},
  {Cioni}, {Crosta}, {Crowley}, {del Peloso}, {del Pozo}, {Drimmel}, {Esquej},
  {Fienga}, {Fraile}, {Gai}, {Garcia-Reinaldos}, {Guerra}, {Hambly}, {Hauser},
  {Jan{\ss}en}, {Jordan}, {Kostrzewa-Rutkowska}, {Lattanzi}, {Liao}, {Licata},
  {Lister}, {L{\"o}ffler}, {Marchant}, {Masip}, {Mignard}, {Mints}, {Molina},
  {Mora}, {Morbidelli}, {Murphy}, {Pagani}, {Panuzzo}, {Pe{\~n}alosa Esteller},
  {Poggio}, {Re Fiorentin}, {Riva}, {Sagrist{\`a} Sell{\'e}s}, {Sanchez
  Gimenez}, {Sarasso}, {Sciacca}, {Siddiqui}, {Smart}, {Souami}, {Spagna},
  {Steele}, {Taris}, {Utrilla}, {van Reeven}, \&
  {Vecchiato}}]{2021A&A...649A...2L}
{Lindegren}, L., {Klioner}, S.~A., {Hern{\'a}ndez}, J., {et~al.}
  2021{\natexlab{b}}, \aap, 649, A2, \dodoi{10.1051/0004-6361/202039709}

\bibitem[{{L{\'o}pez-Corredoira}(2017)}]{2017ApJ...836..218L}
{L{\'o}pez-Corredoira}, M. 2017, \apj, 836, 218,
  \dodoi{10.3847/1538-4357/836/2/218}

\bibitem[{{L{\'o}pez-Corredoira} {et~al.}(2002){L{\'o}pez-Corredoira},
  {Cabrera-Lavers}, {Garz{\'o}n}, \& {Hammersley}}]{2002A&A...394..883L}
{L{\'o}pez-Corredoira}, M., {Cabrera-Lavers}, A., {Garz{\'o}n}, F., \&
  {Hammersley}, P.~L. 2002, \aap, 394, 883, \dodoi{10.1051/0004-6361:20021175}

\bibitem[{{Lucey} {et~al.}(2020){Lucey}, {Ting}, {Ramachandra}, \&
  {Hawkins}}]{2020MNRAS.495.3087L}
{Lucey}, M., {Ting}, Y.-S., {Ramachandra}, N.~S., \& {Hawkins}, K. 2020,
  \mnras, 495, 3087, \dodoi{10.1093/mnras/staa1226}

\bibitem[{{Luo} {et~al.}(2015){Luo}, {Zhao}, {Zhao}, {Deng}, {Liu}, {Jing},
  {Wang}, {Zhang}, {Shi}, {Cui}, {Chu}, {Li}, {Bai}, {Wu}, {Cai}, {Cao}, {Cao},
  {Carlin}, {Chen}, {Chen}, {Chen}, {Chen}, {Chen}, {Chen}, {Chen},
  {Christlieb}, {Chu}, {Cui}, {Dong}, {Du}, {Fan}, {Feng}, {Fu}, {Gao}, {Gong},
  {Gu}, {Guo}, {Han}, {He}, {Hou}, {Hou}, {Hou}, {Hu}, {Hu}, {Hu}, {Huo},
  {Jia}, {Jiang}, {Jiang}, {Jiang}, {Jin}, {Kong}, {Kong}, {Lei}, {Li}, {Li},
  {Li}, {Li}, {Li}, {Li}, {Li}, {Li}, {Li}, {Li}, {Li}, {Li}, {Liang}, {Lin},
  {Liu}, {Liu}, {Liu}, {Liu}, {Lu}, {Luo}, {Mao}, {Newberg}, {Ni}, {Qi}, {Qi},
  {Shen}, {Shi}, {Song}, {Song}, {Su}, {Su}, {Tang}, {Tao}, {Tian}, {Wang},
  {Wang}, {Wang}, {Wang}, {Wang}, {Wang}, {Wang}, {Wang}, {Wang}, {Wang},
  {Wang}, {Wang}, {Wang}, {Wang}, {Wang}, {Wang}, {Wang}, {Wang}, {Wang},
  {Wang}, {Wei}, {Wei}, {Wu}, {Wu}, {Wu}, {Wu}, {Xing}, {Xu}, {Xu}, {Xu},
  {Yan}, {Yang}, {Yang}, {Yang}, {Yang}, {Yao}, {Yu}, {Yuan}, {Yuan}, {Yuan},
  {Yuan}, {Zhai}, {Zhang}, {Zhang}, {Zhang}, {Zhang}, {Zhang}, {Zhang},
  {Zhang}, {Zhang}, {Zhao}, {Zhou}, {Zhou}, {Zhu}, {Zhu}, {Zou}, \&
  {Zuo}}]{2015RAA....15.1095L}
{Luo}, A.~L., {Zhao}, Y.-H., {Zhao}, G., {et~al.} 2015, Research in Astronomy
  and Astrophysics, 15, 1095, \dodoi{10.1088/1674-4527/15/8/002}

\bibitem[{{Mackereth} {et~al.}(2017){Mackereth}, {Bovy}, {Schiavon},
  {Zasowski}, {Cunha}, {Frinchaboy}, {Garc{\'\i}a Perez}, {Hayden}, {Holtzman},
  {Majewski}, {M{\'e}sz{\'a}ros}, {Nidever}, {Pinsonneault}, \&
  {Shetrone}}]{2017MNRAS.471.3057M}
{Mackereth}, J.~T., {Bovy}, J., {Schiavon}, R.~P., {et~al.} 2017, \mnras, 471,
  3057, \dodoi{10.1093/mnras/stx1774}

\bibitem[{{Madore}(1982)}]{1982ApJ...253..575M}
{Madore}, B.~F. 1982, \apj, 253, 575, \dodoi{10.1086/159659}

\bibitem[{{Majewski} {et~al.}(2017){Majewski}, {Schiavon}, {Frinchaboy},
  {Allende Prieto}, {Barkhouser}, {Bizyaev}, {Blank}, {Brunner}, {Burton},
  {Carrera}, {Chojnowski}, {Cunha}, {Epstein}, {Fitzgerald}, {Garc{\'\i}a
  P{\'e}rez}, {Hearty}, {Henderson}, {Holtzman}, {Johnson}, {Lam}, {Lawler},
  {Maseman}, {M{\'e}sz{\'a}ros}, {Nelson}, {Nguyen}, {Nidever}, {Pinsonneault},
  {Shetrone}, {Smee}, {Smith}, {Stolberg}, {Skrutskie}, {Walker}, {Wilson},
  {Zasowski}, {Anders}, {Basu}, {Beland}, {Blanton}, {Bovy}, {Brownstein},
  {Carlberg}, {Chaplin}, {Chiappini}, {Eisenstein}, {Elsworth}, {Feuillet},
  {Fleming}, {Galbraith-Frew}, {Garc{\'\i}a}, {Garc{\'\i}a-Hern{\'a}ndez},
  {Gillespie}, {Girardi}, {Gunn}, {Hasselquist}, {Hayden}, {Hekker}, {Ivans},
  {Kinemuchi}, {Klaene}, {Mahadevan}, {Mathur}, {Mosser}, {Muna}, {Munn},
  {Nichol}, {O'Connell}, {Parejko}, {Robin}, {Rocha-Pinto}, {Schultheis},
  {Serenelli}, {Shane}, {Silva Aguirre}, {Sobeck}, {Thompson}, {Troup},
  {Weinberg}, \& {Zamora}}]{2017AJ....154...94M}
{Majewski}, S.~R., {Schiavon}, R.~P., {Frinchaboy}, P.~M., {et~al.} 2017, \aj,
  154, 94, \dodoi{10.3847/1538-3881/aa784d}

\bibitem[{{McWilliam} \& {Zoccali}(2010)}]{2010ApJ...724.1491M}
{McWilliam}, A., \& {Zoccali}, M. 2010, \apj, 724, 1491,
  \dodoi{10.1088/0004-637X/724/2/1491}

\bibitem[{{Nataf} {et~al.}(2021){Nataf}, {Cassisi}, {Casagrande}, {Yuan}, \&
  {Riess}}]{2021ApJ...910..121N}
{Nataf}, D.~M., {Cassisi}, S., {Casagrande}, L., {Yuan}, W., \& {Riess}, A.~G.
  2021, \apj, 910, 121, \dodoi{10.3847/1538-4357/abe530}

\bibitem[{{Nataf} {et~al.}(2010){Nataf}, {Udalski}, {Gould}, {Fouqu{\'e}}, \&
  {Stanek}}]{2010ApJ...721L..28N}
{Nataf}, D.~M., {Udalski}, A., {Gould}, A., {Fouqu{\'e}}, P., \& {Stanek},
  K.~Z. 2010, \apjl, 721, L28, \dodoi{10.1088/2041-8205/721/1/L28}

\bibitem[{{Nataf} {et~al.}(2013){Nataf}, {Gould}, {Fouqu{\'e}}, {Gonzalez},
  {Johnson}, {Skowron}, {Udalski}, {Szyma{\'n}ski}, {Kubiak},
  {Pietrzy{\'n}ski}, {Soszy{\'n}ski}, {Ulaczyk}, {Wyrzykowski}, \&
  {Poleski}}]{2013ApJ...769...88N}
{Nataf}, D.~M., {Gould}, A., {Fouqu{\'e}}, P., {et~al.} 2013, \apj, 769, 88,
  \dodoi{10.1088/0004-637X/769/2/88}

\bibitem[{{Nataf} {et~al.}(2016){Nataf}, {Gonzalez}, {Casagrande}, {Zasowski},
  {Wegg}, {Wolf}, {Kunder}, {Alonso-Garcia}, {Minniti}, {Rejkuba}, {Saito},
  {Valenti}, {Zoccali}, {Poleski}, {Pietrzy{\'n}ski}, {Skowron},
  {Soszy{\'n}ski}, {Szyma{\'n}ski}, {Udalski}, {Ulaczyk}, \&
  {Wyrzykowski}}]{2016MNRAS.456.2692N}
{Nataf}, D.~M., {Gonzalez}, O.~A., {Casagrande}, L., {et~al.} 2016, \mnras,
  456, 2692, \dodoi{10.1093/mnras/stv2843}

\bibitem[{{Ness} {et~al.}(2016){Ness}, {Hogg}, {Rix}, {Martig}, {Pinsonneault},
  \& {Ho}}]{2016ApJ...823..114N}
{Ness}, M., {Hogg}, D.~W., {Rix}, H.~W., {et~al.} 2016, \apj, 823, 114,
  \dodoi{10.3847/0004-637X/823/2/114}

\bibitem[{{Nidever} {et~al.}(2015){Nidever}, {Holtzman}, {Allende Prieto},
  {Beland}, {Bender}, {Bizyaev}, {Burton}, {Desphande}, {Fleming}, {Garc{\'\i}a
  P{\'e}rez}, {Hearty}, {Majewski}, {M{\'e}sz{\'a}ros}, {Muna}, {Nguyen},
  {Schiavon}, {Shetrone}, {Skrutskie}, {Sobeck}, \&
  {Wilson}}]{2015AJ....150..173N}
{Nidever}, D.~L., {Holtzman}, J.~A., {Allende Prieto}, C., {et~al.} 2015, \aj,
  150, 173, \dodoi{10.1088/0004-6256/150/6/173}

\bibitem[{{Nidever} {et~al.}(2020){Nidever}, {Hasselquist}, {Hayes}, {Hawkins},
  {Povick}, {Majewski}, {Smith}, {Anguiano}, {Stringfellow}, {Sobeck}, {Cunha},
  {Beers}, {Bestenlehner}, {Cohen}, {Garcia-Hernandez}, {J{\"o}nsson},
  {Nitschelm}, {Shetrone}, {Lacerna}, {Allende Prieto}, {Beaton}, {Dell'Agli},
  {Fern{\'a}ndez-Trincado}, {Feuillet}, {Gallart}, {Hearty}, {Holtzman},
  {Manchado}, {Mu{\~n}oz}, {O'Connell}, \& {Rosado}}]{2020ApJ...895...88N}
{Nidever}, D.~L., {Hasselquist}, S., {Hayes}, C.~R., {et~al.} 2020, \apj, 895,
  88, \dodoi{10.3847/1538-4357/ab7305}

\bibitem[{{Nishiyama} {et~al.}(2009){Nishiyama}, {Tamura}, {Hatano}, {Kato},
  {Tanab{\'e}}, {Sugitani}, \& {Nagata}}]{2009ApJ...696.1407N}
{Nishiyama}, S., {Tamura}, M., {Hatano}, H., {et~al.} 2009, \apj, 696, 1407,
  \dodoi{10.1088/0004-637X/696/2/1407}

\bibitem[{{Nogueras-Lara} {et~al.}(2020){Nogueras-Lara}, {Sch{\"o}del},
  {Gallego-Calvente}, {Gallego-Cano}, {Shahzamanian}, {Dong}, {Neumayer},
  {Hilker}, {Najarro}, {Nishiyama}, {Feldmeier-Krause}, {Girard}, \&
  {Cassisi}}]{2020NatAs...4..377N}
{Nogueras-Lara}, F., {Sch{\"o}del}, R., {Gallego-Calvente}, A.~T., {et~al.}
  2020, Nature Astronomy, 4, 377, \dodoi{10.1038/s41550-019-0967-9}

\bibitem[{{Onozato} {et~al.}(2019){Onozato}, {Ita}, {Nakada}, \&
  {Nishiyama}}]{2019MNRAS.486.5600O}
{Onozato}, H., {Ita}, Y., {Nakada}, Y., \& {Nishiyama}, S. 2019, \mnras, 486,
  5600, \dodoi{10.1093/mnras/stz1192}

\bibitem[{{Paczy{\'n}ski} \& {Stanek}(1998)}]{1998ApJ...494L.219P}
{Paczy{\'n}ski}, B., \& {Stanek}, K.~Z. 1998, \apjl, 494, L219,
  \dodoi{10.1086/311181}

\bibitem[{{Pietrukowicz} {et~al.}(2015){Pietrukowicz}, {Koz{\l}owski},
  {Skowron}, {Soszy{\'n}ski}, {Udalski}, {Poleski}, {Wyrzykowski},
  {Szyma{\'n}ski}, {Pietrzy{\'n}ski}, {Ulaczyk}, {Mr{\'o}z}, {Skowron}, \&
  {Kubiak}}]{2015ApJ...811..113P}
{Pietrukowicz}, P., {Koz{\l}owski}, S., {Skowron}, J., {et~al.} 2015, \apj,
  811, 113, \dodoi{10.1088/0004-637X/811/2/113}

\bibitem[{{Pietrzy{\'n}ski} {et~al.}(2010){Pietrzy{\'n}ski}, {G{\'o}rski},
  {Gieren}, {Laney}, {Udalski}, \& {Ciechanowska}}]{2010AJ....140.1038P}
{Pietrzy{\'n}ski}, G., {G{\'o}rski}, M., {Gieren}, W., {et~al.} 2010, \aj, 140,
  1038, \dodoi{10.1088/0004-6256/140/4/1038}

\bibitem[{{Plevne} {et~al.}(2020){Plevne}, {{\"O}nal Ta{\c{s}}}, {Bilir}, \&
  {Seabroke}}]{2020ApJ...893..108P}
{Plevne}, O., {{\"O}nal Ta{\c{s}}}, {\"O}., {Bilir}, S., \& {Seabroke}, G.~M.
  2020, \apj, 893, 108, \dodoi{10.3847/1538-4357/ab80bb}

\bibitem[{{Poggio} {et~al.}(2020){Poggio}, {Drimmel}, {Andrae}, {Bailer-Jones},
  {Fouesneau}, {Lattanzi}, {Smart}, \& {Spagna}}]{2020NatAs...4..590P}
{Poggio}, E., {Drimmel}, R., {Andrae}, R., {et~al.} 2020, Nature Astronomy, 4,
  590, \dodoi{10.1038/s41550-020-1017-3}

\bibitem[{{Poggio} {et~al.}(2018){Poggio}, {Drimmel}, {Lattanzi}, {Smart},
  {Spagna}, {Andrae}, {Bailer-Jones}, {Fouesneau}, {Antoja}, {Babusiaux},
  {Evans}, {Figueras}, {Katz}, {Reyl{\'e}}, {Robin}, {Romero-G{\'o}mez}, \&
  {Seabroke}}]{2018MNRAS.481L..21P}
{Poggio}, E., {Drimmel}, R., {Lattanzi}, M.~G., {et~al.} 2018, \mnras, 481,
  L21, \dodoi{10.1093/mnrasl/sly148}

\bibitem[{{Queiroz} {et~al.}(2020){Queiroz}, {Chiappini}, {Perez-Villegas},
  {Khalatyan}, {Anders}, {Barbuy}, {Santiago}, {Steinmetz}, {Cunha},
  {Schultheis}, {Majewski}, {Minchev}, {Minniti}, {Cohen}, {da Costa},
  {Fern{\'a}ndez-Trincado}, {Garcia-Hern{\'a}ndez}, {Geisler}, {Hasselquist},
  {. Lane}, {Rojas-Arriagada}, {Roman-Lopes}, \& {Smith}}]{2020arXiv200712915Q}
{Queiroz}, A.~B.~A., {Chiappini}, C., {Perez-Villegas}, A., {et~al.} 2020,
  arXiv e-prints, arXiv:2007.12915.
\newblock \doarXiv{2007.12915}

\bibitem[{{Rasmussen} \& {Williams}(2006)}]{2006gpml.book.....R}
{Rasmussen}, C.~E., \& {Williams}, C. K.~I. 2006, {Gaussian Processes for
  Machine Learning} (MIT press Cambridge, MA)

\bibitem[{{Rattenbury} {et~al.}(2007){Rattenbury}, {Mao}, {Sumi}, \&
  {Smith}}]{2007MNRAS.378.1064R}
{Rattenbury}, N.~J., {Mao}, S., {Sumi}, T., \& {Smith}, M.~C. 2007, \mnras,
  378, 1064, \dodoi{10.1111/j.1365-2966.2007.11843.x}

\bibitem[{{Ren} {et~al.}(2021){Ren}, {Chen}, {Zhang}, {de Grijs}, {Deng}, \&
  {Huang}}]{2021ApJ...911L..20R}
{Ren}, F., {Chen}, X., {Zhang}, H., {et~al.} 2021, \apjl, 911, L20,
  \dodoi{10.3847/2041-8213/abf359}

\bibitem[{{Rizzi} {et~al.}(2007){Rizzi}, {Held}, {Saviane}, {Tully}, \&
  {Gullieuszik}}]{2007MNRAS.380.1255R}
{Rizzi}, L., {Held}, E.~V., {Saviane}, I., {Tully}, R.~B., \& {Gullieuszik}, M.
  2007, \mnras, 380, 1255, \dodoi{10.1111/j.1365-2966.2007.12196.x}

\bibitem[{{Romero-G{\'o}mez} {et~al.}(2019){Romero-G{\'o}mez}, {Mateu},
  {Aguilar}, {Figueras}, \& {Castro-Ginard}}]{2019A&A...627A.150R}
{Romero-G{\'o}mez}, M., {Mateu}, C., {Aguilar}, L., {Figueras}, F., \&
  {Castro-Ginard}, A. 2019, \aap, 627, A150,
  \dodoi{10.1051/0004-6361/201834908}

\bibitem[{{Ruiz-Dern} {et~al.}(2018){Ruiz-Dern}, {Babusiaux}, {Arenou},
  {Turon}, \& {Lallement}}]{2018A&A...609A.116R}
{Ruiz-Dern}, L., {Babusiaux}, C., {Arenou}, F., {Turon}, C., \& {Lallement}, R.
  2018, \aap, 609, A116, \dodoi{10.1051/0004-6361/201731572}

\bibitem[{{Salaris} \& {Girardi}(2002)}]{2002MNRAS.337..332S}
{Salaris}, M., \& {Girardi}, L. 2002, \mnras, 337, 332,
  \dodoi{10.1046/j.1365-8711.2002.05917.x}

\bibitem[{{Sarajedini}(1999)}]{1999AJ....118.2321S}
{Sarajedini}, A. 1999, \aj, 118, 2321, \dodoi{10.1086/301112}

\bibitem[{{Sit} \& {Ness}(2020)}]{2020ApJ...900....4S}
{Sit}, T., \& {Ness}, M.~K. 2020, \apj, 900, 4,
  \dodoi{10.3847/1538-4357/ab9ff6}

\bibitem[{{Skowron} {et~al.}(2019){Skowron}, {Skowron}, {Mr{\'o}z}, {Udalski},
  {Pietrukowicz}, {Soszy{\'n}ski}, {Szyma{\'n}ski}, {Poleski}, {Koz{\l}owski},
  {Ulaczyk}, {Rybicki}, \& {Iwanek}}]{2019Sci...365..478S}
{Skowron}, D.~M., {Skowron}, J., {Mr{\'o}z}, P., {et~al.} 2019, Science, 365,
  478, \dodoi{10.1126/science.aau3181}

\bibitem[{{Skowron} {et~al.}(2021){Skowron}, {Skowron}, {Udalski},
  {Szyma{\'n}ski}, {Soszy{\'n}ski}, {Wyrzykowski}, {Ulaczyk}, {Poleski},
  {Koz{\l}owski}, {Pietrukowicz}, {Mr{\'o}z}, {Rybicki}, {Iwanek}, {Wrona}, \&
  {Gromadzki}}]{2021ApJS..252...23S}
{Skowron}, D.~M., {Skowron}, J., {Udalski}, A., {et~al.} 2021, \apjs, 252, 23,
  \dodoi{10.3847/1538-4365/abcb81}

\bibitem[{{Skrutskie} {et~al.}(2006){Skrutskie}, {Cutri}, {Stiening},
  {Weinberg}, {Schneider}, {Carpenter}, {Beichman}, {Capps}, {Chester},
  {Elias}, {Huchra}, {Liebert}, {Lonsdale}, {Monet}, {Price}, {Seitzer},
  {Jarrett}, {Kirkpatrick}, {Gizis}, {Howard}, {Evans}, {Fowler}, {Fullmer},
  {Hurt}, {Light}, {Kopan}, {Marsh}, {McCallon}, {Tam}, {Van Dyk}, \&
  {Wheelock}}]{2006AJ....131.1163S}
{Skrutskie}, M.~F., {Cutri}, R.~M., {Stiening}, R., {et~al.} 2006, \aj, 131,
  1163, \dodoi{10.1086/498708}

\bibitem[{{Stanek} {et~al.}(1998){Stanek}, {Zaritsky}, \&
  {Harris}}]{1998ApJ...500L.141S}
{Stanek}, K.~Z., {Zaritsky}, D., \& {Harris}, J. 1998, \apjl, 500, L141,
  \dodoi{10.1086/311420}

\bibitem[{{Stassun} \& {Torres}(2021)}]{2021ApJ...907L..33S}
{Stassun}, K.~G., \& {Torres}, G. 2021, \apjl, 907, L33,
  \dodoi{10.3847/2041-8213/abdaad}

\bibitem[{{Ting} {et~al.}(2018){Ting}, {Hawkins}, \&
  {Rix}}]{2018ApJ...858L...7T}
{Ting}, Y.-S., {Hawkins}, K., \& {Rix}, H.-W. 2018, \apjl, 858, L7,
  \dodoi{10.3847/2041-8213/aabf8e}

\bibitem[{{Twarog} {et~al.}(1999){Twarog}, {Anthony-Twarog}, \&
  {Bricker}}]{1999AJ....117.1816T}
{Twarog}, B.~A., {Anthony-Twarog}, B.~J., \& {Bricker}, A.~R. 1999, \aj, 117,
  1816, \dodoi{10.1086/300810}

\bibitem[{{Udalski}(2000)}]{2000ApJ...531L..25U}
{Udalski}, A. 2000, \apjl, 531, L25, \dodoi{10.1086/312513}

\bibitem[{{van Helshoecht} \& {Groenewegen}(2007)}]{2007A&A...463..559V}
{van Helshoecht}, V., \& {Groenewegen}, M.~A.~T. 2007, \aap, 463, 559,
  \dodoi{10.1051/0004-6361:20052721}

\bibitem[{{Wan} {et~al.}(2015){Wan}, {Liu}, {Deng}, {Cui}, {Zhang}, {Hou},
  {Yang}, \& {Wu}}]{2015RAA....15.1166W}
{Wan}, J.-C., {Liu}, C., {Deng}, L.-C., {et~al.} 2015, Research in Astronomy
  and Astrophysics, 15, 1166, \dodoi{10.1088/1674-4527/15/8/006}

\bibitem[{{Wang} {et~al.}(2020{\natexlab{a}}){Wang}, {L{\'o}pez-Corredoira},
  {Huang}, {Carlin}, {Chen}, {Wang}, {Chang}, {Zhang}, {Xiang}, {Yuan}, {Sun},
  {Li}, {Yang}, \& {Deng}}]{2020MNRAS.491.2104W}
{Wang}, H.~F., {L{\'o}pez-Corredoira}, M., {Huang}, Y., {et~al.}
  2020{\natexlab{a}}, \mnras, 491, 2104, \dodoi{10.1093/mnras/stz3113}

\bibitem[{{Wang} {et~al.}(2020{\natexlab{b}}){Wang}, {L{\'o}pez-Corredoira},
  {Huang}, {Chang}, {Zhang}, {Carlin}, {Chen}, {Chrob{\'a}kov{\'a}}, \&
  {Chen}}]{2020ApJ...897..119W}
---. 2020{\natexlab{b}}, \apj, 897, 119, \dodoi{10.3847/1538-4357/ab93ad}

\bibitem[{{Wang} \& {Chen}(2019)}]{2019ApJ...877..116W}
{Wang}, S., \& {Chen}, X. 2019, \apj, 877, 116,
  \dodoi{10.3847/1538-4357/ab1c61}

\bibitem[{{Wang} {et~al.}(2017){Wang}, {Jiang}, {Zhao}, {Chen}, \& {de
  Grijs}}]{2017ApJ...848..106W}
{Wang}, S., {Jiang}, B.~W., {Zhao}, H., {Chen}, X., \& {de Grijs}, R. 2017,
  \apj, 848, 106, \dodoi{10.3847/1538-4357/aa8db7}

\bibitem[{{Wang} {et~al.}(2020{\natexlab{c}}){Wang}, {Zhang}, {Jiang}, {Zhao},
  {Chen}, {Chen}, {Gao}, \& {Liu}}]{2020A&A...639A..72W}
{Wang}, S., {Zhang}, C., {Jiang}, B., {et~al.} 2020{\natexlab{c}}, \aap, 639,
  A72, \dodoi{10.1051/0004-6361/201936868}

\bibitem[{{Wegg} \& {Gerhard}(2013)}]{2013MNRAS.435.1874W}
{Wegg}, C., \& {Gerhard}, O. 2013, \mnras, 435, 1874,
  \dodoi{10.1093/mnras/stt1376}

\bibitem[{{Wright} {et~al.}(2010){Wright}, {Eisenhardt}, {Mainzer}, {Ressler},
  {Cutri}, {Jarrett}, {Kirkpatrick}, {Padgett}, {McMillan}, {Skrutskie},
  {Stanford}, {Cohen}, {Walker}, {Mather}, {Leisawitz}, {Gautier}, {McLean},
  {Benford}, {Lonsdale}, {Blain}, {Mendez}, {Irace}, {Duval}, {Liu}, {Royer},
  {Heinrichsen}, {Howard}, {Shannon}, {Kendall}, {Walsh}, {Larsen}, {Cardon},
  {Schick}, {Schwalm}, {Abid}, {Fabinsky}, {Naes}, \&
  {Tsai}}]{2010AJ....140.1868W}
{Wright}, E.~L., {Eisenhardt}, P. R.~M., {Mainzer}, A.~K., {et~al.} 2010, \aj,
  140, 1868, \dodoi{10.1088/0004-6256/140/6/1868}

\bibitem[{{Xiang} {et~al.}(2019){Xiang}, {Ting}, {Rix}, {Sandford}, {Buder},
  {Lind}, {Liu}, {Shi}, \& {Zhang}}]{2019ApJS..245...34X}
{Xiang}, M., {Ting}, Y.-S., {Rix}, H.-W., {et~al.} 2019, \apjs, 245, 34,
  \dodoi{10.3847/1538-4365/ab5364}

\bibitem[{{Xiang} {et~al.}(2015){Xiang}, {Liu}, {Yuan}, {Huang}, {Wang}, {Ren},
  {Chen}, {Sun}, {Zhang}, {Huo}, \& {Rebassa-Mansergas}}]{2015RAA....15.1209X}
{Xiang}, M.-S., {Liu}, X.-W., {Yuan}, H.-B., {et~al.} 2015, Research in
  Astronomy and Astrophysics, 15, 1209, \dodoi{10.1088/1674-4527/15/8/009}

\bibitem[{{Zasowski} {et~al.}(2009){Zasowski}, {Majewski}, {Indebetouw},
  {Meade}, {Nidever}, {Patterson}, {Babler}, {Skrutskie}, {Watson}, {Whitney},
  \& {Churchwell}}]{2009ApJ...707..510Z}
{Zasowski}, G., {Majewski}, S.~R., {Indebetouw}, R., {et~al.} 2009, \apj, 707,
  510, \dodoi{10.1088/0004-637X/707/1/510}

\bibitem[{{Zhao} {et~al.}(2012){Zhao}, {Zhao}, {Chu}, {Jing}, \&
  {Deng}}]{2012RAA....12..723Z}
{Zhao}, G., {Zhao}, Y.-H., {Chu}, Y.-Q., {Jing}, Y.-P., \& {Deng}, L.-C. 2012,
  Research in Astronomy and Astrophysics, 12, 723,
  \dodoi{10.1088/1674-4527/12/7/002}

\end{thebibliography}
\bibliographystyle{aasjournal}

\appendix

\setcounter{table}{0}
\renewcommand{\thetable}{A\arabic{table}}
\setcounter{figure}{0}
\renewcommand{\thefigure}{A\arabic{figure}}

\section{Supplementary Figures}

\begin{figure*}[ht]
\centering
\subfigure{\includegraphics[width=3.5in]{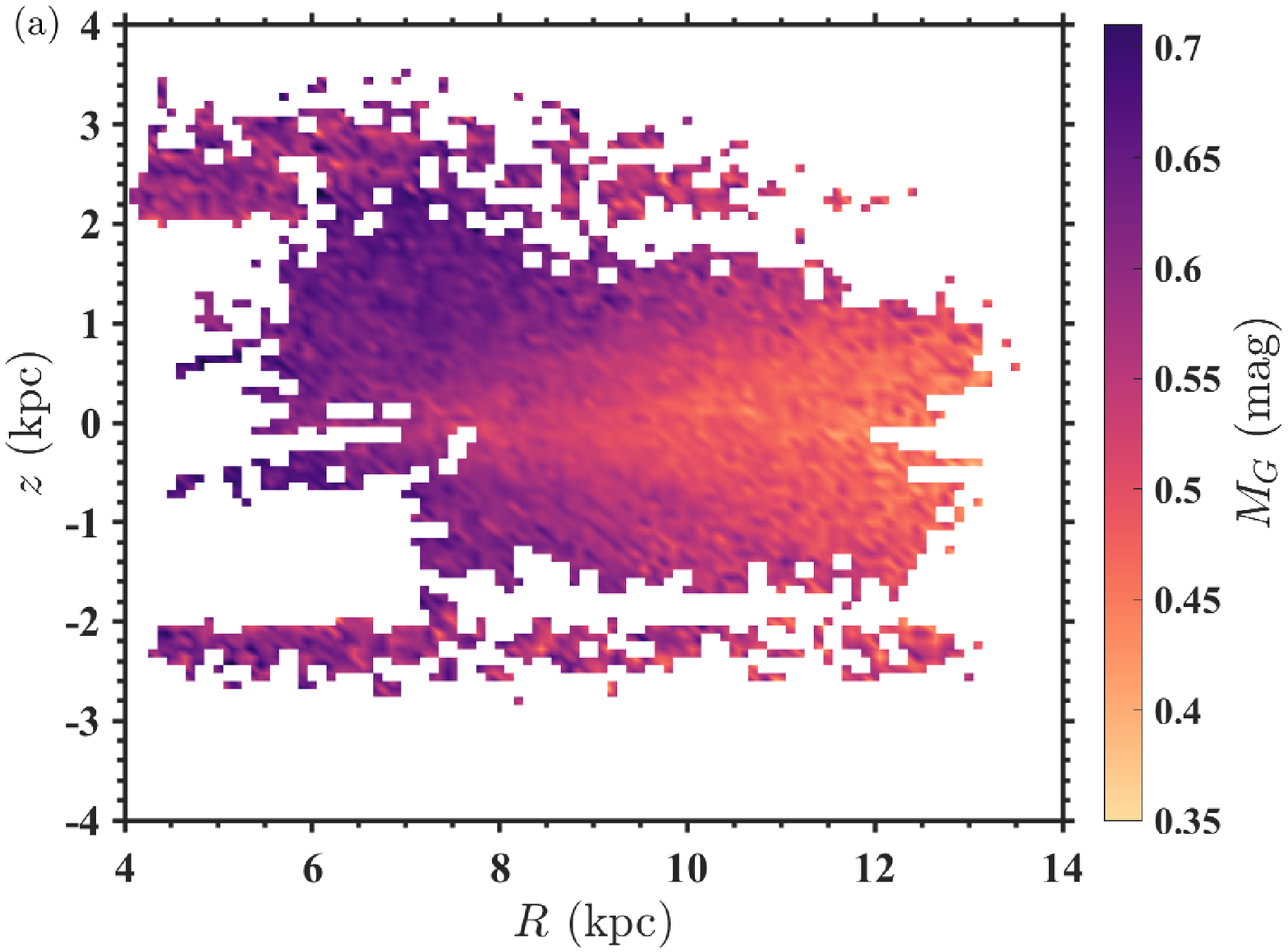}}
\subfigure{\includegraphics[width=3.5in]{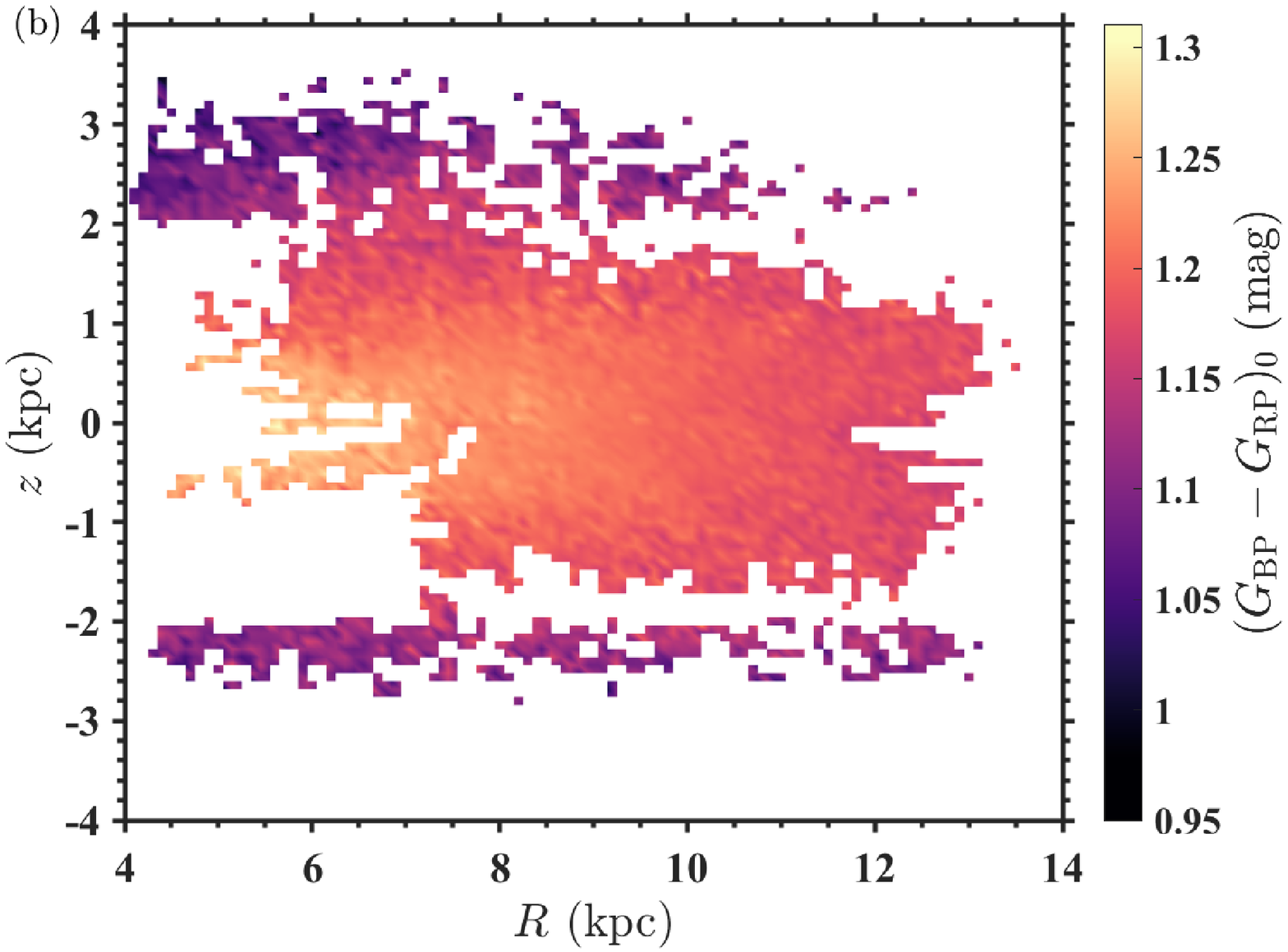}}
\quad
\subfigure{\includegraphics[width=3.5in]{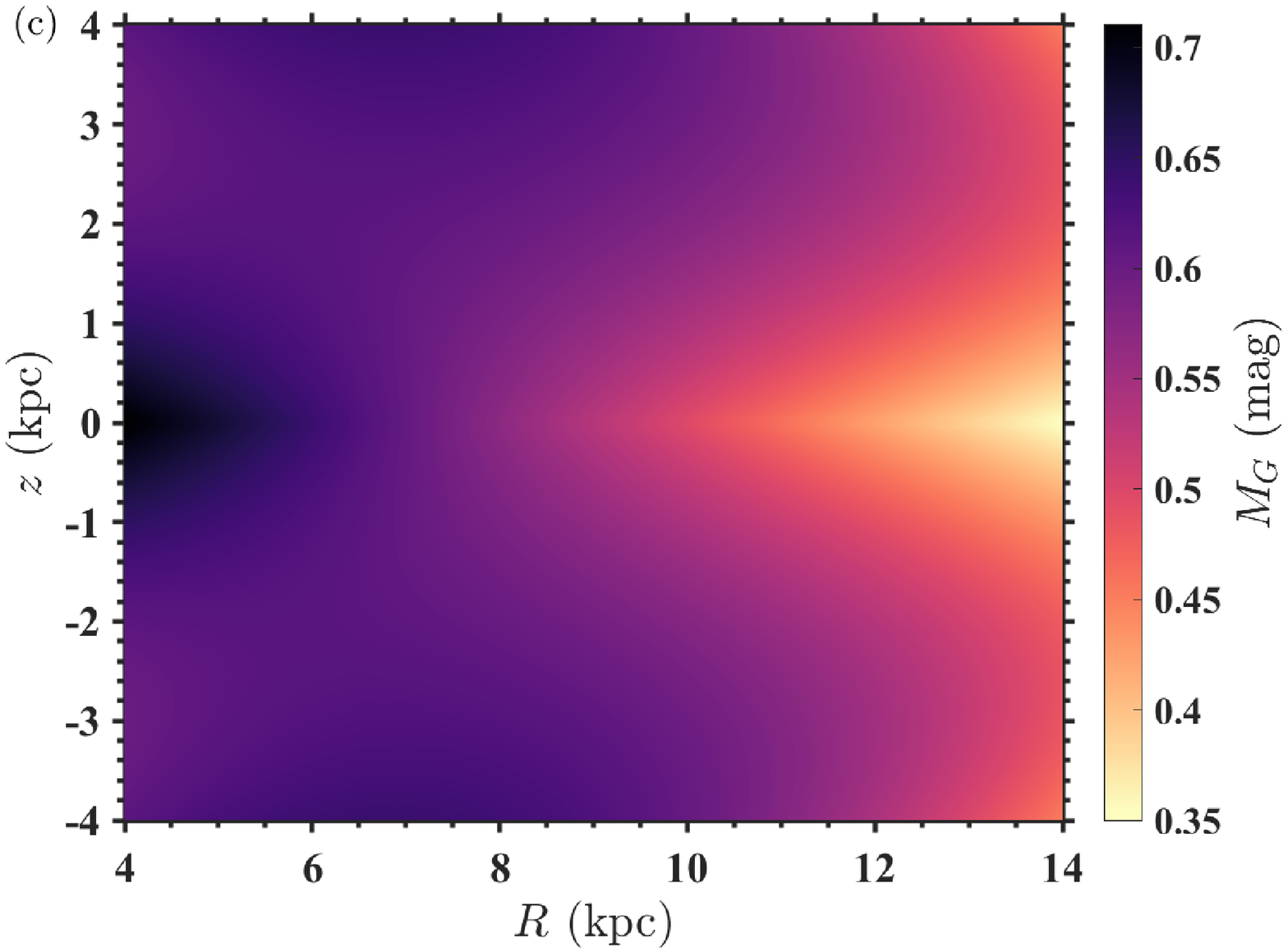}}
\subfigure{\includegraphics[width=3.5in]{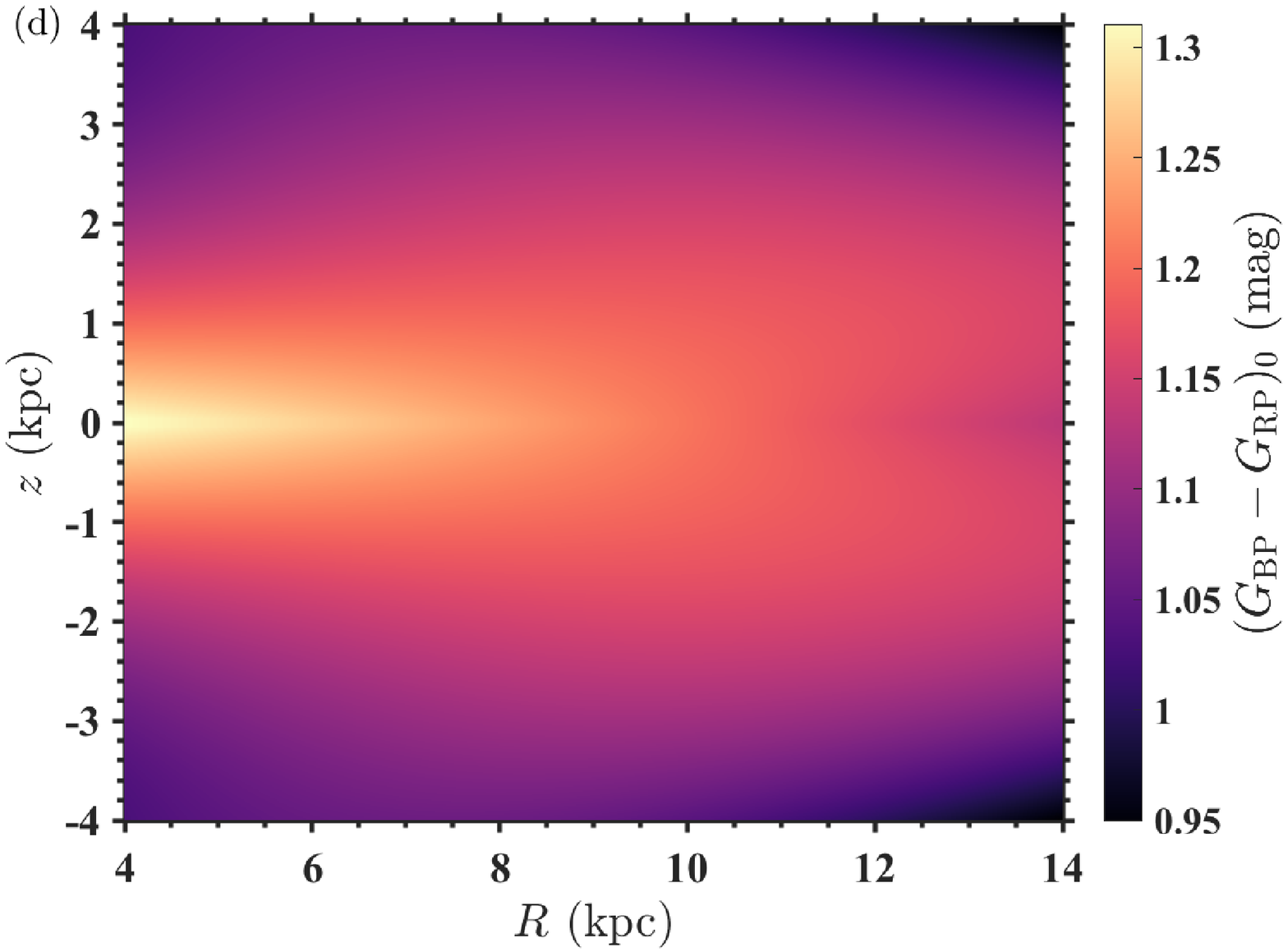}}
\caption{Absolute magnitude and intrinsic color distributions of RCs in the $R-z$ plane. Similar to Figure~\ref{Fig_jk_kmag} but for $\MG$ and $(\GBP-\GRP)_0$. The colorbars denote the values of $\MG$ and $(\GBP-\GRP)_0$. $\MG$ and $(\GBP-\GRP)_0$ derived from observations are plotted in panels (a) and (b), while $\MG$ and $(\GBP-\GRP)_0$ from our parameter maps are plotted in panels (c) and (d).}\label{Fig_bprp_Gmag}
\end{figure*}

\begin{figure*}[ht]
\centering
\includegraphics[width=\hsize]{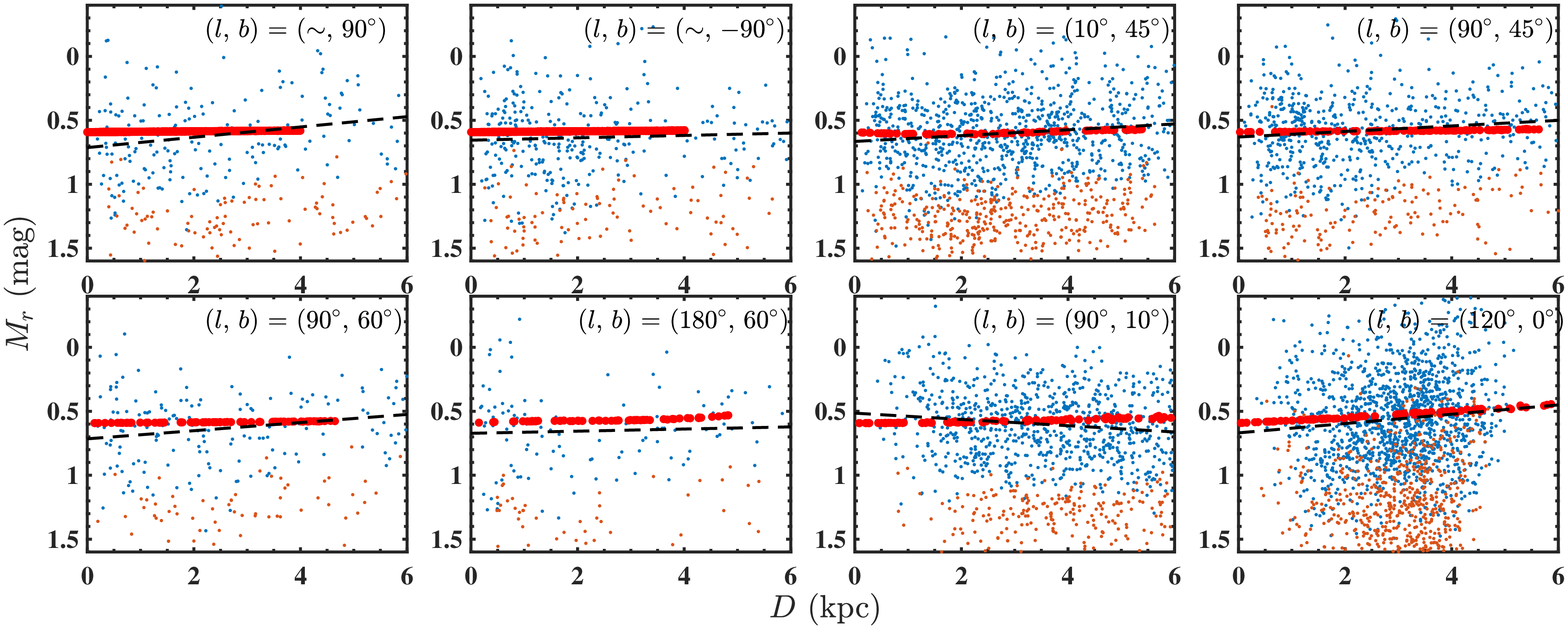}
\caption{The absolute magnitude $\Mr$ distributions of RCs along distance $D$ at different lines of sight.
Blue dots are distribution of the external test RC sample (spectral RCs have been excluded), and orange dots are excluded contamination. 
Large red dots are predicted mean absolute magnitudes from our 3D maps. The black dashed lines are linear fits of the blue dots, indicating the variation of the mean absolute magnitude with distance.}
\label{Fig_pred_r}
\end{figure*}

\begin{figure*}[ht]
\centering
\includegraphics[width=\hsize]{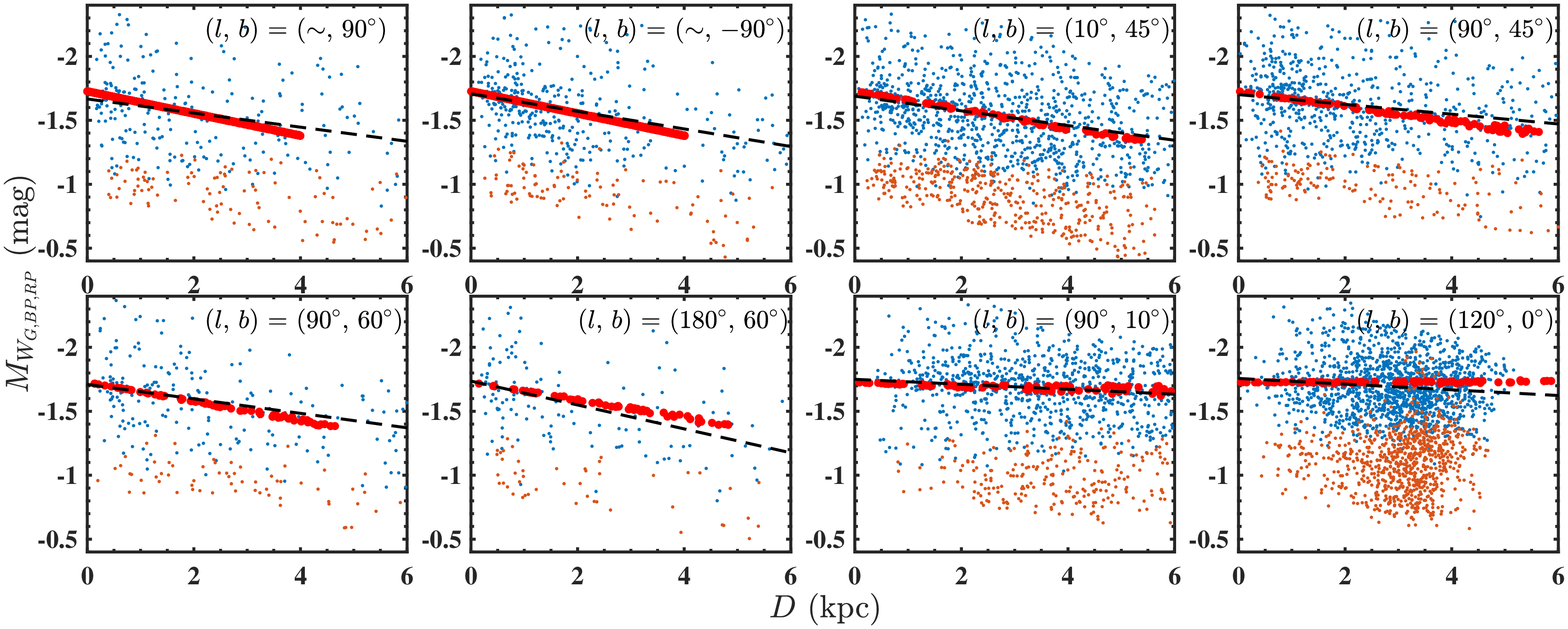}
\caption{Similar diagram as Figure~\ref{Fig_pred_r} but for the absolute Weisenheit magnitude $M_{W_{G,\GBP,\GRP}}$.}
\label{Fig_pred_wg}
\end{figure*}

\begin{figure*}[ht]
\centering
\includegraphics[width=\hsize]{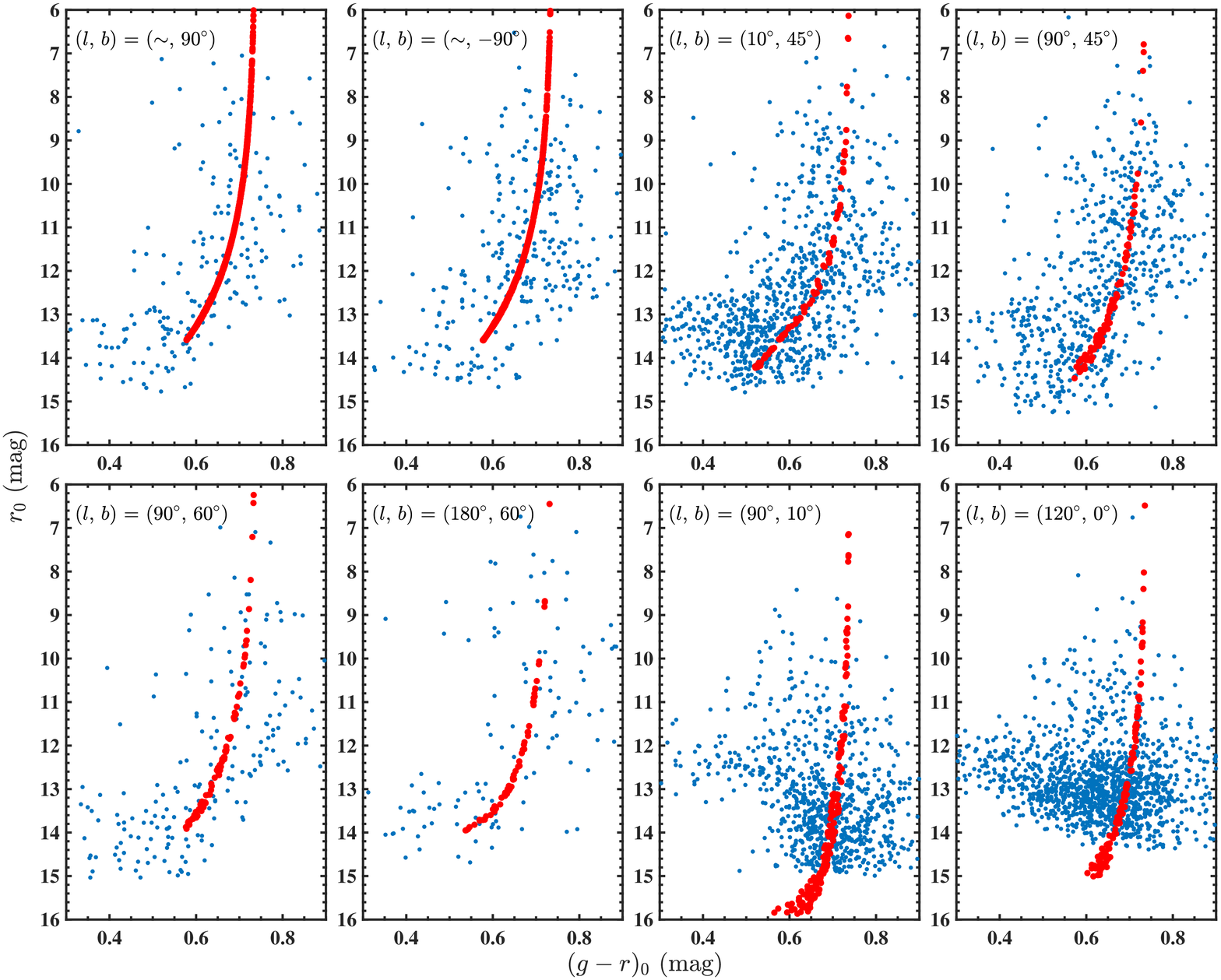}
\caption{The extinction-corrected CMDs of RCs in PS1 bands $r_0$ vs. $(g-r)_0$ at different lines of sight.
Blue dots are distributions of the external test RC sample, while red dots show the distributions of predicted mean magnitudes and mean colors from our 3D maps.}
\label{Fig_pred_gr}
\end{figure*}

\begin{figure*}[ht]
\centering
\includegraphics[width=\hsize]{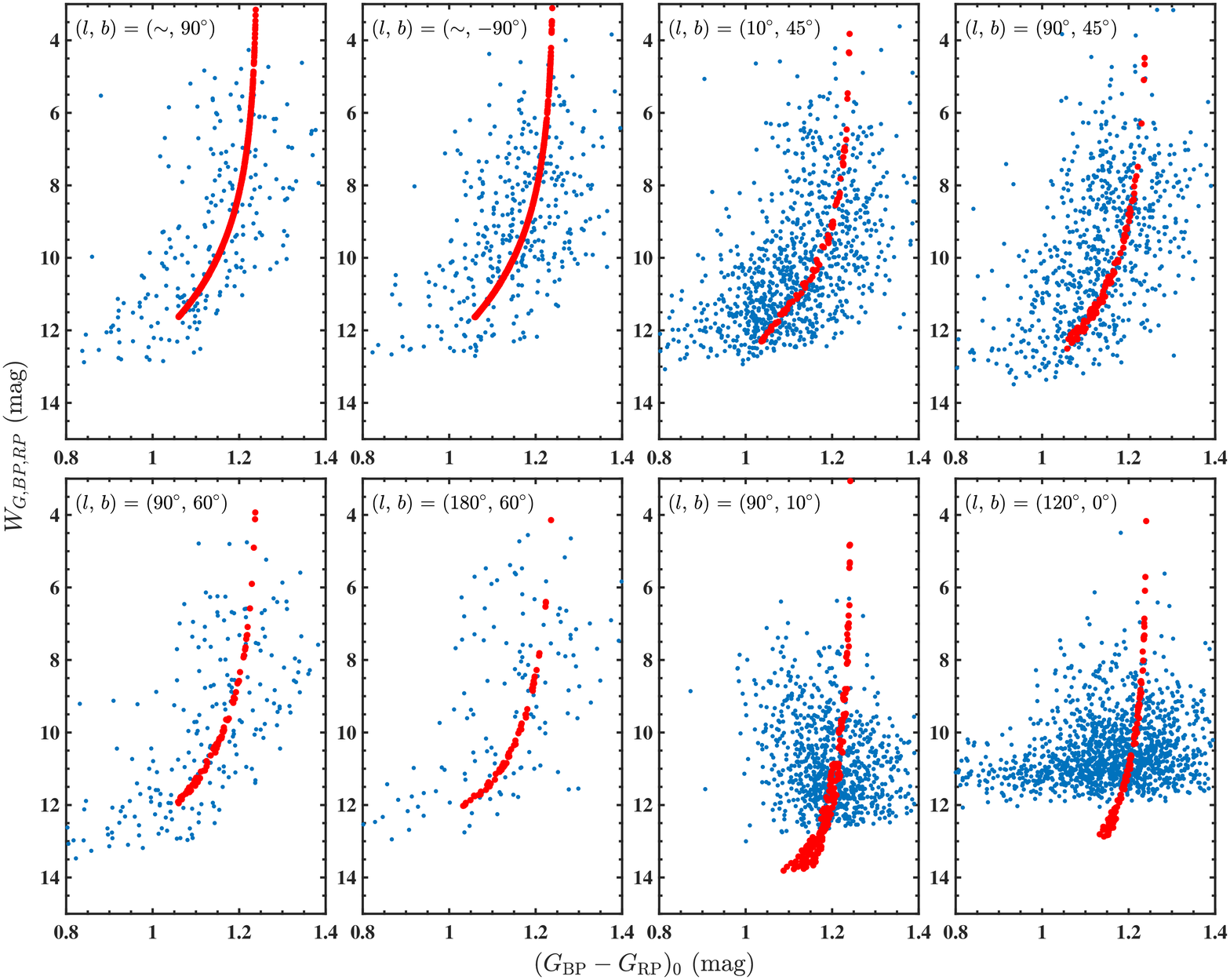}
\caption{Similar extinction-corrected CMDs as Figure~\ref{Fig_pred_gr} but in {\it Gaia} bands $W_{G,\GBP,\GRP}$ vs. $(\GBP-\GRP)_0$.}
\label{Fig_pred_bprp}
\end{figure*}

\end{document}